\documentclass{aa}  

\usepackage{graphicx}
\usepackage{txfonts}
\usepackage[colorlinks=true, citecolor=blue]{hyperref}
\begin{document} 
   
   \title{Fitting spectral energy distributions of FMOS-COSMOS emission-line galaxies at z$\sim$1.6: Star formation rates, dust attenuation\textbf{,} and [OIII]$\lambda$5007 emission-line luminosities}   

  \author{J. A. Villa-V\'elez
          \inst{1}\fnmsep\thanks{E-mail: \url{jorge.villa@lam.fr}}
           \and
          V. Buat\inst{1,2}\and
          P. Theulé\inst{1}\and
          M. Boquien\inst{3}\and
          D. Burgarella\inst{1}}

   \institute{Aix Marseille Univ, CNRS, CNES, LAM, Marseille, France
            %   \email{jorge.villa@lam.fr}
              \and
              Institute Universitaire de France (IUF)
              \and
              Centro de Astronomía (CITEVA), Universidad de Antofagasta, Avenida Angamos 601, Antofagasta, Chile
             }

   \date{Received March 25, 2021; accepted August 9, 2021}

% \abstract{}{}{}{}{} 
% 5 {} token are mandatory

  \abstract
   {We perform a spectral energy distribution fitting analysis on a COSMOS photometric sample covering the ultra-violet up to the far-infrared wavelengths and including emission lines from the Fiber Multi-Object Spectrograph (FMOS) survey. The sample consists of 182 objects with H$\alpha$ and [OIII]$\lambda5007$ emission line measurements lying in a redshift range of $1.40 < \mathrm{z} < 1.68$. We obtain robust estimates of the stellar mass and star-formation rate spanning over a range of $10^{9.5}-10^{11.5}~\mathrm{M_\odot}$ and $10^1-10^3~\mathrm{M_\odot}~\mathrm{yr}^{-1}$ from the Bayesian analysis performed with CIGALE and using continuum photometry and H$\alpha$ fluxes. Combining photometry and spectroscopy gives secure estimations of the amount of dust attenuation for both continuum and line emissions. We obtain a median attenuation of A$_\mathrm{H\alpha} = 1.16\pm0.19$~mag and A$_\mathrm{[OIII]} = 1.41\pm0.22$~mag. H$\alpha$ and [OIII]$\lambda5007$ attenuations are found to increase with stellar mass, confirming previous findings with H$\alpha$. A difference of $57$\% in the attenuation experienced by emission lines and continuum is found to be in agreement with the emission lines being more attenuated than the continuum emission. Implementation of new CLOUDY HII-region models in CIGALE enables good fits of the H$\alpha$, H$\beta$, [OIII]$\lambda5007$ emission lines with discrepancies smaller than $0.2$~dex in the predicted fluxes. Fitting the [NII]$\lambda6584$ line is found challenging due to well-known discrepancies in the locus of galaxies in the [NII]-BPT diagram at intermediate and high redshifts. We find a positive correlation between SFR and L$_\mathrm{[OIII]\lambda5007}$ after correcting for dust attenuation and we derive the linear relation $\log_{10} \mathrm{(SFR/\mathrm{M}_\odot~\mathrm{yr}^{-1})} = \log_{10} (\mathrm{L}_{[\mathrm{OIII]}}/\mathrm{ergs~s^{-1}}) - (41.20\pm0.02)$. Leaving the slope as a free parameter leads to $\log_{10} \mathrm{(SFR/\mathrm{M}_\odot~\mathrm{yr}^{-1})} = (0.83\pm0.06)\log_{10} (\mathrm{L}_{[\mathrm{OIII]}}/\mathrm{ergs~s^{-1}}) - (34.01\pm2.63)$. The spread in the relation is driven by differences in the gas-phase metallicity and ionization parameter accounting for a $0.24$~dex and $1.1$~dex of the dispersion, respectively. We report an average value of $\log\mathrm{U}\approx-2.85$ for this sample of galaxies. Including HII-region models to fit simultaneously photometric data and emission line fluxes is paramount to analyses of upcoming data sets from  large spectroscopic surveys of the future, such as MOONS and PFS.}


   \titlerunning{FMOS-COSMOS star formation rates, dust attenuation, and [OIII]$\lambda$5007 emission-line luminosities}

   \maketitle
%
%-------------------------------------------------------------------

\section{Introduction}

The spectral energy distribution (SED) of a galaxy from the ultra-violet (UV) to the far-infrared (far-IR) reflects its stellar populations and the interplay of their emitted light with gas and dust in the interstellar medium (ISM). The UV to near-infrared (NIR) emission is the result of stellar populations created over a galaxy's lifetime. Dust absorption of the stellar light leaves its imprint on the gas and the stellar continuum and also via the re-emission of the heated dust in the mid and far-IR. The ionized gas component produced by the high-energy flux of massive stars is studied through the nebular emission of the galaxy, providing access to the very recent star formation of massive stars and physical conditions inside HII-regions.

As a consequence, SEDs over a large range of wavelengths and combining photometric data and emission lines give us access to crucial quantities, such as the current star formation rate (SFR), stellar mass, star formation history (SFH) over different timescales,  dust attenuation, dust content, or stellar and gas metallicities. However, the coupling of stars, gas, and dust in a galaxy makes difficult the extraction of the information for each component. The comparison of models predicting the full SED to observed fluxes from the continuum and line emission has been proven to be very powerful to infer these physical parameters in star-forming galaxy populations \citep{Fossati2018, Buat2018, Corre2018, Yuan2019}. For such comparisons, different categories of models are used. The SFH is either modeled with simple analytical functions or non-parametric forms, or it comes from numerical simulations (semi-analytical or hydrodynamical). The interplay between dust and stars is described either using a radiation transfer modeling with more or less complex configurations or with simpler, phenomenological laws that can sometimes be combined with an energy budget \citep[][]{Silva1998, Popescu2000, daCunha2008, Boquien2019}. The nebular component is added either using physical modeling with codes such as CLOUDY or MAPPINGS \citep{Ferland2017, Allen2008}, as in Code Investigating GALaxy Emission (CIGALE\footnote{\url{https://cigale.lam.fr/}}) \citep{Boquien2019}, BayEsian Analysis of GaLaxy sEds (BEAGLE) \citep{Chevallard16}, \textit{ProSpect} package \citep{Robotham2020}, Python code for Stellar Population Inference from Spectra and SEDs (Prospector) \citep{Leja2017, Johnson2021}, Bayesian Analysis of Galaxies for Physical Inference and Parameter EStimation (Bagpipes) \citep{Carnall2018}, \textit{MCSED} \citep{Bowman2020}, or with empirical relations relating different emissions as in PHotometric Analysis for Redshift Estimations (lePHARE) \citep{Arnouts1999, Ilbert2006}. Some SED fitting codes are capable of combining continuum and emission line fluxes when the analysis of data obtained with new facilities will require the handling of very large data sets and the modeling of very different galaxy populations \citep{Ellis2017, Thorne2021}.

The IR emission is crucial with regard to putting constraints on stellar obscuration \citep{Buat2018} and to estimate physical quantities as the SFR of galaxies from the energy budget \citep{Smith2012, Malek2018, Narsesian2019, Dobbels2020}.
We note that the modeling of the infrared (IR) dust emission is not included in all the SED fitting codes; CIGALE includes three different sets of models from \cite{Draine&Li2007}, with updates of \cite{Draine2014}, \cite{Casey2012}, and \cite{Dale2014}, while Prospector, Bagpipes, ProSpect and BEAGLE only include one of the aforementioned sets. Moreover, active galactic nuclei (AGN) templates from \cite{Fritz2006}, \cite{Casey2012}, and \cite{Dale2014} are also available in CIGALE and ProSpect where the former also includes \cite{Andrews2018} templates. Nevertheless, emission line models are still not available for use with the AGN templates. 
Large spectro-photometric surveys on $8$-meter telescopes will
provide thousands to millions of spectra in the deepest photometric fields, such as with the Multi-Object Optical and Near-infrared Spectrograph (MOONS) for the Very Large Telescope (VLT) and the Subaru Prime Focus Spectrograph (PFS), and larger facilities (e.g., Extremely Large Telescope) such as the Multi-Object Spectrograph for Astrophysics, Intergalactic-medium studies, and Cosmology (MOSAIC) . The James Webb Space Telescope (JWST) will push the observations of both rest-frame UV-optical continuum and emission lines to very high redshift in particular the H$\alpha$ and [OIII]$\lambda$5007 emission lines for which different physical conditions and configurations will need to be taken into account in their modeling \citep{Schaerer2009, Wright2015, Wells2015, Marquez2019, Chevallard2019}. While treating both photometric and emission line information simultaneously, it is important to account for the potential contribution of emission lines in photometric bands, which can substantially modify the observed fluxes increasing the uncertainties in the parameter estimations \citep{Schaerer2010, Stark2013, Tang2021}. The coupling of multi-wavelength data sets in the SED fitting including emission lines and a good treatment of the IR emission remains paramount in the work to characterize, derive, and accurately measure  the properties of galaxies.

In this paper, we attempt to simultaneously fit photometric and spectroscopic data with CIGALE in the Cosmic Evolution Survey (COSMOS) field using photometry from \cite{Laigle2016} and emission line fluxes from the Fiber Multi-Object Spectrograph (FMOS-COSMOS) survey \citep{kashino2013, Silverman2015}. This work has multiple aims as we want to test the consistency of both types of data (photometric and spectroscopic) to derive SFR, stellar mass, and dust attenuation for both the lines and continuum. The conservation of the energy budget by CIGALE allows us to get robust estimates of the amount of dust attenuation. In order to get both H$\alpha$ and [OIII]$\lambda5007$ fluxes with good quality, the redshift range is reduced to $1.40<\mathrm{z}<1.68$. We fit the photometry and H$\alpha$ flux emission to constrain the attenuation and correct emission line luminosities for dust effects. We focus our study on understanding the SFR-[OIII]$\lambda5007$ relation, and the influence of metallicity and ionization field. In the FMOS-COSMOS catalog, we have access to other emission lines as H$\beta$ and [NII]$\lambda6584,$ which are used to emphasize the difficulties in HII-region modeling and coupling these models with SED fitting methods.

The data selection is presented in Sect. \ref{sec:Data}. The wavelength coverage of the photometric data goes from the UV to the far-IR including emission lines fluxes from the FMOS-COSMOS survey. We fit the H$\alpha$ emission line flux and photometric data to measure SFR and stellar mass and put our sample in a SFR-M$_\mathrm{star}$ diagram, the SED fitting analysis, and its results are presented in Sect. \ref{sec:CIGALE}. Dust attenuation for both stellar continuum and emission lines are discussed in Sect. \ref{sec:BD_analysis} and we study the SFR-[OIII]$\lambda5007$ relation in Sect. \ref{sec:Calibration}. In Sect. \ref{sec:models}, we add other lines to the fitting analysis and discuss the ability of extended CLOUDY models to fit the full set of data. Finally, in Sect. \ref{sec:conclusions}, we summarize the main parts of this work and conclusions.
This work assumes a flat seven-year Wilkinson Microwave Anisotropy Probe (WMAP7) cosmology \citep{Komatsu2011} with $h = 0.704$, $\Omega_\Lambda = 0.727$, and $\Omega_\mathrm{M} = 0.273$, AB magnitudes, and a \cite{Chabrier2003} initial mass function (IMF).

%--------------------------------------------------------------------
\section{Data selection}\label{sec:Data}

\subsection{UV-to-NIR Photometry}\label{seubsec:photodata}

The COSMOS field covers 2 deg$^2$ and is centered at $\alpha$(J2000) = $10^h0^m27.9^s$ and $\delta$(J2000) = $0^h8^m50.3^s$ \citep{Scoville07}. We adopt the multi-wavelength catalog of \cite{Laigle2016}, COSMOS2015, containing photometry from Galaxy Evolution Explorer near ultra-violet (GALEX NUV) as well as U, B, V, r, i, z, y, J, H, and K$_s$, and the \textit{Spitzer} Infrared Array Camera (IRAC) $3.6$, $4.5$, $5.8$, and $8.0$ $\mu$m photometry from Canada France Hawaii Telescope (CFHT) MegaCam and Wide-field InfraRed Camera (WIRCam), SUBARU Prime Focus Camera (Suprime-Cam) and Hyper Suprime-Cam (HSC), and United Kingdom Infra-Red Telescope (UKIRT) Wide Field Infrared Camera (WFC), and \textit{Spitzer}, respectively. The NUV fluxes overlapping the Lyman break are discarded. This corresponds to sources with $\mathrm{z} > 1.5$ at the GALEX NUV filter's effective wavelength of $\lambda_\mathrm{eff} = 2304.74$~\AA.

\begin{figure}
  \includegraphics[width=0.48\textwidth,clip]{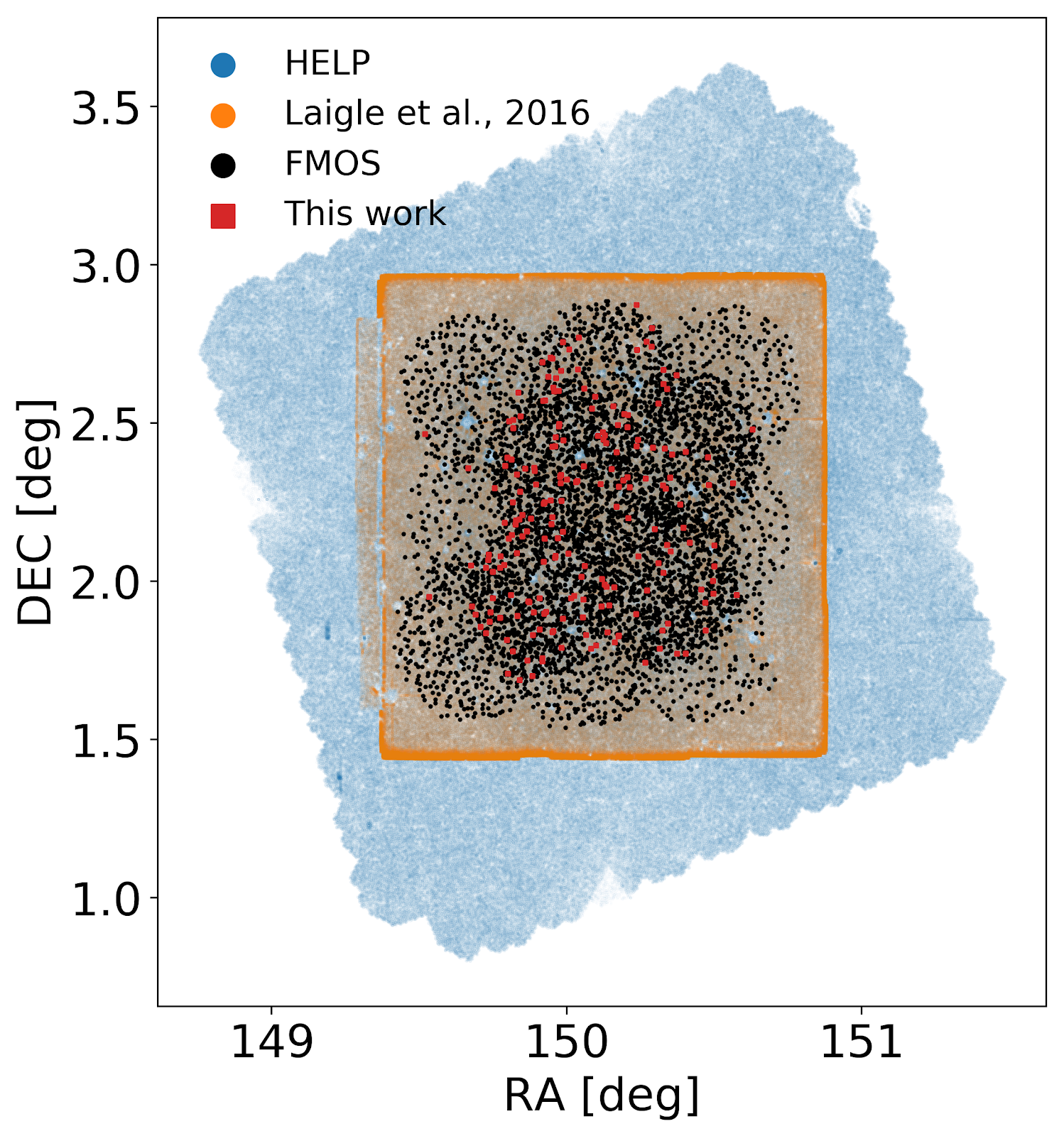}
  \caption{Spectro-photometric sample. In blue, we show the position for objects in the HELP-project catalog. In orange, the main bulk of photometry from the COSMOS2015 catalog \citep{Laigle2016}. In black, the FMOS-COSMOS emission line targets. In red, the final selection for this work.}
  \label{fig0}
\end{figure}

\begin{figure*}
    %\resizebox{\hsize}{!}
     \includegraphics[width=0.99\textwidth,clip]{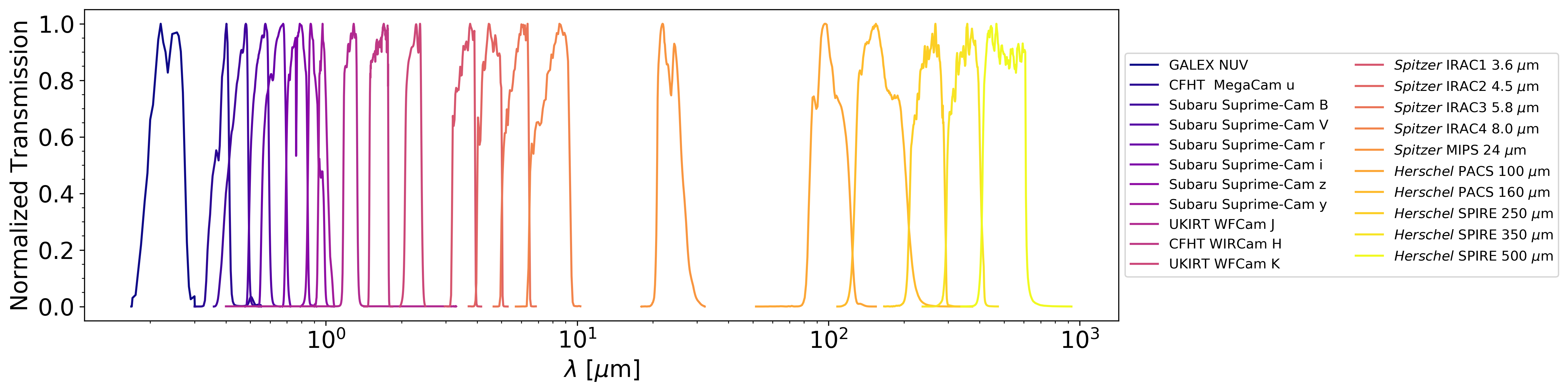}
    \caption{Normalized transmission curves of photometric filters used in CIGALE to estimate the modeled flux densities in each band for the FMOS-COSMOS sample. The list of the filters is shown on the right side.} 
    \label{fig1}
\end{figure*}

\subsection{\textit{Spitzer} MIPS and \textit{Herschel} PACS and SPIRE data}\label{ssec:Herschel_data}

For the IR range, we accessed data from the \textit{Herschel} Extragalactic Legacy Project (HELP\footnote{\url{https://herschellegacyproject.wordpress.com/}}). We used the \textit{Spitzer} Multi-Band Imaging Photometer (MIPS) $24$ $\mu$m, \textit{Herschel} Photodetector Array Camera and Spectrometer (PACS) $100$ and $160$ $\mu$m, and \textit{Herschel} Spectral and Photometric Imaging REceiver  (SPIRE) $250$, $350$, $500$ $\mu$m data products.

\textit{Spitzer} MIPS and \textit{Herschel} PACS and SPIRE fluxes are the result of the HELP XID+ extraction based on the position of priors; XID+ uses a Bayesian probabilistic framework that includes priors to measure fluxes and their uncertainties \citep{Hurley2017}.

Both \textit{Spitzer} MIPS $24$ $\mu$m and \textit{Herschel} PACS $100$, and $160$ $\mu$m were obtained using \textit{Spitzer} IRAC sources from the Spitzer Large Area Survey with the Hyper-Suprime-Cam \citep[SPLASH;][]{Capak2012} as positional priors. In the case of \textit{Herschel} SPIRE $250$, $350$, $500$ $\mu$m, XID+ was run using $24$ $\mu$m priors from \cite{LeFloch2009} catalog with the addition of the Submillimetre Common-User Bolometer Array 2 (SCUBA-2) radio and Atacama Large Millimeter/submillimeter Array (ALMA) data. The products corresponds to \textit{Spitzer} MIPS `dmu26\_XID+MIPS\_COSMOS', \textit{Herschel} PACS `dmu26\_XID+PACS\_COSMOS', and \textit{Herschel} SPIRE `dmu26\_XID+SPIRE\_COSMOS' available in the merged version `dmu32\_COSMOS' at the HELP-repository\footnote{\url{http://hedam.lam.fr/HELP/dataproducts/dmu32/dmu32\_COSMOS/}}. Further information can be found in the Github repository of the project\footnote{\url{https://github.com/H-E-L-P/dmu_products}.}.

\subsection{FMOS-COSMOS emission lines}\label{subsec:FMOS-COMSOS_lines}

The Fiber Multi-Object Spectrograph (FMOS) is an instrument located at the Subaru telescope at the Mauna Kea Observatory in Hawaii. The instrument is a fiber-fed system that allows for wide-field spectroscopy and enabling NIR spectroscopy. We used data from the FMOS-COSMOS survey as described in \cite{kashino2013} and \cite{Silverman2015} for the emission line fluxes. These observations were performed using FMOS in high-resolution (HR) mode. It contains H$\alpha$, H$\beta$, [NII]$\lambda$6584, [OIII]$\lambda$5007, [SII]$\lambda$6717, and [SII]$\lambda$6731\footnote{The [SII]$\lambda$6717 and [SII]$\lambda$6731 data are a private communication of the authors (Kashino \& Silverman)} emission line fluxes.  We selected objects with at least H$\alpha$ and [OIII]$\lambda$5007 flux measurements, limiting the redshift range to $1.38 < \mathrm{z} < 2.6$ in which both lines are observable by FMOS. Aperture corrections were applied following the prescriptions of \cite{Kashino2019}. To guarantee a good level of quality for the H$\alpha$, [NII]$\lambda$6584, H$\beta$, and [OIII]$\lambda$5007 spectral lines, we selected catalog sources with spectral fits obtained using a single Gaussian model \citep{Silverman2015}. We only considered emission lines detected with a signal-to-noise ratio (S/N) larger than 3.
The measurement error in the emission line corresponds to the formal error provided by the line fitting process. The aperture correction value provided for each object is the best value derived from three different methods \citep[see][]{Kashino2019} and is used to correct in-fiber measurements. We keep only objects for which the flux loss is lower than 70\%. To calculate the full uncertainty (fitting process and aperture correction) the formal errors on the observed emission line must be added up in quadrature, with a suggested factor of 1.5 from \cite{Kashino2019}, to account for the uncertainty introduced by the aperture correction.

\subsection{Final sample selection}\label{sec:finalsec}

We started with a sample of $1182108$ galaxies from the COSMOS2015 catalog, which was reduced to $199$ once it was cross-matched to the available IR data from the HELP-project and the FMOS-COSMOS catalog. The different data sets used in this work are represented in Fig. \ref{fig0}. In Fig. \ref{fig1}, the normalized transmission curves of each photometric filter used to estimate the modeled flux densities are presented (see Sect. \ref{seubsec:photodata} and Sect. \ref{subsec:FMOS-COMSOS_lines}).

\begin{figure}
  \resizebox{\hsize}{!}{\includegraphics{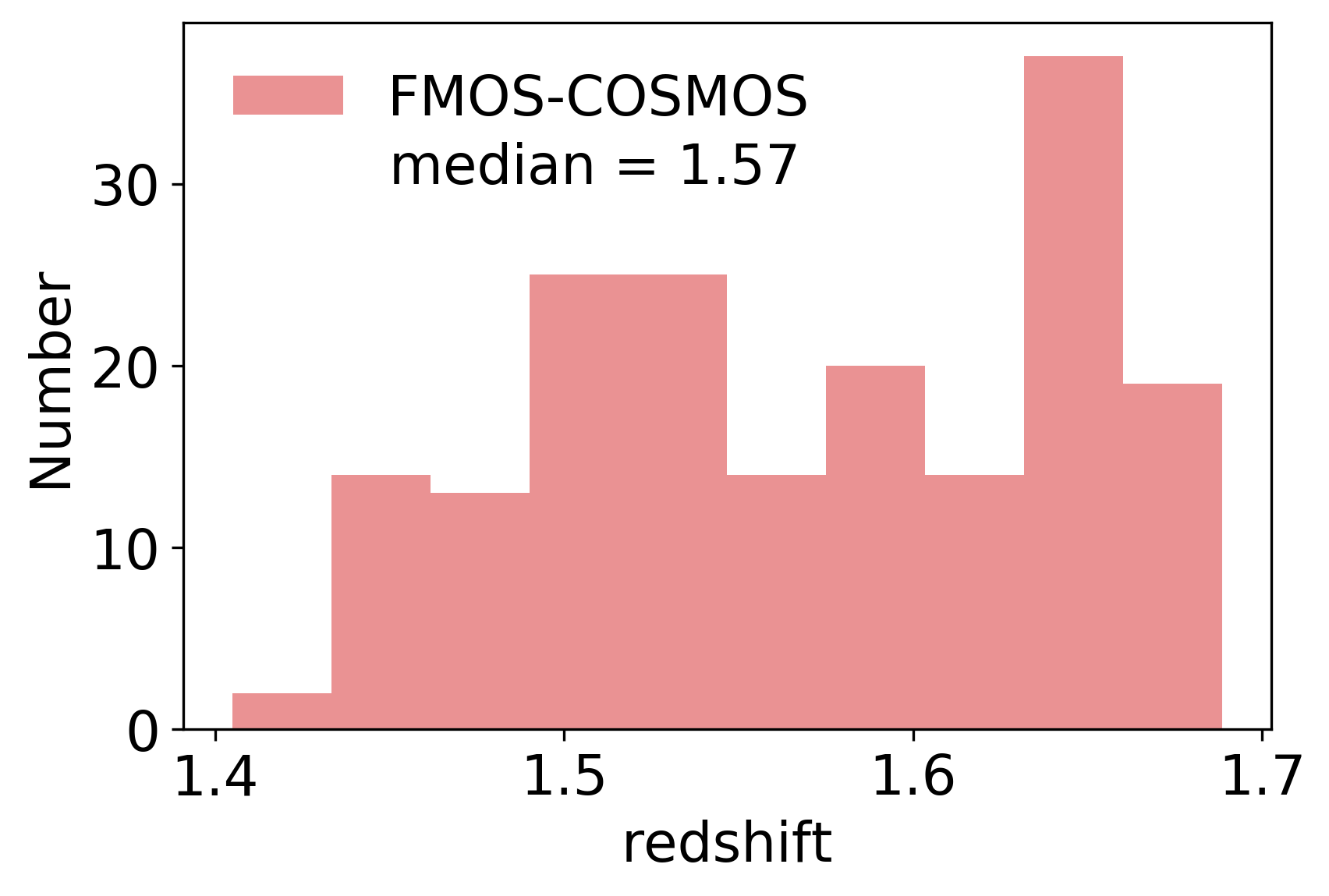}}
  \caption{Redshift distribution of the 182 FMOS-COSMOS selected sources. The sample has a median of 1.57 for the redshift. }
  \label{fig0_1}
\end{figure}

A total of 182 objects were selected on the basis of their flux measurements for both H$\alpha$ and [OIII]$\lambda$5007 with S/N $> 3$. In Fig. \ref{fig0_1}, the final sample redshift distribution covering $1.4 < \mathrm{z} < 1.68$ is presented. This selection is shown in Fig. \ref{fig0} as red dots over imposed on the FMOS-COSMOS sample (black dots), the COSMOS2015 catalog from \cite{Laigle2016}, and the HELP-project data. The number of objects per band is shown in Table \ref{table:1}. S/N drops for \textit{Herschel} PACS and \textit{Herschel} SPIRE bands, leading to 43 and 15 objects with S/N $> 3$ at 100 $\mu$m, and 160 $\mu$m, and 50, 35, and 0 objects for the \textit{Herschel} SPIRE bands $250$, $350$, and $500$ $\mu$m, respectively. Our entire sample has a S/N $> 3$ in the \textit{Spitzer} MIPS 24 $\mu$m, \textit{Spitzer} IRAC1 3.6 $\mu$m, and \textit{Spitzer} IRAC2 4.5 $\mu$m bands, but drops for \textit{Spitzer} IRAC3 5.8 $\mu$m and \textit{Spitzer} IRAC4 8.0 $\mu$m with 95 and 44 objects, respectively. Individual poorly measured photometric data does not affect the SED fitting result because the weight of these bands in the overall fit is small due to their large measurement error.

Most of the sample includes UV-to-MIR photometry and H$\beta$, [NII]$\lambda$6584, [SII]$\lambda$6717, and [SII]$\lambda$6731 emission line fluxes. The quality of the measure of the flux is assessed by the S/N. Of the $139$ objects that have data for the H$\beta$ and [NII]$\lambda$6584 emission lines, $114$ and $112$ sources have a S/N $> 3$, respectively. [SII]$\lambda$6717 and [SII]$\lambda$6731 emission line measurements are of very poor quality, leading only to 22 and 15 objects with a S/N $> 3$. The quality of the H$\beta$ measurements in terms of the Balmer Decrement (BD) will be addressed in Sect. \ref{sec:BD_analysis}. In Sect. \ref{sec:models}, we present a more detailed study of the emission lines.  

We need to avoid any possible contamination due to an AGN contribution. \cite{Kashino2017} flagged AGN in the FMOS-COSMOS sample as point sources with an associated X-ray emission provided by the \textit{Chandra}-COSMOS Legacy survey catalog \citep{Elvis09,Civano2016} with $\mathrm{L}_\mathrm{X-ray} \gtrsim 10^{42}$erg s$^{-1}$ at $0.5-0.7$ keV. Any source flagged as being associated with X-ray emission was discarded from our main sample. Obscured AGN contamination is still possible even if X-ray detected sources are excluded. \cite{Kashino2017} identified 39 objects consistent with obscured (type-II) AGN based on narrow line ratios, with only four of those objects having an X-ray counterpart. Nevertheless, we did not exclude these objects from our analysis because our results are robust enough even if these objects are included probably because they are not extreme cases. We used the color criteria of \cite{Donley2012} to separate AGN from star-forming galaxies and verify that none of our objects lie inside the AGN region.

\begin{table}
\caption{Final sample. Number of objects per band and S/N $> 3$. Emission lines marked with a star are the result of private communication with the authors (Kashino \& Silverman).}              % title of Table
\label{table:1}      % is used to refer this table in the text
\centering                                      % used for centering table
\begin{tabular}{c c c c}          % centered columns (4 columns)
\hline\hline                        % inserts double horizontal lines
Bands & Data  & $> 3\sigma$\\    % table heading
\hline                                   % inserts single horizontal line
                                         % inserting body of the table
    GALEX NUV                     & 51  &  13\\ 
    CFHT  MegaCam u                & 182  &  181\\
    Subaru Suprime-Cam B              & 183  &  183\\
    Subaru Suprime-Cam V              & 183  &  183\\
    Subaru Suprime-Cam r              & 183  &  183\\
    Subaru Suprime-Cam i              & 183  &  183\\
    Subaru Suprime-Cam z              & 183  &  183\\  
    Subaru Suprime-Cam y              & 183  &  182\\
    UKIRT WFCam J                 & 182  &  182\\
    CFHT WIRCam H                 & 182  &  182\\
    UKIRT WFCam K                 & 182  &  182\\
    \textit{Spitzer} IRAC1 3.6 $\mu$m       & 183  &  183\\
    \textit{Spitzer} IRAC2 4.5 $\mu$m       & 183  &  183\\
    \textit{Spitzer} IRAC3 5.8 $\mu$m       & 173  &  95\\
    \textit{Spitzer} IRAC4 8.0 $\mu$m       & 128  &  44\\
    \textit{Spitzer} MIPS 24 $\mu$m         & 183  &  183\\
    \textit{Herschel} PACS 100 $\mu$m       & 183  &  43\\
    \textit{Herschel} PACS 160 $\mu$m       & 183  &  15\\
    \textit{Herschel} SPIRE 250 $\mu$m      & 81   &  50\\
    \textit{Herschel} SPIRE 350 $\mu$m      & 81   &  35\\
    \textit{Herschel} SPIRE 500 $\mu$m      & 81   &  0\\ \\
    \hline
    Emission line\\
    \hline
    FMOS H$\alpha$              & 183  &  182\\
    FMOS H$\beta$               & 139  &  114\\  
    FMOS [NII]$\lambda$6584     & 139  &  112\\
    FMOS [OIII]$\lambda$5007    & 183  &  182\\
    FMOS [SII]$\lambda$6717$^*$ & 40   &  22\\
    FMOS [SII]$\lambda$6731$^*$ & 35   &  15\\

\hline                                             %inserts single line
\end{tabular}
\end{table}

%--------------------------------------------------------------------
\section{SED fitting with CIGALE}\label{sec:CIGALE}

In this section, we present the SED fitting process carried out with the code CIGALE \citep{Boquien2019}. Here, we describe the modules that we use and their direct impact on the analysis of our data set. The structure of the code is addressed by different authors in the literature \citep[e.g.,][]{Ciesla2017, Malek2018, Buat2018, Boquien2019}. CIGALE is based on the energy balance principle in which dust partially absorbs emission of all origins in the UV-optical wavelength range and re-emits this energy in the IR. The code creates millions of models to be compared with the observations and estimates physical parameters of galaxies (such as SFR, stellar mass, dust luminosity, dust attenuation, AGN fraction) while applying a Bayesian statistical analysis approach. 

Different modules were chosen from the CIGALE library to model the star-formation history (SFH), the nebular emission, the dust attenuation, and re-emission. The parameters and their uncertainties are computed using the goodness of fit for all the models as a likelihood-weighted mean and likelihood-weighted standard deviation, respectively. A global indicator of the quality of the fits is given by the reduced $\chi^2$ ($\chi^2_r$). In general, the $\chi^2_r$ must account for the degrees of freedom, however, the non-linearity of the equations \citep{Chevallard16} linking the parameters and their non-trivial dependence on each other makes it difficult to include them. Therefore, the $\chi^2_r$ corresponds to the $\chi^2$ divided by the total number of input fluxes.

The combination of H$\alpha$ fluxes with UV-to-IR data with the preservation of the energy budget allows us to get reliable estimates of dust attenuation (see Sect. \ref{subsec_BC03}) without resorting to other methods such as the Balmer Decrement (BD). It also allows for secure estimations of the SFR to be obtained. Although we have H$\beta$, [NII]$\lambda$6584, [OIII]$\lambda$5007, [SII]$\lambda$6717, and [SII]$\lambda$6731 information, we did not fit these lines due to current discrepancies concerning the HII-region models (see Sect. \ref{sec:models}) and to the poor quality of the data, as presented in Sect. \ref{sec:finalsec} and Table \ref{table:1}. In particular, the H$\beta$ fluxes were discarded either because of their low S/N or a H$\alpha$/H$\beta$ ratio below the canonical value of 2.86 (see Sect. \ref{sec:BD_analysis}). [NII]$\lambda$6584, [SII]$\lambda$6717, and [SII]$\lambda$6731 only satisfy the S/N criterion for 112, and 22, and 15 objects, respectively.
The relevant CIGALE modules selected for this work are introduced in the next sections.

\subsection{Star-formation history}\label{subsect:SFH-cigale}

We used a delayed SFH with the functional form presented in Eq. \ref{eq:0}. This form depends on the time of the star-formation onset, $\mathrm{t}_0$, and the e-folding time of the stellar population, $\tau_\mathrm{main}$. A  recent burst of constant star formation that has been going for at most $70$ Myr is superimposed to the delayed SFH. This form allows us to have a variation of the SFH where the SFR increases from the onset of star-formation until its peak at $\tau_\mathrm{main}$. After that point, the SFR starts declining:

\begin{equation}
    \mathrm{SFR_\mathrm{main}}(\mathrm{t}) \propto \frac{\mathrm{t}}{\mathrm{\tau^2_\mathrm{main}}}\times \exp(-\mathrm{t}/\tau_{main})~\mathrm{for}~0\leq\mathrm{t}\leq\mathrm{t}_0.
    \label{eq:0}
\end{equation}

The delayed component allows us to add a burst that we define with constant amplitude over the last years \citep{Corre2018, Carnall2019, Leja2019, Chevallard2019, Buat2019a}. The amplitude of the burst is fixed by f$_\mathrm{burst}$, defined as a mass ratio between the stars formed during the burst and the total mass of stars. As we select objects with emission lines, we expect the amplitude of the burst to have a direct impact on the derivation of SFR for at least some objects in our sample. The SFR is averaged over $10$ Myr compatible to the timescale traced by H$\alpha$. The input parameters are described in Table \ref{table:2}, where the e-folding time and the age of the main stellar population are left as free parameters as well as the age and the mass fraction of the burst. We used a \cite{Chabrier2003} IMF at solar metallicity $0.02$ for the stellar population. Effects on the variation of the estimated parameters due to a fixed stellar metallicity are discussed in Sect. \ref{sec:Fitting}.

\subsection{Nebular emission lines}\label{subsubsec: Nebular_CIGALE}

CIGALE models the galaxy's emission of the ionized gas by stellar generations as an effective HII-region, encompassing the ensemble of HII-regions and the diffuse ionized gas. The nebular emission lines are pre-computed in the nebular module of CIGALE and re-scaled with the number of Lyman continuum photons from the stellar emission of the model galaxy.

The radiation field intensity is given by the dimensionless ionization parameter $\log\mathrm{U} \equiv \log(\mathrm{n}_\gamma/\mathrm{n_H})$, where $\mathrm{n}_\gamma$ is the  number density of  photons capable of ionizing hydrogen and $\mathrm{n_H}$ the number density of hydrogen. The photo-ionizing field is generated with the single stellar population (SSP) model library of \cite{BC03}, using a constant SFH over 10Myr. In Sect. \ref{sec:models}, we detail the major changes made in modeling nebular emission as compared to the previous version described in \cite{Boquien2019}. The updated version of the HII-region models is used alongside this work to fit photometry and emission line fluxes. Our main purpose is to measure dust attenuation and SFR and to compare the latter to [OIII]$\lambda5007$ luminosities. The new photoionization models allow us to interpret the locus of our galaxies in the excitation diagrams in an attempt to understand the underlying physics of the ISM.

\subsection{Dust attenuation recipe and dust emission}\label{subsec_BC03}

We adopted the recipe proposed by \cite{CF00} (CIGALE module called dustatt\_modified\_CF00; hereafter CF00), where two stellar populations are considered: young stars (age $< 10^7$ years) that are still located in a birth cloud (BC), while older stars (age $> 10^7$ years) have already moved into the interstellar medium (ISM). Both populations experience a different dust attenuation: the emission of young stars goes through the BC and the ISM while the emission of older stars is only attenuated  by dust located in the  ISM. The dust attenuation in the ISM and the BC are both modeled as power laws normalized to the amount of attenuation in the V-band ($\lambda_\mathrm{V} = 0.55\mu \mathrm{m}$): 

\begin{equation}
    \mathrm{A}^{\mathrm{BC}}_\lambda = \mathrm{A}^{\mathrm{BC}}_\mathrm{V}(\lambda/\lambda_\mathrm{V})^{\mathrm{n}^{\mathrm{BC}}},
    \label{eq:1}
\end{equation}
\begin{equation}
    \mathrm{A}^{\mathrm{ISM}}_\lambda = \mathrm{A}^{\mathrm{ISM}}_\mathrm{V}(\lambda/\lambda_\mathrm{V})^{\mathrm{n}^{\mathrm{ISM}}}.
    \label{eq:2}
\end{equation}

The ratio of the attenuation in the V-band experimented by the young and old stars is given by:

\begin{equation}
    \mu = \mathrm{A}^{\mathrm{ISM}}_\mathrm{V}/(\mathrm{A}^{\mathrm{ISM}}_\mathrm{V}+\mathrm{A}^{\mathrm{BC}}_\mathrm{V}),
    \label{eq:3}
\end{equation}
where $\mu$ is considered as a free parameter in our fitting process \citep{daCunha2008, Battisti2019}. Nebular lines coming from the Lyman continuum photons due to very young stars are attenuated as young stars (BC+ISM). The range for each parameter used in the CIGALE's attenuation module is given in Table \ref{table:2}.

The slopes of the two power-laws (i.e., BC and ISM) are fixed to $-0.7$ following \cite{CF00}. This value reproduces the observed mean relation between the IRX and UV spectral slope of nearby starburst galaxies. \cite{LoFaro2017}, \cite{Wild2007}, \cite{daCunha2008}, and \cite{Battisti2019} set n$^{\mathrm{BC}} = -1.3$ to account for effects introduced in the absorption curve due to the optical properties of dust grains as those present in the Milky Way and the Large, and Small Magellanic clouds. However, it has been shown from HII-region studies \citep[see][]{Caplan1986, Liu2013} that grayer values closer to the one chosen in this work are more suitable for reproducing the effective attenuation in dusty galaxies. Letting n$^{\mathrm{BC}}$ free is not suitable because we use only one emission line and its value would be poorly constrained. In Sect. \ref{sec:BD_analysis}, we explore the effects of changing $-0.7$ to $-1.3$ for n$^\mathrm{BC}$. The only free parameters in our recipe are $\mu$ and A$_{\mathrm{v}}^\mathrm{ISM}$. The $\mu$ parameter relates the undergone attenuation in the V-band by the ISM (i.e., old stars, continuum) and the BC+ISM (i.e., young stars, emission lines). Leaving $\mu$ free to vary allows for some variation on the effective attenuation law \citep{Battisti2016, Buat2018, Malek2018, Chevallard2019}. Moreover, introducing this flexibility in the SED fitting process allows for a better quality fit of the H$\alpha$ emission. Variations of the attenuation law will be explored as further work. The dust emission was fitted with the \cite{Draine&Li2007} models based on a set of parameters to constrain the starlight intensity and link dust to star-formation including updates of \cite{Draine2014}. We used these models because we have $24$ $\mu\mathrm{m}$ information making them better suited and more flexible for our purpose.

\subsection{Spectral energy distribution fitting results}\label{sec:Fitting}

\begin{table*}
\caption{CIGALE modules and input parameters used for the SED fitting process as presented in \cite{Boquien2019}}             
\label{table:2}      
\centering 
\scalebox{0.72}{
\begin{tabular}{c c c}     % 3 columns 
\hline\hline       
                      % To combine 4 columns into a single one 
\textbf{Parameter} & \textbf{Symbol} & \textbf{Range}\\ 
\hline                    
    & \textbf{Delayed Star Formation History and Recent Burst} &  \\
e-folding time of main stellar population~(Myr) &  $\tau_\mathrm{main}$ & 1000.0, 3000.0, 4000.0 \\
age of main stellar population~(Myr)    & age$_\mathrm{main}$ & 2000.0, 2500.0, 3500.0, 4000.0 \\  
age of the late burst~(Myr)    & age$_\mathrm{burst}$ & 10.0, 40.0, 70.0 \\
%e-folding time of the late burst & $\tau_\mathrm{burst}$ & 10000.0\\
mass fraction of the late burst population    & f$_\mathrm{burst}$ & 0.0, 0.001, 0.05, 0.1, 0.15 \\  \hline
    & \textbf{Stellar Populations} &  \\ 
    &Stellar population synthesis models from \cite{BC03}& \\ \hline
 initial mass function   & IMF & Chabrier \\  
metallicity    & Z$_\mathrm{star}$ & 0.02 \\ \hline
    & \textbf{Nebular Emission} &  \\  
ionization parameter   & $\log$U & -4.0, -3.5, -3.0, -2.5, -2.0, -1.5, -1.0 \\
gas metallicity   & Z$_\mathrm{gas}$ & 0.008, 0.006, 0.01, 0.02, 0.03, 0.04, 0.05 \\ \hline
    & \textbf{Dust attenuation} &  \\   
    &Templates based on values adopted by \cite{CF00}; \cite{Buat2018}&\\ \hline
ISM attenuation in the V-band~(mag)   & A$_\mathrm{V}^\mathrm{ISM}$ & 0.0, 0.3, 0.5, 0.7, 0.9, 1.1, 1.3, 1.5, 1.7, 1.9 \\  
A$^{\mathrm{ISM}}_\mathrm{V}$/(A$^{\mathrm{ISM}}_\mathrm{V}+$A$^{\mathrm{BC}}_\mathrm{V}$)    & $\mu$ &  0.1, 0.3, 0.5, 0.7, 1.0\\
slope of the attenuation in the ISM   & n$_\mathrm{ISM}$ & -0.7 \\  
slope of the attenuation in the BC    & n$_\mathrm{BC}$ &  -0.7\\ \hline   
    & \textbf{Dust Emission} &  \\
    &Templates based on average values following \cite{Malek2018} and references therein&\\ \hline
mass fraction of PAH    & qPAH & 0.47, 1.12, 1.77, 2.5  \\   
minimum radiation field    & U$_\mathrm{min}$ & 5.0, 10.0, 25.0 \\  
powerlaw slope $\mathrm{dU/dM}\propto\mathrm{U}^\alpha$    & $\alpha$ & 2.0 \\
  fraction illuminated from $\mathrm{U_{min}}$ to $\mathrm{U_{max}}$  & $\gamma$ & 0.02 \\  
\hline                  
\end{tabular}
}
\end{table*}

\begin{figure}
\centering
    \includegraphics[width=0.47\textwidth,clip]{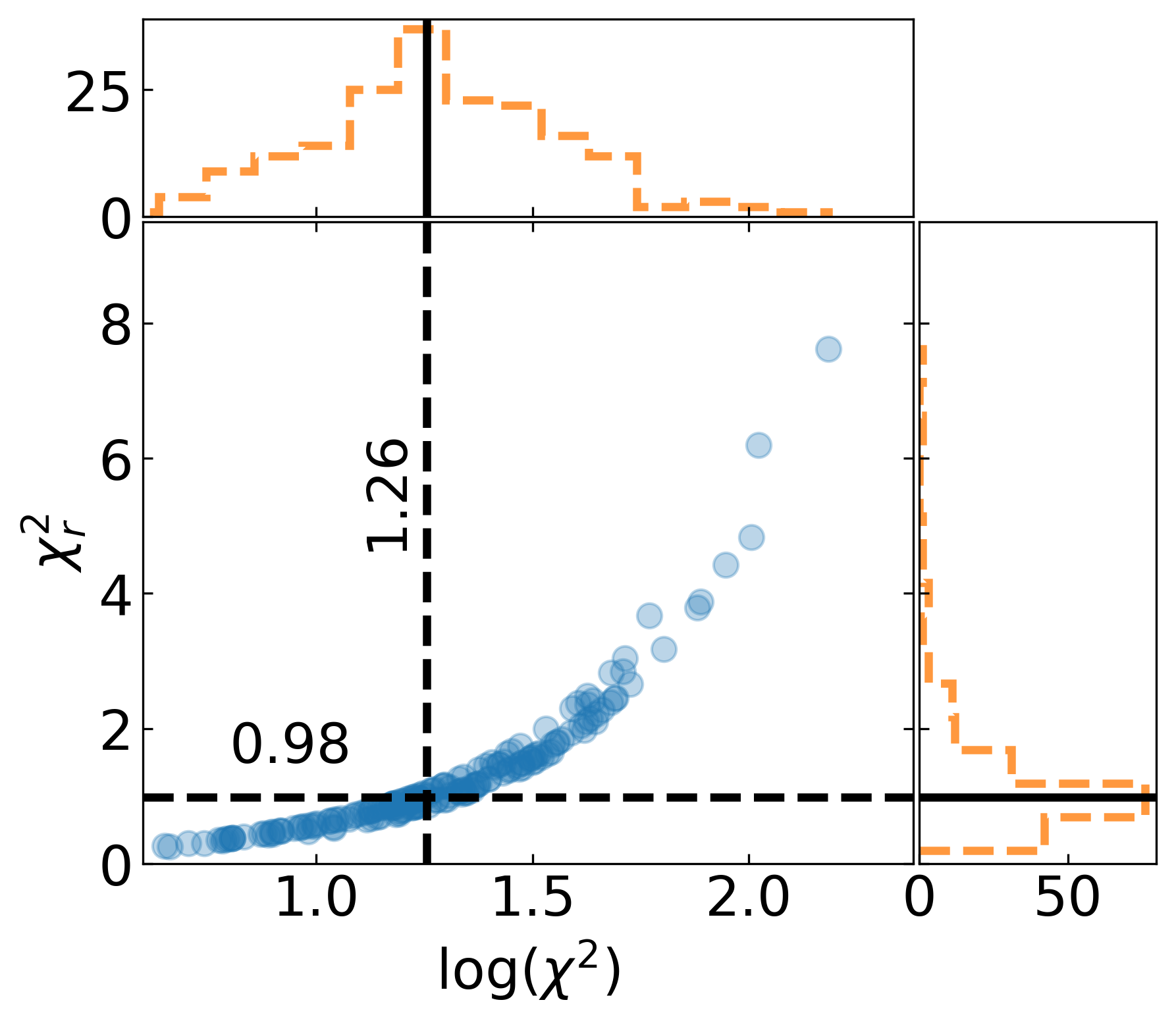}%
    \caption{Reduced $\chi^2_\mathrm{r}$ versus the $\chi^2$. The upper and right panels show the distribution of each parameter while the blue points corresponds to objects in our sample selection. The black lines correspond to the mean value of $0.98$ and $1.26$ in each case.}
    \label{fig6}
\end{figure}

\begin{figure*}
    \centering
     \includegraphics[width=0.99\textwidth,clip]{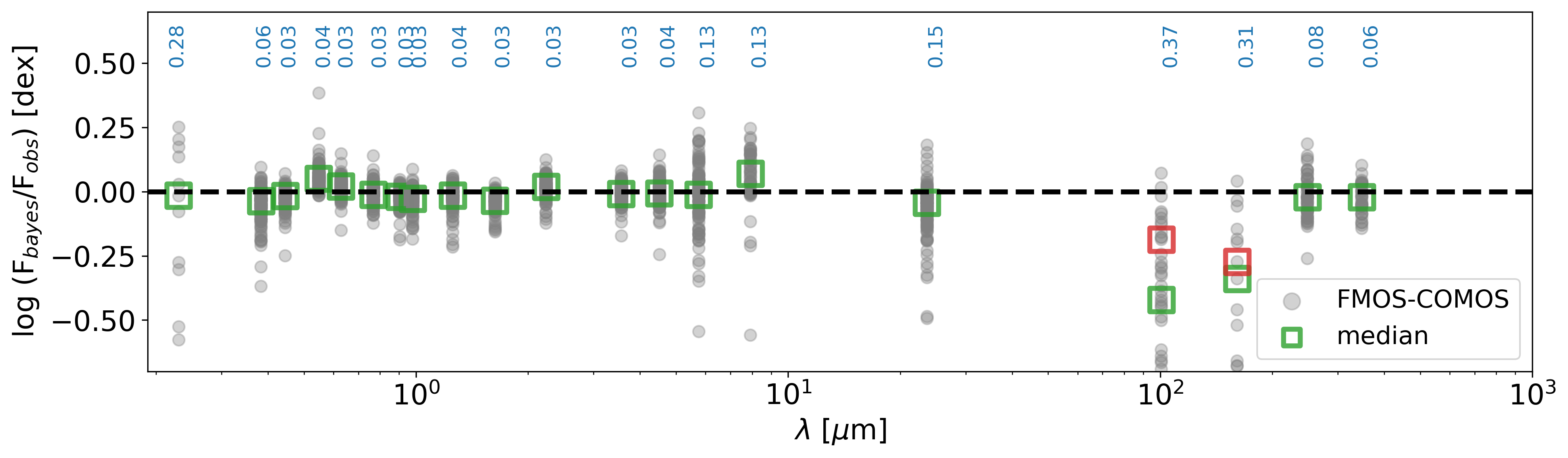}
    \caption{Median distribution of the SED estimated and observed fluxes. On the x-axis, each wavelength corresponds to the photometric bands used in the SED fitting. The difference between the computed Bayesian flux with CIGALE and the observations is shown in the y-axis. Each object in our sample is drawn as a gray dot with a $\mathrm{S/N}>3$ in each band. Median values are shown as green squares. The median values for objects with $\mathrm{S/N}>3$ in both \textit{Herschel} PACS 100 $\mu$m and \textit{Herschel} PACS 160 $\mu$m are presented as red squares. Dispersion on the flux difference at each wavelength is reported on top.}
    \label{fig3}
\end{figure*}

\begin{figure}
  \resizebox{\hsize}{!}{\includegraphics{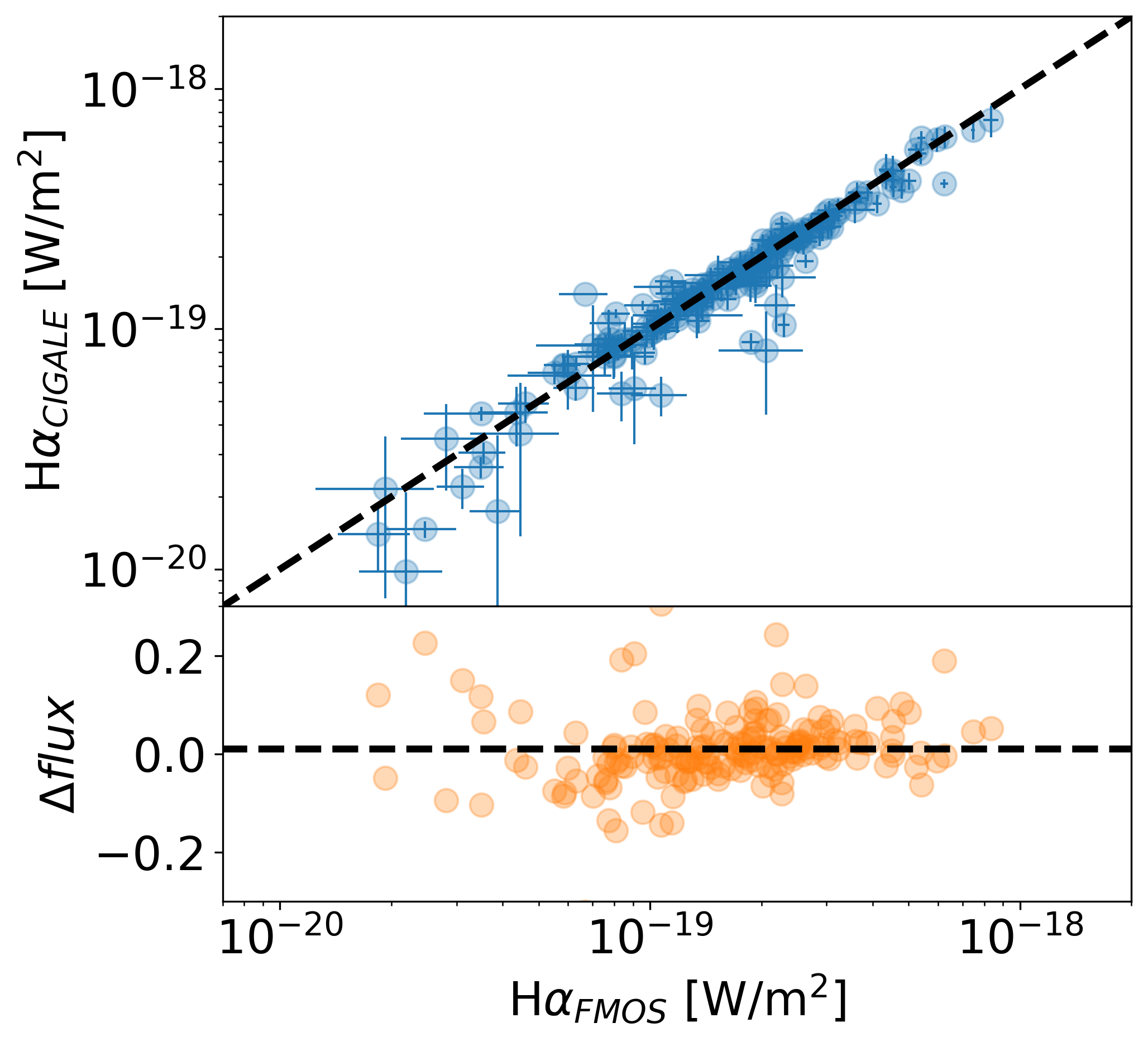}}
  \caption{Fit quality of the H$\alpha$ emission line. The upper panel shows the computed Bayesian flux using CIGALE on the y-axis vs the FMOS measured flux in the x-axis. The 1:1 relation is shown as a dashed line. The relative difference in flux is presented in the lower panel in logarithmic scale. Bayesian and measured errors are shown for CIGALE and FMOS H$\alpha$ fluxes, respectively. The differences between the computed flux with CIGALE and the observed data are not larger than $0.2$~dex.}
  \label{fig4}
\end{figure}

We fit the SEDs of our sample using a set of parameters obtained from previous works and summarized in Table \ref{table:2}. We analyzed the quality of the fits, using the global $\chi^2$ and also comparing the Bayesian estimate of each flux with the observed values. In Fig. \ref{fig6}, the distribution of the obtained $\chi^2$ and $\chi_r^2$ are shown. For each filter, we computed the median value of the difference between the Bayesian estimate from CIGALE and the observed values of fluxes with $\mathrm{S/N}>3$. Their distribution is presented in Fig. \ref{fig3}. The SED-modeled H$\alpha$ fluxes are presented in Fig. \ref{fig4} as a function of the observed flux as well as the flux difference ($\mathrm{H}\alpha_\mathrm{CIGALE}$-$\mathrm{H}\alpha_\mathrm{FMOS}$). 

From Fig. \ref{fig3}, it is clear that the fluxes are well fitted by CIGALE with a flux difference lower than $0.1$~dex and dispersion lower than $0.13$~dex except for GALEX NUV for which the dispersion reaches $0.28$~dex. \textit{Herschel} PACS 100 $\mu$m and  160 $\mu$m are underestimated and exhibit a larger dispersion of $0.37$~dex and $0.31$~dex, respectively.  ONLY  43 objects have $\mathrm{S/N}>3$ in \textit{Herschel} PACS 100 $\mu$m. 23 objects have 100 $\mu$m fluxes larger than their respective flux at 160 $\mu$m. This is an unrealistic configuration at $\mathrm{z}\sim1.5$ with rest-frame fluxes 40 and 60 $\mu$m, respectively. The dispersion of the results in \textit{Herschel} PACS bands can be reduced (e.g., red squares in Fig. \ref{fig3}) if we consider the sample of 9 objects with $\mathrm{S/N}>3$ in both \textit{Herschel} PACS 100 $\mu$m and \textit{Herschel} PACS 160 $\mu$m. Therefore, the overall bad results at \textit{Herschel} PACS bands can be explained as a combined effect of low S/N and non-realistic configuration at the IR bands. We used a deblended IR catalog in the COSMOS field provided by \citep{Jin2018} to check if the situation for these bands could be improved. Both data sets \citep[i.e., HELP and][]{Jin2018} are consistent for our sample with the same trends in terms of the unrealistic configuration of \textit{Herschel} PACS 100 $\mu$m and \textit{Herschel} PACS 160 $\mu$m. The \cite{Jin2018} data have even lower S/N than our final sample selection (described in Sect. \ref{sec:finalsec}). Using HELP data for our sample leads to better results for the fits. CIGALE weighs the data as the inverse square of the error, which means that even if flux with a low S/N is badly fitted, its impact on the overall SED fitting output is minimal. We also fit our sources without the \textit{Herschel} PACS and SPIRE bands to check whether SFR, stellar mass, and attenuation estimations are affected when IR data are excluded. The retrieved estimates with and without IR information are consistent with each other within 1$\sigma$ uncertainty. The impact of \textit{Herschel} SPIRE data on the SFR as compared to the \textit{Herschel} PACS bands is larger (i.e., $0.15$ dex). In both cases, the difference in attenuation is larger ($3\times$) for [OIII]$\lambda5007$ than for H$\alpha$. We kept the \textit{Herschel} data for the whole sample in the overall fit. In Fig. \ref{fig4}, the Bayesian values of the H$\alpha$ fluxes and the FMOS measurements follow the 1:1 relation, indicating that we are able to fit the line flux very well within a $0.2$~dex scatter, as already found by \cite{Buat2018} on a smaller sample in the same field using spectroscopic information from the 3D-HST survey.

\begin{figure}
  \resizebox{\hsize}{!}{\includegraphics{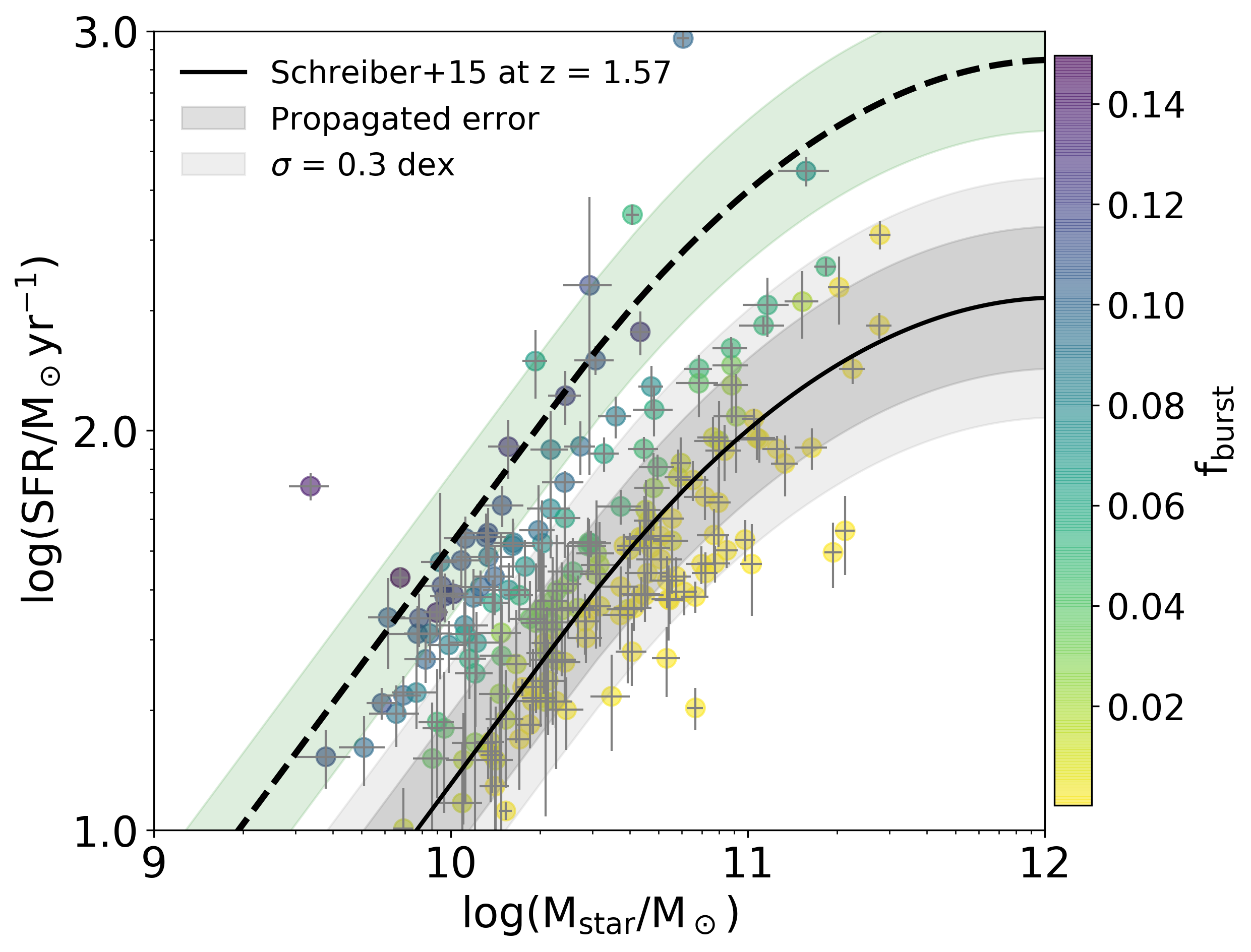}}
  \caption{Star-formation rate versus stellar mass diagram. The dashed line corresponds to the \cite{Schreiber2015} curve evaluated at a median sample redshift of $\mathrm{z}=1.5$. The propagated error dispersion from the fitted errors of the function and a 0.3 dex scatter usually found for the relation in literature are shown as shades. Objects are color-coded by the mass fraction of the late burst population (f$_\mathrm{burst}$). Galaxies four times above the main-sequence (black solid line) are consistent with a SED validated starburst population ($\mathrm{f}_\mathrm{burst}\sim0.1$) within the $1\sigma$ dispersion shown as green shade area. Bayesian error for SFR and stellar mass are reported.}
  \label{fig5}
\end{figure}

Accurate determination of SFRs is crucial as we want to compare them with the [OIII]$\lambda$5007 luminosities. Stellar masses and SFRs retrieved from the SED fitting cover a range of $10^{9.5}-10^{11.5}~\mathrm{M}_\odot$, and $10^1-10^3$~M$_\odot$ yr$^{-1}$, respectively. Current high SFRs can be expected for some objects since our sample is selected on the detection of recombination lines. To confirm the need to add a burst the delayed SFH we performed a fit with a burst amplitude set to $0$. We compared both fits by calculating the Bayesian Information Criterion (BIC) of each of them as introduced by \cite{Schwarz1978} and defined as: $$\mathrm{BIC} = -2\times\ln(\mathrm{likelihood}) + \kappa\times\ln(\mathrm{N}),$$ where $\mathrm{N}$ stands for the number of observations namely, the number of bands fitted, $\kappa$ the number of degrees of freedom, and $\chi^2 = -2\times\ln(\mathrm{likelihood})$ the non-reduced $\chi^2$ obtained from CIGALE \citep{Ciesla2017,Buat2018}. A significant difference and evidence against a model is characterized by the higher BIC. We computed the difference of BIC values $\Delta(\mathrm{BIC})$. As the number of bands fitted is the same in both cases, any difference arises from the number of degrees of freedom ($\kappa_\mathrm{fix} = 14$ and $\kappa_\mathrm{free} = 16$) and the $\chi^2$. $\Delta(\mathrm{BIC}) > 2$ for 90 ($\Delta(\mathrm{BIC}) > 6$ for 69) objects, meaning that for $1/3$ to $1/2$ of our sample the introduction of a burst improves the fit. In Fig. \ref{fig5}, our sample is presented in the SFR-mass diagram color-coded with f$_\mathrm{burst}$. The \cite{Schreiber2015} main-sequence relation for a redshift of $\mathrm{z}=1.5$ is drawn with a 0.3 dex dispersion \citep{Karim2011,Rodighiero2011}. The dispersion computed by error propagation from the originally fitted parameter errors is also shown as a darker shaded contour. The bulk of galaxies is found to lie within the 0.3 dex curves. Galaxies four times above the main-sequence are usually considered starbursting \citep{Sargent2012}. A total of 31 objects in our sample are consistent to be starbursts with f$_\mathrm{burst} > 0.06$ within a 1$\sigma$ dispersion. The median value of f$_\mathrm{burst}$ for these objects is $0.10\pm0.03$. We fit the data using a stellar continuum metallicity of $0.008$ and confirm that the results are consistent with the $0.02$ case. Variations are within $1\sigma$ dispersion for the stellar mass, SFR, and attenuation estimates.

%--------------------------------------------------------------------
\section{Dust attenuation}\label{sec:BD_analysis}
% \label{sec:sample_char}

As noted above, we simultaneously fit  the photometric and spectroscopic data (i.e., H$\alpha$ fluxes) using the \cite{CF00} modified recipe. Emission lines are produced by excited gas inside the BC as a result of the emitted radiation by young stars. Therefore, the lines suffer from attenuation due to the BC and by the ISM surrounding the HII-region.

The combination of the energy budget principle of stellar photons being absorbed and re-emitted by dust in the Mid-and-far-IR with the attenuation recipe allows us to measure the net effect of dust obscuration (i.e., the effective attenuation) in our galaxy sample. CIGALE calculates the resulting attenuation at any wavelength and for any emission line allowing us to obtain the amount of attenuation in the H$\alpha$ and [OIII]$\lambda$5007 emission lines directly from the fits using photometry and H$\alpha$ emission line fluxes. A comparison to the attenuations estimated with the BD method is addressed in this section.

\begin{figure*}
\centering
    \includegraphics[width=0.95\textwidth,clip]{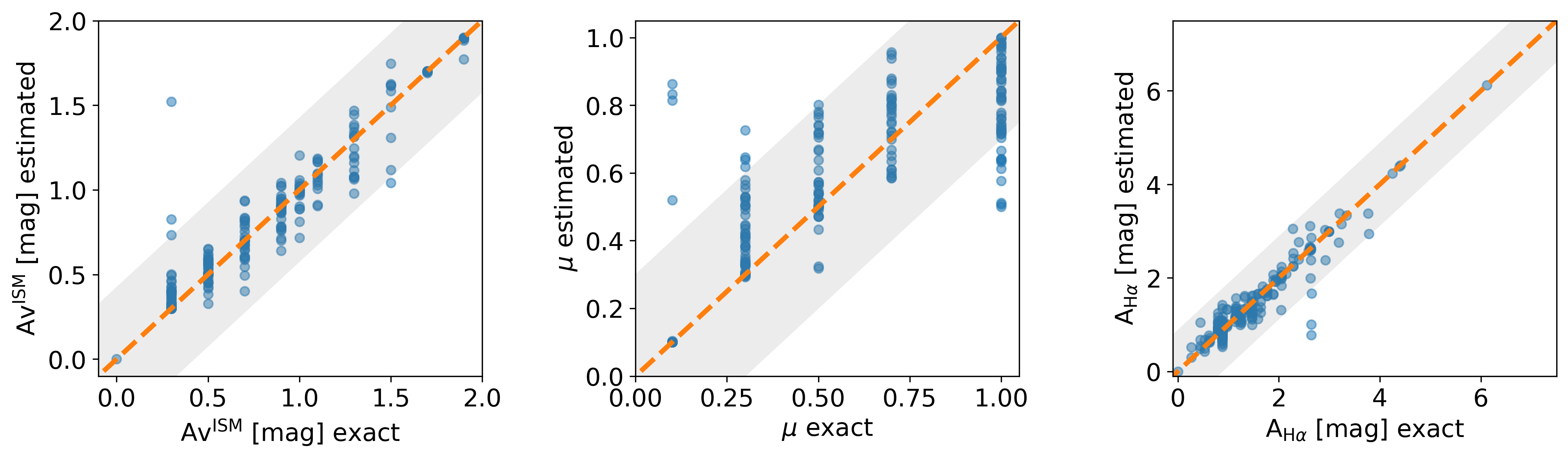}%
    \caption{Mock data sets comparison for the CF00 attenuation recipe intrinsic parameters ($\mathrm{A}_\mathrm{V}^\mathrm{ISM}$ and $\mu$), and the ${\mathrm{H}\alpha}$ attenuation. On the y-axis, the estimated values from the mock catalogs are shown. The x-axis corresponds to the exact value retrieved by CIGALE. The dashed line shows the 1:1 relation. The 1$\sigma$ levels computed as the standard deviation of the estimated parameters are drawn as gray shades.}
    \label{fig8_1}
\end{figure*}

The reliability of the estimation of the effective attenuation has to be checked carefully. We used a CIGALE option to create and analyze mock catalogs and to verify if the main parameters involved in our derivation can be trusted and ensure that our analysis is robust enough to constrain the attenuation. Each flux in the mock catalog corresponds to the value of the best model and is modified by adding a value taken from a Gaussian distribution centered at $0$ and with the same standard deviation as the observations. The mock catalog of fluxes contains the same number of sources and is fitted in the same way as the original catalog. The estimated and exact values for the two free parameters in the \cite{CF00} recipe and the H$\alpha$ attenuation are compared in Fig. \ref{fig8_1}. The value for $\mathrm{A}_\mathrm{V}^\mathrm{ISM}$ is very well reproduced with a Spearman's correlation coefficient of $\rho_\mathrm{s}=0.95$. The $\mu$ parameter is more difficult to constrain even if a clear positive correlation is found between exact and estimated values (Spearman's correlation coefficient of $\rho_\mathrm{s}=0.84$). The ratio of the attenuation in the V-band experienced by the young and old stars characterized by $\mu$ is difficult to constrain and as a consequence $\mathrm{A}_\mathrm{V}^\mathrm{BC}$ is also not very well constrained. Moreover, A$_{\mathrm{H}\alpha}$ is well-constrained with a Spearman's correlation coefficient of $\rho_\mathrm{s}=0.93,$ reflecting the effects of the BC+ISM combination.

\subsection*{Attenuation robustness: Mocks, Balmer decrement, and stellar mass trends}

The retrieved attenuation for the H$\alpha$ and [OIII]$\lambda$5007 emission lines is a crucial output of the SED fitting process for this work. We can use these values to correct line luminosities as the attenuation is constrained by the energy budget method through the UV-to-IR photometry coverage. Nevertheless, the measure of attenuation based on the BD method is widely used to compute the amount of attenuation when both H$\alpha$ and H$\beta$ are available. Attenuation in the H$\alpha$ line can be obtained directly using the observed H$\alpha$ and H$\beta$ fluxes and assuming a given extinction curve following:

\begin{equation}
\mathrm{A}_{\mathrm{H}\alpha} = \frac{-2.5\kappa_{\mathrm{H}\alpha}}{\kappa_{\mathrm{H}\beta} - \kappa_{\mathrm{H}\alpha}}\log_{10}\left(\frac{2.86}{{\mathrm{H}\alpha}/{\mathrm{H}\beta}}\right),
    \label{eq:BD}
\end{equation}

where $\kappa_{\mathrm{H}\alpha}$ and $\kappa_{\mathrm{H}\beta}$ represent a particular extinction curve evaluated at H$\alpha$ and H$\beta$ wavelengths, respectively. The attenuation for the hydrogen lines (i.e., H$\alpha$) can be translated into an attenuation of the [OIII]$\lambda$5007 emission using the same extinction curve at the respective wavelength.

In order to compare BD derived attenuations with CIGALE's attenuation estimates, we selected, from a total of $139$ objects with both H$\alpha$ and H$\beta$ measurements, a sub-sample of 80 objects satisfying the H$\alpha$/H$\beta > 2.86$\footnote{The intrinsic value of $2.86$ is assumed to be consistent with a temperature T$ = 10^4$K and electron density n$_\mathrm{e} = 100\mathrm{cm}^{-3}$ for case B recombination \cite{Osterbrock1989} and does not vary significantly with the physical conditions.} criterion. Here, 23 of the excluded sources have a Balmer decrement that is consistent with $2.86$ within 1$\sigma$ uncertainty. The rest that are not consistent have overestimated H$\beta$ fluxes as compared to H$\alpha$. We did not include any of these sources in the analysis so that we could remain consistent with the photoionization models implemented in CIGALE. Excluding values below the canonical value of the ratio allows us to exclude negative attenuation values inconsistent with the case B recombination assumption for the emission lines. We computed the attenuation applying Eq. \ref{eq:BD}, along with the Milky Way curve ($\kappa_{\mathrm{H}\alpha}$ = 2.52; and $\kappa_{\mathrm{H}\beta}$ = 3.66; $\kappa_{\mathrm{[OIII]}}$ = 3.52) of \cite{Cardelli1989} updated by \cite{ODonnell1994}. For comparison and to show the effects in the derived attenuation due to the choice of an extinction curve, we also considered the \cite{Calzetti2000} (C00) curve ($\kappa_{\mathrm{H}\alpha}$ = 3.33; and $\kappa_{\mathrm{H}\beta}$ = 4.59; $\kappa_{\mathrm{[OIII]}}$ = 4.46)\footnote{This curve is only valid for stellar continuum and not appropriate to describe nebular emission line attenuation \citep{Pannella2015, Reddy2015, Puglisi2016, Battisti2016, Theios2019, Qin2019, Shivaei2020}.}. 

\begin{table}
\caption{H$\alpha$ and [OIII]$\lambda5007$ attenuation comparison for selected sample of 80 galaxies. The median values of attenuation and uncertainties are presented. The single-star value is computed using n$^\mathrm{BC} = -1.3$ instead of $-0.7$. The dagger value is obtained including H$\alpha$, H$\beta$, and [OIII]$\lambda$5007 in the fit (see Sect. \ref{subsec:fit_all}).}              % title of Table
\label{table:3}      % is used to refer this table in the text
\centering                                      % used for centering table
\begin{tabular}{c c c c}          % centered columns (4 columns)
\hline\hline                        % inserts double horizontal lines
Method & A$_{\mathrm{H}_\alpha}$~[mag] & A$_{[\mathrm{OIII}]}$~[mag] &  A$_{[\mathrm{OIII}]}$-A$_{\mathrm{H}_\alpha}$ \\    % table heading
\hline                                   % inserts single horizontal line
                                         % inserting body of the table
    CIGALE fit                    & 1.16$\pm$0.19 & 1.41$\pm$0.22 & 0.25\\
    CIGALE fit$^\star$                    & 1.17$\pm$0.14 & 1.53$\pm$0.20 & 0.36\\ 
    CIGALE fit$^{\dagger}$                    & 1.11$\pm$0.18 & 1.34$\pm$0.22 & 0.23\\ 
    BD (MW)         & 0.82$\pm$0.19 & 1.14$\pm$0.26 & 0.31\\
    BD (C00)        & 0.97$\pm$0.22 & 1.30$\pm$0.30 & 0.33\\
    % CIGALE inferred  (MW)         & 0.61$\pm$0.22 & 0.85$\pm$0.31 \\  
    % CIGALE fit  (All)             & 1.48 & 1.78 \\  
\hline                                             %inserts single line
\end{tabular}\\
{\raggedright $^{\star}$ with n$^\mathrm{BC} = -1.3$ \par}
{\raggedright $^{\dagger}$including H$\alpha$, H$\beta$, and [OIII]$\lambda$5007 in the fit \par}
\end{table}

\begin{figure}
\centering
    \includegraphics[width=0.5\textwidth,clip]{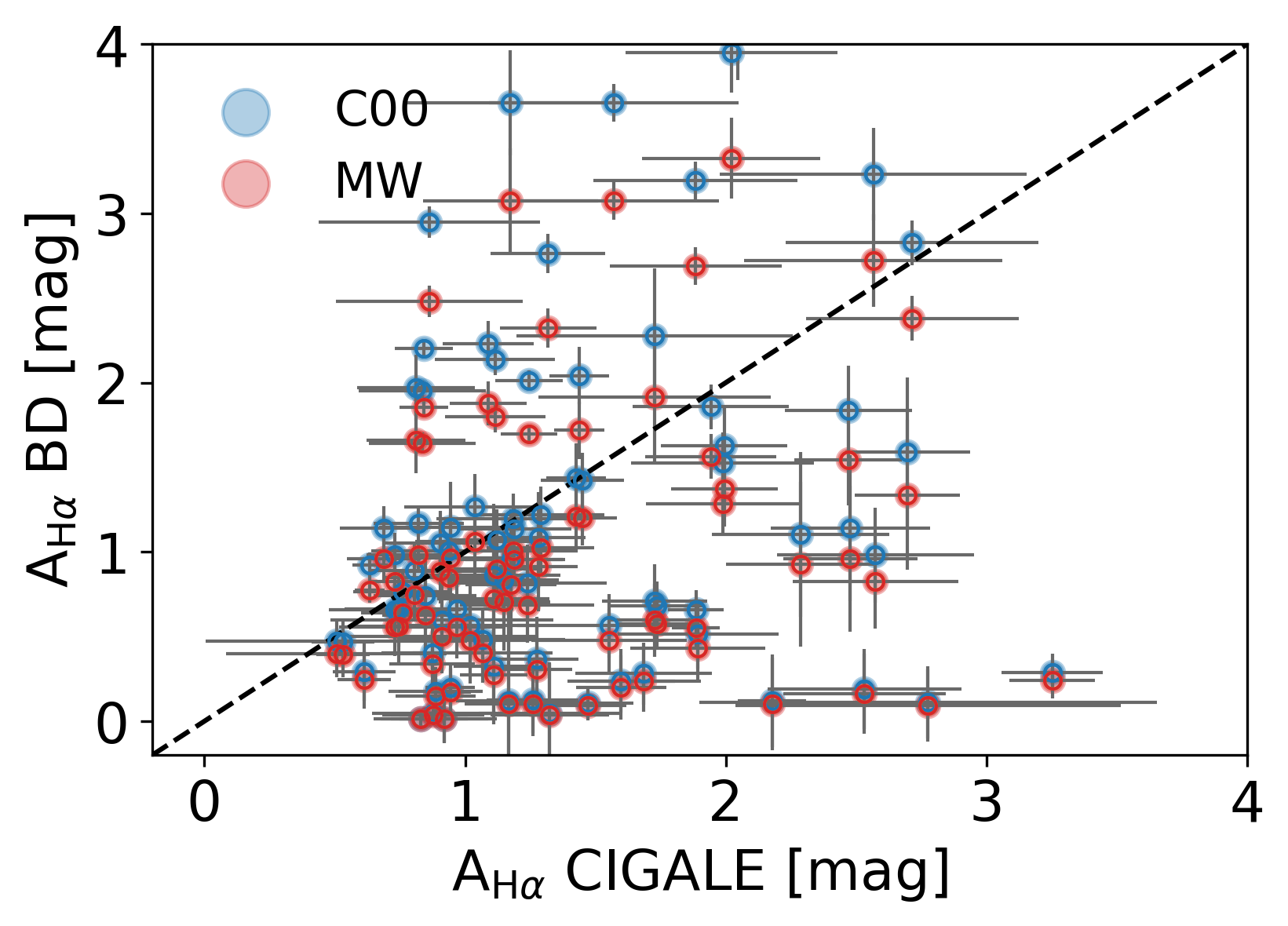}%
    \caption{Balmer Decrement and CIGALE's A$_{\mathrm{H}\alpha}$ comparison. Blue and red dots correspond to the BD derived attenuations computed using Eq. \ref{eq:BD} and a \cite{Calzetti2000} and Milky Way extinction curve, respectively. The H$\alpha$ attenuation obtained directly from CIGALE is shown in the x-axis. The 1:1 relation is shown as a dashed black line.}
    \label{figBD}
\end{figure}

\begin{figure}
\centering
    \includegraphics[width=0.5\textwidth,clip]{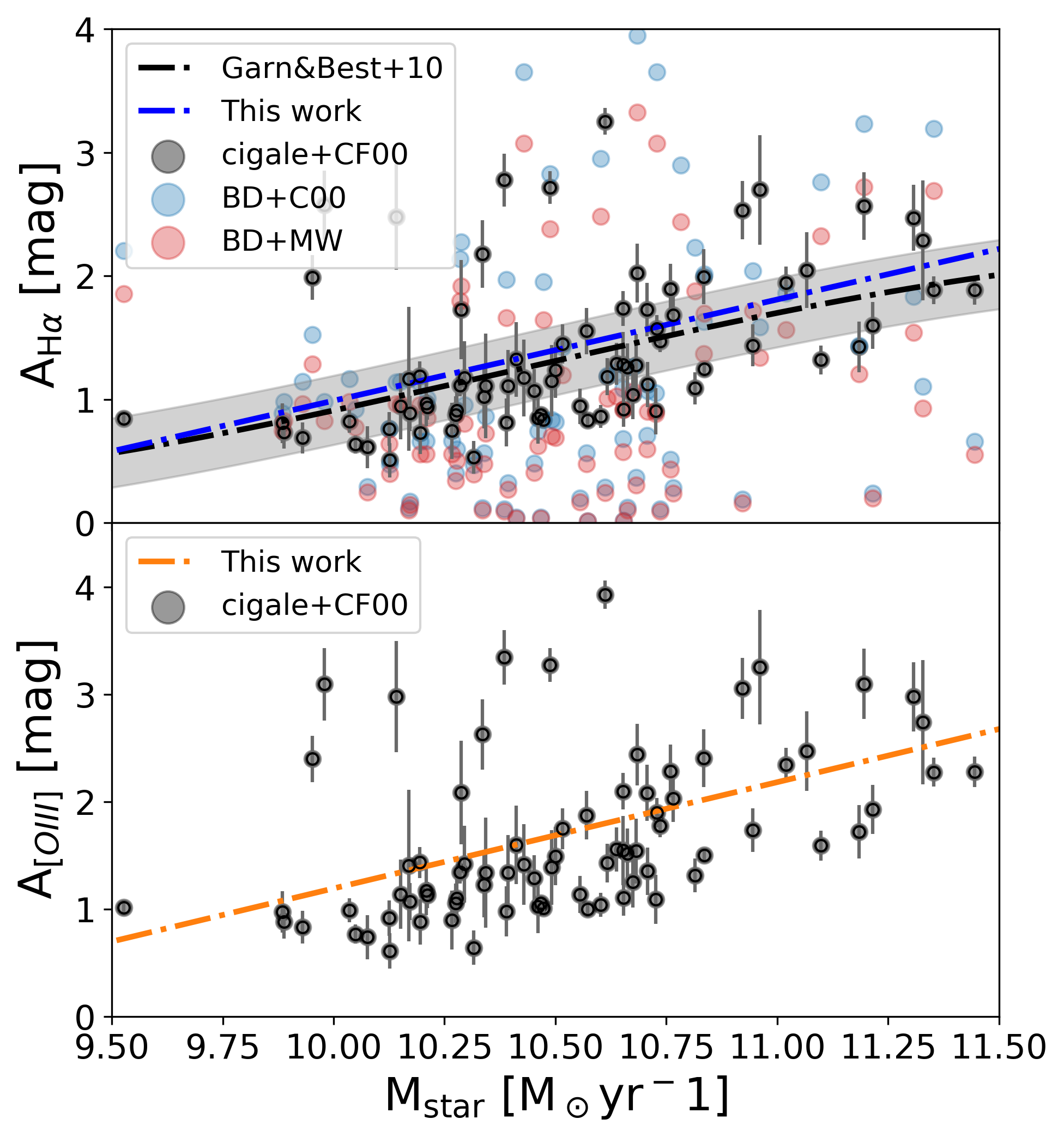}%
    \caption{Variation of the A$_{\mathrm{H}\alpha}$, and A$_{\mathrm{[OIII]}}$ attenuation with the stellar mass. The attenuation computed with CIGALE is presented as black dots with their respective uncertainties. Milky Way and C00 are shown in red and blue, respectively without uncertainties for clarity on the upper panel. The dashed line corresponds to the A$_{\mathrm{H}\alpha}$-M$_\mathrm{star}$ relation obtained by \cite{GarnBest2010} in the local universe. The shaded area indicates the relation's $\pm$1$\sigma$ distribution width. The blue line represents a linear fit to the black dots. The lower panel corresponds to the attenuation retrieved for the [OIII]$5007$ emission line. A linear fit is shown in orange.}
    \label{figGarn}
\end{figure}

In Table \ref{table:3}, we present median attenuation values for H$\alpha$ and [OIII]$\lambda$5007, as well as the A$_{[\mathrm{OIII}]}$-A$_{\mathrm{H}_\alpha}$ ratio. All values are derived with a fixed BC slope n$^\mathrm{BC} = -0.7$ except for the starred case in which n$^\mathrm{BC} = -1.3$ was implemented. The fit with n$^\mathrm{BC} = -1.3$ is as good as the n$^\mathrm{BC} = -0.7$ case. In Fig. \ref{figBD}, the derived attenuations using the BD method are compared to CIGALE's estimates. We check the distribution of CIGALE's attenuation in the [OIII]$\lambda5007$ emission line as a function of stellar mass and fit a linear relation shown in orange in the lower panel of Fig. \ref{figGarn}. Thus, we obtain:$$\mathrm{A_{[OIII]}} = (1.19\pm0.15)+(0.98\pm0.22)\log_{10}\mathrm{(M_*/10^{10}M_\odot)}.$$

We go on to explore the effects of using this relation to correct [OIII]$\lambda$5007 luminosities in Sect. \ref{subsec:Abscence_AOIII}. 

Applying either the Milky Way or C00 reddening curve leads to lower median attenuations than those derived using \cite{CF00} recipe in CIGALE, as shown in Fig. \ref{figGarn}. CIGALE's attenuation estimates are globally in agreement with the A$_{\mathrm{H}\alpha}$-mass relation of \cite{GarnBest2010}. BD derived attenuations (i.e., blue and red dots) are not entirely consistent with this relation placing the majority of the objects below the relation. Objects above \cite{GarnBest2010} relation have a SFR $> 10^{1.5}$ M$_\odot$ yr$^{-1}$ but they are not classified as starbursting from our analysis. The H$\alpha$ attenuation increases with stellar mass in agreement with previous works \citep[][]{Garn2010, GarnBest2010, Ibar2013, Zahid2017, Shivaei2020}, following the \cite{GarnBest2010} relation and confirming that in more massive galaxies nebular regions have a higher attenuation \citep{Koyama2019}. This also confirms that this relation is not just valid for the local universe but  at higher redshift as well. From the attenuation ratio between [OIII]$\lambda$5007 and H$\alpha$ presented in Table \ref{table:3}, we can see that our results using CF00 lead to a flatter effective attenuation curve \citep{Chevallard16, LoFaro2017, Buat2018}, as compared to the Milky Way or C00 recipe. The difference arises from the choice in fixing both slopes in the CF00 to the same value of $-0.7$. We performed the same analysis using n$^\mathrm{BC}=-1.3$ and verified that the effective curve is steeper than in the n$^\mathrm{BC}=-0.7$ case and closer (and even slightly steeper) than the Milky Way and C00 case.

\begin{figure}
\centering
    \includegraphics[width=0.5\textwidth,clip]{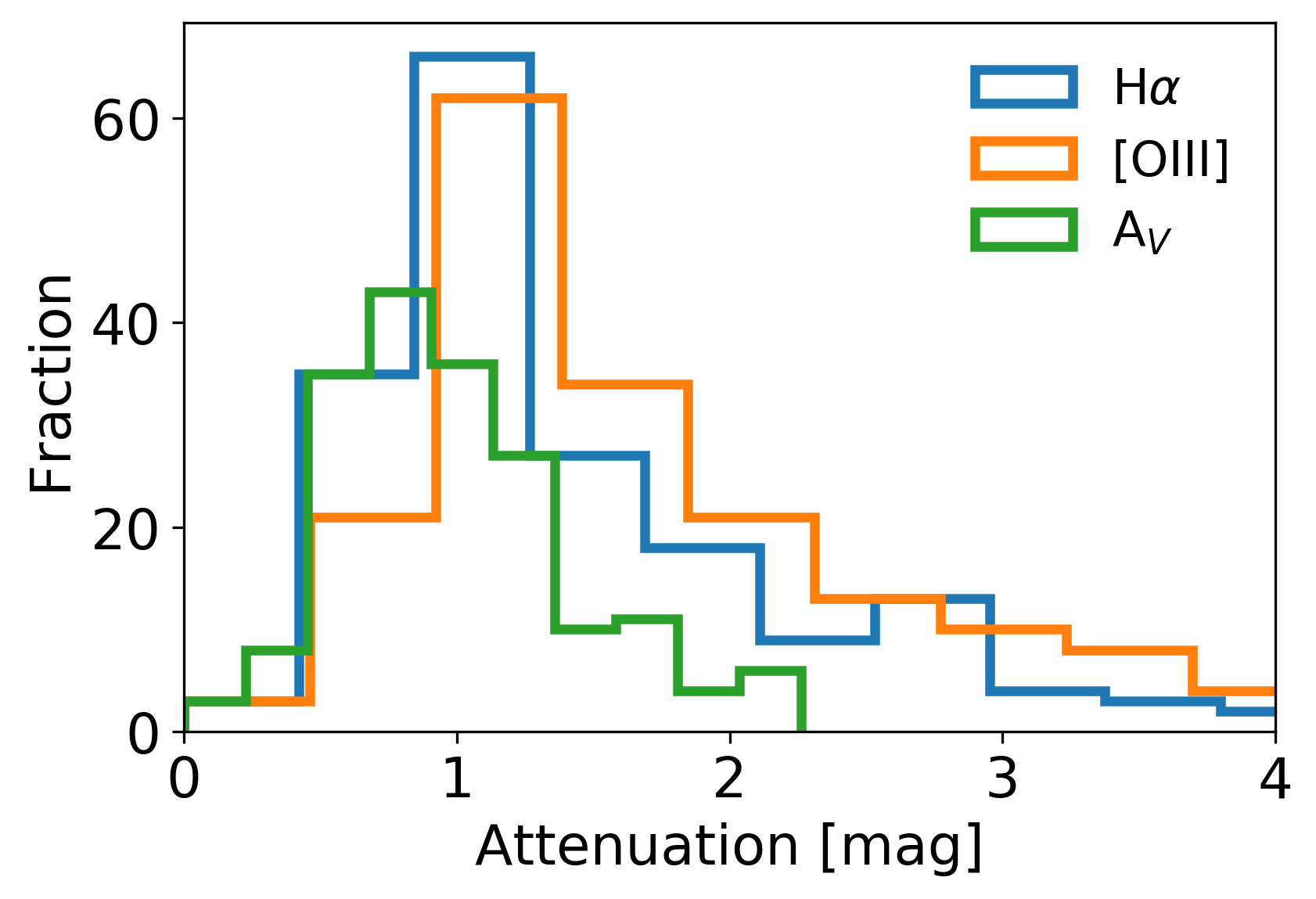}%
    \caption{Distribution of the total attenuation in the H$\alpha$ and [OIII]$\lambda5007$ emission lines and the V-band are shown in blue, orange, and green, respectively. The amount of attenuation obtained for the emission lines is similar.}
    \label{fig7}
\end{figure}

The attenuation distributions obtained for the entire sample from the SED fitting are shown in Fig. \ref{fig7} for the H$\alpha$ and [OIII]$\lambda$5007 emission lines as well as for the V-band. The nebular line attenuation distribution in both cases follows a similar behavior as both lines are produced inside the HII-regions and they are attenuated in a similar way (inherent to CIGALE modeling). Stellar continuum attenuation as traced by A$_\mathrm{v}$ is lower. The differential attenuation suffered by young and old stars is measured by the $\mu$ parameter in the CF00 recipe as explained in Sect. \ref{subsec_BC03}. The original value proposed by \cite{CF00} corresponds to $0.3$ for the nearby universe. We measured a larger value of $\mu = 0.57 \pm 0.14$. In general, higher values are found at higher redshift \citep[see][and references therein]{Buat2018}. This is in agreement with nebular emission being more attenuated than the stellar continuum.

%--------------------------------------------------------------------

\section{\texorpdfstring{$\mathrm{[OIII]}\lambda5007$, $\mathrm{H}\alpha$}{OIII, Ha  fluxes and SFR measurements}  fluxes, and SFR measurements}\label{sec:Calibration}

In Sect. \ref{sec:BD_analysis}, we show that we constrain well the attenuation in our sample with the multi-wavelength plus IR bands coverage and the inclusion of the H$\alpha$ emission line, and the CF00 recipe, which introduces a differential attenuation between young and old stellar populations.

\begin{figure}
\centering
    \includegraphics[width=0.48\textwidth,clip]{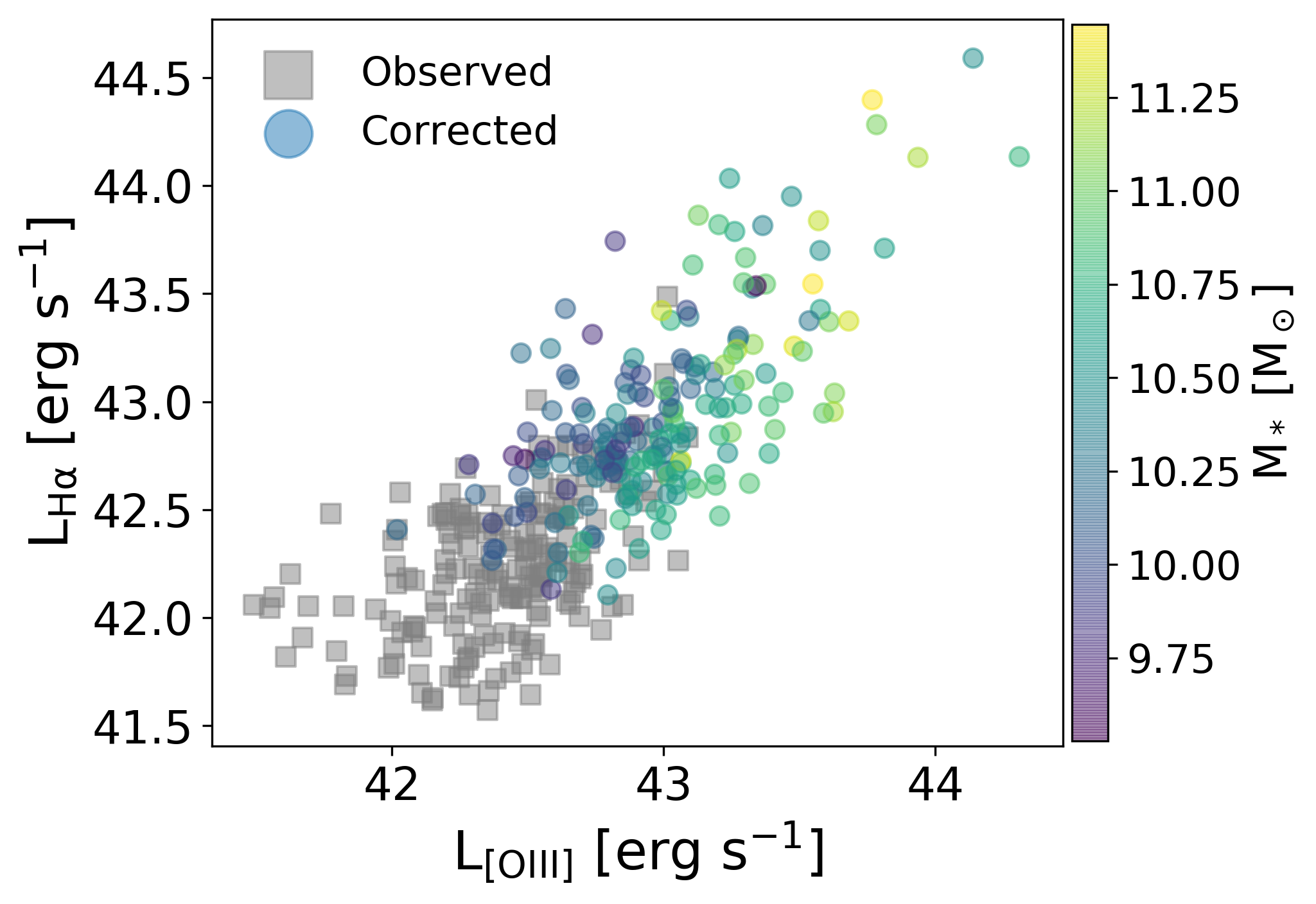}%
    \caption{H$\alpha$ and [OIII]$\lambda5007$ luminosity. The observed luminosities uncorrected for dust are shown as gray dots. The color dots correspond to luminosities de-reddened using A$_\mathrm{H\alpha}$ and A$_\mathrm{[OIII]}$ as constrained by the SED fitting with CIGALE. The corrected data is color-coded by the stellar mass. A slope of 0.99 is measured in the dust corrected sample with a $0.39$~dex dispersion.}
    \label{fig9_1_1}
\end{figure}

We computed the [OIII]$\lambda5007$ line luminosities for our sample of galaxies using FMOS-COSMOS observed fluxes. These luminosities are corrected for dust effects using the Bayesian attenuation $\mathrm{A}_\mathbf{[OIII]}$ presented in Sect. \ref{sec:BD_analysis}. We find that the H$\alpha$ and [OIII]$\lambda$5007 emission lines span a similar range in luminosities and the amount of attenuation is found to increase with luminosity and SFR confirming previous works %This is consistent with the fact that the dust content in galaxies increases with the SFR
\citep[e.g.,][]{Cortese2012, Bourne2012, Santini2014}. In Fig. \ref{fig9_1_1}, the observed and dust corrected H$\alpha$ and [OIII]$\lambda$5007 luminosities are shown in gray and color dots, respectively. Emission lines are not strongly correlated before dust correction. After accounting for dust attenuation both quantities correlate (with a Spearman's coefficient of $\rho_\mathrm{s}=0.6$). In the next subsections, we address the gas-phase metallicity influence on the [OIII]$\lambda$5007/H$\alpha$ line ratio, the SFR measurements, and we compare both [OIII]$\lambda5007$ and [OIII] $88 ~\mu$m line emissions from the photoionization models.

\subsection{Gas-phase metallicity}\label{subsec:gas-phase-metal}

\begin{figure}
\centering
    \includegraphics[width=0.48\textwidth,clip]{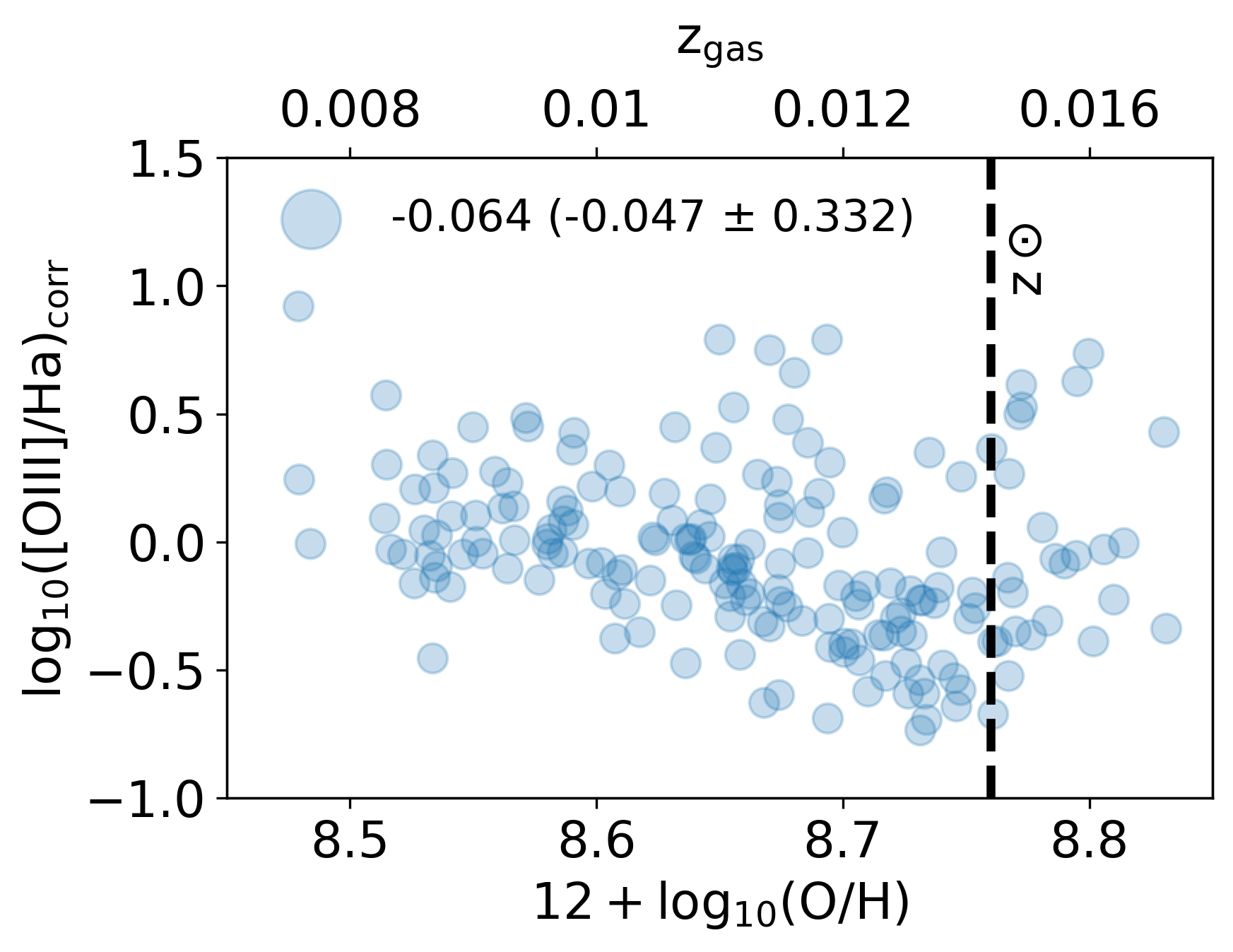}%
    \caption{[OIII]$\lambda5007$/H$\alpha$ ratio corrected for dust attenuation. We present the logarithm of the ratio as a function of oxygen abundance computed using \cite{Curti2020} FMR relation on the x-axis. The secondary axis corresponds to the translation in terms of gas-phase metallicity where the solar values are $0.0142$ or $8.76$. The ratio has been corrected using the attenuation for each line as computed with CIGALE. A linear and logarithmic median and mean ratio of $0.86$ ($-0.065$) and $0.89$ ($-0.047$) are found. The $0.32$ dispersion corresponds to the standard deviation.}
    \label{fig9_1_1_1}
\end{figure}

The [OIII]$\lambda$5007 emission line is sensitive to the ionization parameter \citep[i.e., the ionizing field;][]{Kewley2013, Kewley2013b, Dopita2013, Dopita2016, Nicholls2017} but can be also affected by the gas-phase metallicity \citep{Kennicutt1992, Kennicutt2000, Steidel2014, Gutkin2016, Byler2017}.
Different mass metallicity relations (MZR) based on direct and strong-line methods are reported in literature, sometimes including the effects of the SFR \citep[e.g.,][]{Pettini2004, Mannucci2010, Andrews_Martini2013, Zahid2017, Sanders2021, Sanders2020, Curti2020}. We compute oxygen abundances using the fundamental metallicity relation (FMR) calibration of \cite{Curti2020}, which is based on strong-line oxygen abundance measurements for Sloan Digital Sky Survey (SDSS) galaxies. Their relation is fully consistent with a relation derived for a MOSFIRE Deep Evolution Field (MOSDEF) sample of individual and stacked galaxies at $\mathrm{z}\sim2.3$ in \cite{Sanders2021}. The dust-corrected [OIII]$\lambda$5007/H$\alpha$ ratio is presented in Fig. \ref{fig9_1_1_1} as a function of oxygen abundance. Our galaxy sample covers a sub-solar abundance range of $8.4 < 12+\log(\mathrm{O/H}) < 8.8$ (i.e., gas-phase metallicity $0.006 < \mathrm{Z_{gas}} < 0.016$)\footnote{The solar metallicity is $\mathrm{z_\odot} = 0.0142$ as in \cite{Grevesse2010}.}. The ratio slightly decreases when metallicity increases. The median value of the ratio is $0.86$ (or $-0.065$~dex) with a dispersion of $0.3$~dex as measured by the standard deviation of the data set. The dispersion at a fixed metallicity can be produced by the sensitivity of the oxygen species to the ionizing field. The dependence on both metallicity and ionizing field is likely to produce effects on the ratio which we cannot easily disentangle. In Sect. \ref{subsec:fit_all}, we explore variations of the intensity of the radiation field fixing the gas-phase metallicity in three bins with an equal number of galaxies and simultaneously fitting the H$\alpha$, H$\beta$, and [OIII]$\lambda5007$ line fluxes. 

\subsection{SFR\texorpdfstring{$-\mathrm{[OIII]}\lambda5007$}{SFR OIII calibration}  calibration}\label{subsec:SFR-OIII_calibration}

\begin{figure}
\centering
    \includegraphics[width=0.45\textwidth,clip]{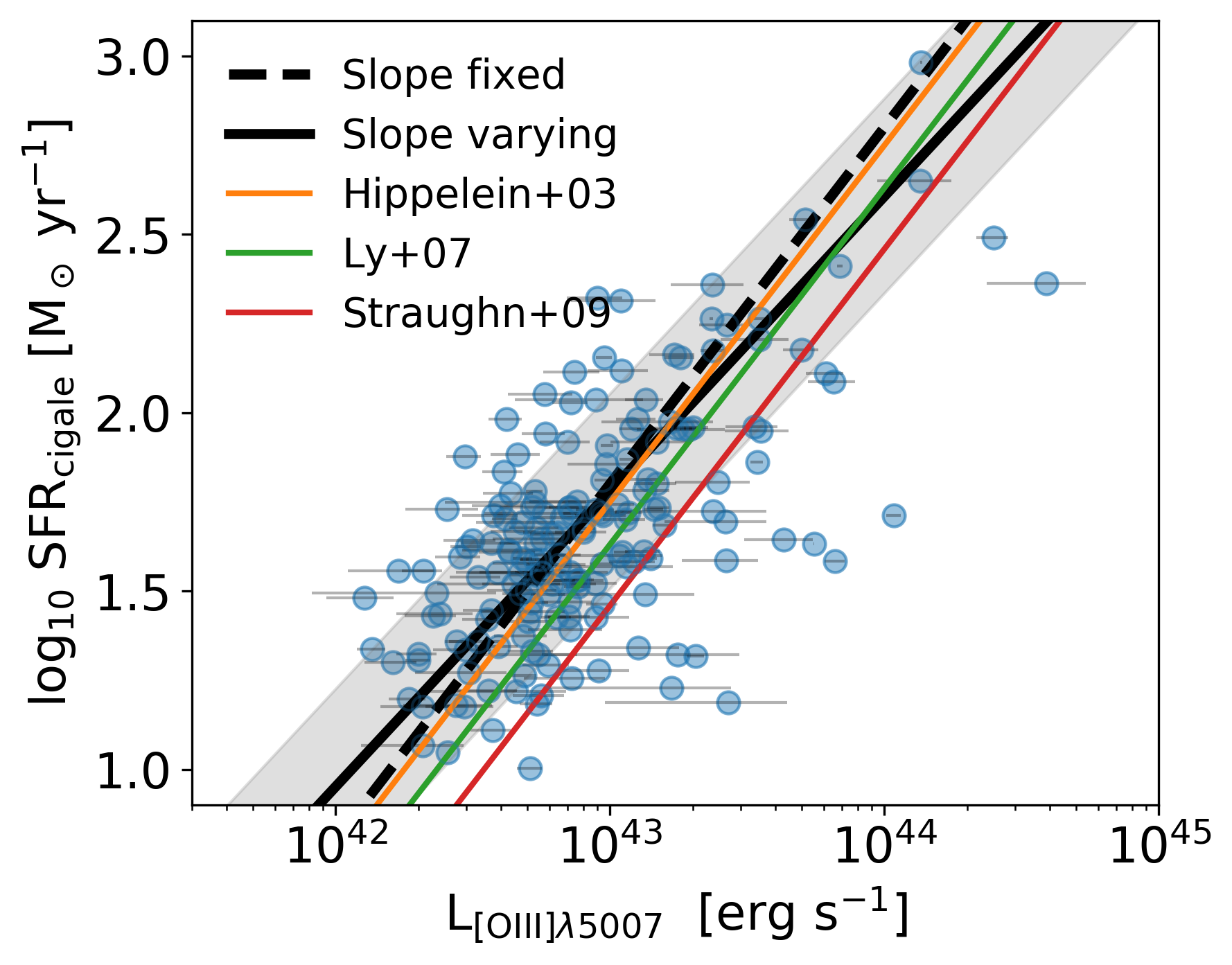}
    \caption{[OIII]$\lambda$5007 corrected luminosity and SFR relation. The dashed and continuous black lines correspond to fits using a bootstrapped orthogonal distance regression method with a Spearman's regression coefficient $\rho_\mathrm{s} = 0.57$. The $0.32$~dex dispersion from the standard deviation is presented as a gray shaded area. The relation shows a positive correlation with scatter significantly higher compared to \cite{Kennicutt1998} H$\alpha$ relation. Relations proposed by \cite{Hippelein2003}, and \cite{Ly2007} are shown in orange and green, respectively. \cite{Straughn2009} is also presented in red for comparison. Relations are converted from Salpeter to Chabrier IMF.}
    \label{fig11}
\end{figure}

In Fig. \ref{fig11}, we present the dust-corrected [OIII]$\lambda5007$ observed luminosities, and SED fitting derived SFR. As we want to make a comparison with the relations in the literature based on a fixed [OIII]$\lambda$5007/H$\alpha$ ratio, we first fit only the intercept using an orthogonal distance regression (ODR) in the $\mathrm{L}_{[\mathrm{OIII}]}$-SFR plane. The quality of the fit is measured by the non-parametric Spearman's and Kendall's correlation coefficients equal to $\rho_\mathrm{s} = 0.57$ and $\rho_\mathrm{s} = 0.40$, respectively. This leads to:

$$\log_{10} \mathrm{SFR} = \log_{10} \mathrm{L}_{[\mathrm{OIII]}} - (41.27\pm0.03),$$

with the luminosity and SFR expressed in units of $\mathrm{ergs~s^{-1}}$ and $\mathrm{M}_\odot~\mathrm{yr}^{-1}$, respectively. 

The presence of outliers can have an impact on the ODR fit. To reduce the effects of extreme values we compute the geometrical distance of each data point to the line drawn by the equation above. We check the geometrical distance distribution and trim $15$ data points with $\mathrm{d} \geq \pm0.45$ that corresponds to $\sim2\sigma$. These outliers are characterized by poor S/N in GALEX NUV, and the \textit{Herschel} bands. We check that the distribution in the SFR-L$_\mathrm{[OIII]\lambda5007}$ plane after discarding these points remains representative of our sample. We implemented a bootstrap method that uses a random iterative replacement technique to estimate the slope and intercept in the fits. Using this method, we drew a new sample of the same size $200$ times and re-fit using the ODR method with a fixed slope to obtain the distribution of the intercept from which we compute the mean value and the uncertainty in the intercept as the standard deviation of the distribution. This leads to:

\begin{equation}
\log_{10} \mathrm{SFR} = \log_{10} \mathrm{L}_{[\mathrm{OIII]}} - (41.20\pm0.02),
    \label{eq:5_fix}
\end{equation} 

with the luminosity in $\mathrm{ergs~s^{-1}}$ and SFR in $\mathrm{M}_\odot~\mathrm{yr}^{-1}$. The relation is consistent within a $2\sigma$ uncertainty with the equation obtained including the outliers. Following the same procedure  described earlier in this paper and allowing the slope and intercept to vary, we obtain:

\begin{equation}
\log_{10} \mathrm{SFR} = (0.83\pm0.06)\log_{10} \mathrm{L}_{[\mathrm{OIII]}} - (34.01\pm2.63),
    \label{eq:5_varying}
\end{equation} 

with the luminosity and SFR given in units similar to Eq. \ref{eq:5_fix}. The slope in Eq. \ref{eq:5_varying} is consistent with unity at $3\sigma$. The relations presented in Eq. \ref{eq:5_fix} and \ref{eq:5_varying} are drawn in Fig. \ref{fig11} as a black dashed and continuous line with a $0.32$~dex dispersion from the standard deviation of the observations as a gray shaded area. Since the H$\alpha$ emission can be considered as proportional to the SFR, the dispersion of the [OIII]$\lambda5007$-SFR correlation is expected to have the same origin as the one affecting the [OIII]$\lambda$5007/H$\alpha$ ratio: the metallicity and radiation field.

Several authors in the past studied the link between the [OIII]$\lambda5007$ emission line and SFR based on a fixed [OIII]$\lambda$5007/H$\alpha$ ratio and the calibration of the H$\alpha$ emission line in terms of SFR. At $\mathrm{z}\sim0.1$ on a sample of 196 SDSS narrow band emitters \cite{Moustakas2006} measured a scatter in the [OIII]$\lambda5007$-SFR relation as large as a factor of $3-4$ uncertainty (1$\sigma$) when compared to other tracers such as [OII]$\lambda 3727$, H$\beta$ emission lines or the U-band.  Based on a sample of 92 galaxies at $0.40<\mathrm{z}<0.64,$ \cite{Hippelein2003} derived an [OIII]$\lambda$5007/H$\alpha$ ratio of ~$0.79$ from line luminosity functions, which is consistent with our median value. Although [OIII]$\lambda5007$ depends on excitation and metallicity, their estimated median values of SFR($\mathrm{[OIII]}$) were consistent with the evolution of SFR density of the universe. \cite{Ly2007} reported a [OIII]$\lambda$5007/H$\alpha$ ratio of $1.05$ at $\mathrm{z}=0.07-1.47$ for a spectroscopic sample of 196 narrow-band emitters. \cite{Teplitz2000} fixed the ratio to unity at $\mathrm{z}>3$ for a sample of five galaxies observed with the near-IR camera on the Keck I telescope. Both works from \cite{Teplitz2000} and \cite{Ly2007} found similar results on the dependence of the line ratio on metallicity and ionization field. \cite{Maschietto2008} studied a sample of 13 [OIII] emitters at $\mathrm{z}\sim3.09-3.16$ and using a [OIII]$\lambda$5007/H$\alpha$ ratio of ~$2.4$ derived a lower limit relation for SFR and [OIII]$\lambda5007$ luminosity. \cite{Straughn2009} also explored the relation based on [OIII]$\lambda$5007/H$\alpha$ ratios in knots of 136 galaxies at $\mathrm{z}\sim0.5$ for which both emission lines were observed.
%lately implemented in \cite{Pharo2020},
Their relation implies a median [OIII]$\lambda$5007/H$\alpha$ ratio of $\sim1.23$. \cite{Bowman2019} found a strong correlation but did not propose any relation for a set of 3D-HST galaxies at $\mathrm{z}\sim2$ (see their Figure 6).

Our calibration relies on quantities estimated for each galaxy (e.g., [OIII]$\lambda5007$ luminosity, SFR). It can be compared to relations also obtained with individual [OIII]$\lambda5007$/H$\alpha$ ratios as those obtained by \cite{Hippelein2003}, and \cite{Ly2007}\footnote{All relations are converted to Chabrier IMF}. Their dust corrected relations are reported in Fig. \ref{fig11} (orange and green lines) and are fully consistent with our result within the $1\sigma$ dispersion error and our median reported [OIII]$\lambda5007$/H$\alpha$ ratio of $0.86$.

\subsection{Impact of the \texorpdfstring{$\mathrm{[OIII]}\lambda5007$}{Impact of the OIII dust attenuation} dust attenuation}\label{subsec:Abscence_AOIII}

In the previous section, we studied the relation between the $\mathrm{[OIII]}\lambda5007$ dust corrected luminosity and SFR. In practice, it is very difficult to correct the [OIII]$\lambda$5007 emission line for dust attenuation. When A$_\mathrm{[OIII]}$ cannot be estimated but stellar masses are known, attenuations can be tentatively estimated by using the A$_\mathrm{[OIII]}$-M$_\mathrm{star}$ relation presented in Sect. \ref{sec:BD_analysis}. We applied this relation to correct our observed [OIII]$\lambda5007$ luminosities and checked how the relation presented in Eq. \ref{eq:5_fix} is modified. Following the same procedure presented in Sect. \ref{subsec:SFR-OIII_calibration}, we find:
\begin{equation}
\log_{10} \mathrm{SFR} = \log_{10} \mathrm{L}_{[\mathrm{OIII]}} - (41.23\pm0.03).
    \label{eq:5_new}
\end{equation}
The average relation is similar to that of  Eq. \ref{eq:5_fix}, but a large dispersion is expected because of the dispersion in the A$_\mathrm{[OIII]}$-M$_\mathrm{star}$ relation.  To test it,  we divided our %dust-corrected (WHY the dust corrected sAMple since you correcT AFTEr) 
sample in mass bins of $0.5$~dex from $10^{9.5}$ to $10^{11.5}~\mathrm{M}_\odot$ and applied both  CIGALE's estimated attenuation and the mass-method attenuation. We measured an average increase in SFR of only $0.15$~dex when we apply Eq. \ref{eq:5_new}, as compared to the SFR estimated by CIGALE. The dispersion of $0.3$~dex in each bin is much larger than the increase in SFR. Thus, Eq. \ref{eq:5_new} can be applied when only an averaged value of SFR is needed when analyzing galaxies whose stellar masses are known.

\subsection{\texorpdfstring{Benchmark for the $\mathrm{[OIII]}\lambda$5007 and $\mathrm{[OIII]} ~{88}~\mu\mathrm{m}$}{OIII and OIII lines benchmark} lines }

The [OIII] also produces a fine-structure emission line at $88.33~\mu\mathrm{m}$ from low excitation and highly ionized gas located in a low-density environment. This line is not affected by dust attenuation and is also used to trace SFR at both low and high redshifts \citep[e.g.,][]{DeLooze2014, Harikane2020}. In this section,  we present our efforts to predict a relation between [OIII] $88~\mu$m and SFR from the relation we measured between [OIII]$\lambda5007$ and SFR and  HII-region models.

\begin{figure}
  \resizebox{\hsize}{!}{\includegraphics{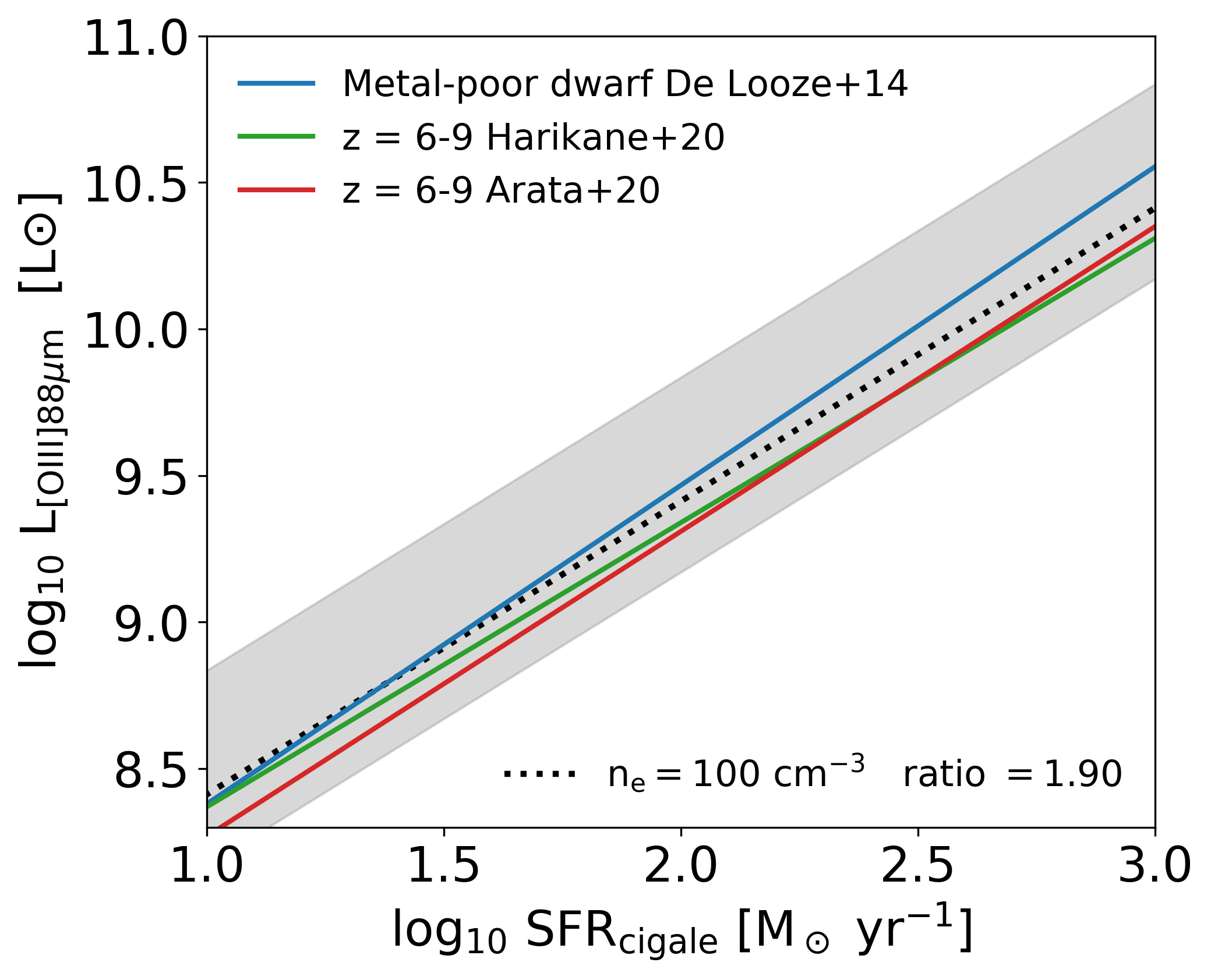}}
  \caption{[OIII] $88 ~\mu$m-SFR relation. The blue line shows the  metal-poor local dwarf galaxies relation from \cite{DeLooze2014}. The green line corresponds to the high-redshift observed relation proposed by \cite{Harikane2020} while the red line corresponds to simulation results from \cite{Arata2020} both at $\mathrm{z}=6-9$. We translate our [OIII]$\lambda5007$ luminosities into [OIII] $88 ~\mu$m using a mean ratio of $1.9$ derived from CLOUDY HII-region models at an electron density of $\mathrm{n_e} = 100~\mathrm{cm}^{-3}$ shown as a black dotted line. The gray area is the dispersion in the translation with gas-phase metallicity.}
  \label{fig13}
\end{figure}

The [OIII]$\lambda$5007-[OIII] $88 ~\mu$m ratio is known to vary with gas-phase metallicity and density \citep{Dinerstein1983, Stacey2010, Ferkinhoff2010, Moriwaki2018}, but it is not considered to be especially sensitive to the ionization parameter \citep{Moriwaki2018}. To select the photoionization models that are relevant to our analysis, we restricted the ionization parameter to $-3.0 < \log\mathrm{U} < -2.0,$ in agreement with our results (presented in Sect. \ref{subsec:fit_all}). For the electron
density,  \citet{Kashino2017} derived a value of $\sim200~\mathrm{cm}^{-3}$  for the FMOS-COSMOS sample. Three different values of the electron density are available in CIGALE (i.e., $\mathrm{n_e} = 10, 100,$ and $1000~\mathrm{cm}^{-3}$) and we fixed $\mathrm{n_e}$ to $100~\mathrm{cm}^{-3}$. From our estimations of the gas-phase metallicity  (Sect. \ref{subsec:gas-phase-metal}), we restricted the  models to  $0.006 < \mathrm{Z_{gas}} < 0.016$ (or $8.4 < 12+\log(\mathrm{O/H}) < 8.8$). With the selected models, we derived a mean [OIII]$\lambda$5007-[OIII] $88 ~\mu$m ratio of $\sim1.90$ (with a $1\sigma$ dispersion of $0.84$). We used this average value of the ratio  to translate our SFR-[OIII]$\lambda5007$ linear relation (presented in Eq. \ref{eq:5_fix}) into another linear relation between [OIII] $88 ~\mu$m and SFR. This is presented in Fig. \ref{fig13} as a black dotted line with a gray shaded area accounting for the dispersion with the gas-phase metallicity.

Our relation is found to be consistent with the relations proposed by \cite{DeLooze2014} for metal-poor local dwarf galaxies and by \cite{Harikane2020} for high redshift galaxies despite the different gas-phase metallicities of their samples.  \cite{Arata2020} report a L$_\mathrm{[OIII]88\mu\mathrm{m}}-\mathrm{SFR}$ relation derived from simulations at $\mathrm{z}=6-9$ covering a gas-phase metallicity range of $6.6<12+\log(\mathrm{O/H})<8.9$ that matches our metallicity range and is also consistent with different predictions from the literature \citep{DeLooze2014, Moriwaki2018, Harikane2020}. Both electron density and gas-phase metallicity are paramount to properly convert one emission into the other as well as to compare different samples of galaxies. At intermediate and high densities [OIII]$\lambda$5007 line remains an important coolant for ionized gas as compared to [OIII] $88 ~\mu$m. The [OIII]$\lambda$5007 emission line will be targeted in future surveys like MOONS, PFS, and JWST allowing for synergies with observatories like ALMA that can perform observations of [OIII] $88 ~\mu$m at high-redshift providing a direct method to characterize the overall cooling budget and the ISM physics of HII-regions in intermediate- and high-redshift galaxies as well as to refine their calibration in SFR. 

%--------------------------------------------------------------------
\section{Spectro-photometric modeling: HII-region and stellar continuum modeling}\label{sec:models}

Our FMOS-COSMOS sample has H$\alpha$, H$\beta$, [NII]$\lambda$6584, [OIII]$\lambda$5007, [SII]$\lambda$6717, and [SII]$\lambda$6731 fluxes (c.f Sect. \ref{sec:Data}). In our initial SED fitting analysis, we only included the H$\alpha$ emission line because of the current challenges in the production of the photoionization models to reproduce observations. We discuss these issues in this section and introduce more line intensities to the fits.

The nebular emission lines module was briefly introduced in Sect. \ref{subsubsec: Nebular_CIGALE}; here, we add a more detailed description of the photoionization calculations. We use the CLOUDY 17.01 photoionization code \citep{Ferland2017} to predict the relative intensities of 124 atomic lines originating from the far-ultraviolet (FUV) to FIR. Line emissivity models for a plane-parallel geometry are built with the intensity of the photo-ionizing radiation field varying from $-4.0 < \log\mathrm{U} < -1.0$ and using a sampling of $0.1$~dex. The electron density is set to three different values of $\mathrm{n}_\mathrm{e} = 10, 100, 1000~\mathrm{cm^{-3}}$ covering 25 gas-phase metallicities spanning over $0.0001<\mathrm{Z_{gas}}<0.05$ disconnected from stellar metallicity $(0.0001, 0.0004, 0.004, 0.008, 0.02, 0.05)$ as presented in Table \ref{tab:metallicity_models}. The closest stellar metallicity to the gas-phase metallicity is chosen as the input ionizing spectrum in the models. The cosmic abundance standard and the derivation of element abundances implemented in our models are discussed in \cite{Nicholls2017}. They are based on a scaling developed by \cite{Nieva2012} derived from observed metallicities of 29 early B-type stars in the Milky Way. The depletion onto grains is not taken into account since the abundances are based on gas-phase measurements. Our models are built on the so-called local Galactic concordance, $12+\log(\mathrm{O/H})_\mathrm{GC}=8.76$, which is close to the $8.73$ estimated primordial solar abundance derived by \cite{Asplund2009} and \cite{Lodders2010}, with (O/H)$_\mathrm{GC}$=5.76$\times$10$^{-4}$, (N/H)$_\mathrm{GC}$=6.17$\times$10$^{-5}$, (C/O)$_\mathrm{GC}$=0.46, and Z$_\mathrm{GC}$=0.0142. Super-solar and sub-solar metallicities are scaled following the latter definitions. As compared to the previous version implemented in CIGALE, these models disconnect the stellar and gas-phase metallicities and implement current abundance measurements, making them more appropriate for reproducing the line fluxes of our sample of galaxies. We used these new models to fit both photometry and other emission lines similar to Sect. \ref{sec:Calibration}, where photometry and only H$\alpha$ emission fluxes were fitted together.

\begin{table}
\caption{Connection between the stellar and gas-phase metallicity.}   
\label{tab:metallicity_models}      
\centering 
\scalebox{0.9}{
\begin{tabular}{c  c }     % 7 columns 
\hline \hline      
$\mathrm{Z_{star}}$ & $\mathrm{Z_{gas}}$\\ 
\hline                    
   0.05   & 0.051, 0.046, 0.041  \\  
   0.02   & 0.037, 0.033, 0.030, 0.025, 0.022, 0.019, 0.016, 0.014  \\
   0.008  & 0.012, 0.011, 0.009, 0.008 ,0.007\\
   0.004  & 0.006, 0.005, 0.004, 0.003 \\
   0.0004 & 0.0025, 0.002, 0.001, 0.0004\\
   0.0001 & 0.0001\\
\hline                  
\end{tabular}
}
\end{table}

\subsection{Baldwin-Phillips-Terlevich and [SII]\texorpdfstring{$\lambda\lambda$}6717,31 excitation diagrams}\label{subsec:BPT_diagrams}

\begin{figure*}
\centering
    \includegraphics[width=0.98\textwidth,clip]{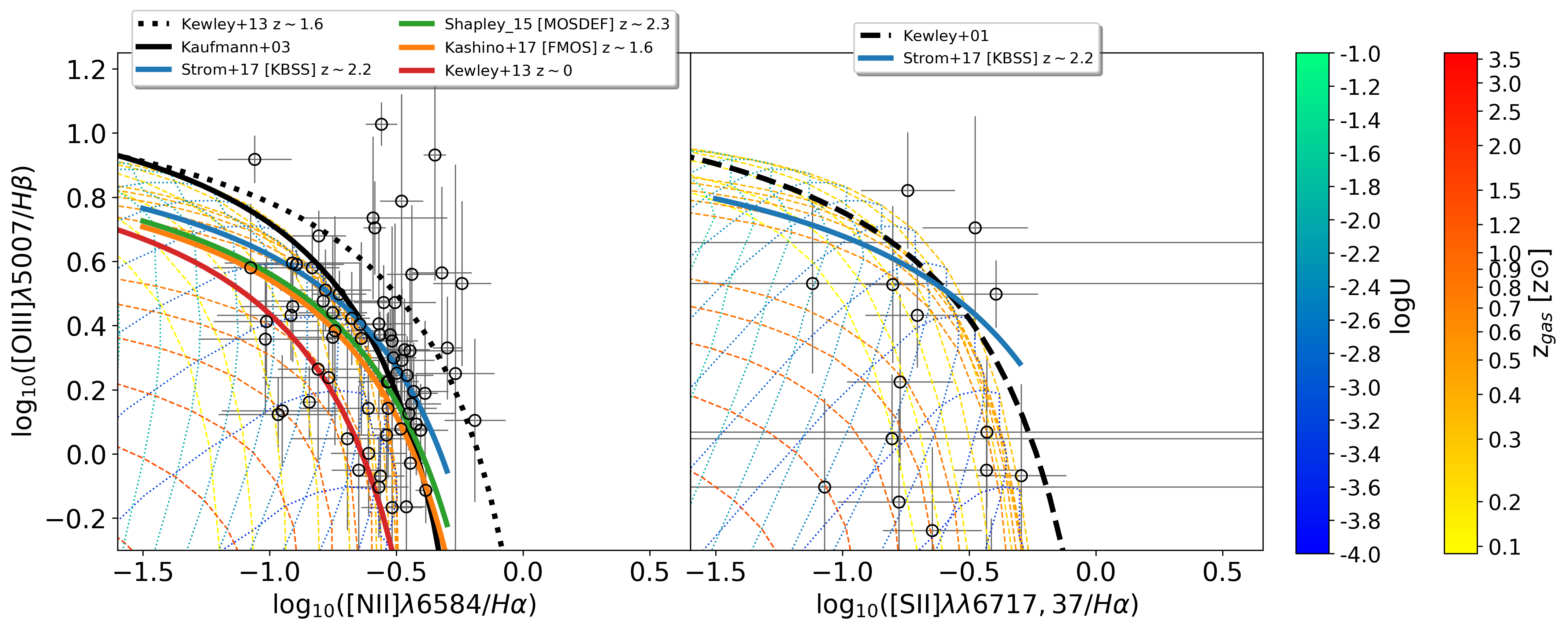}%
    \caption{Excitation diagnostic diagrams. \textbf{Left:} Baldwin–Phillips–Terlevich (BPT) diagram ([OIII]$\lambda$5007/H$\beta$ versus [NII]$\lambda$6584/H$\alpha$). The new models implemented in CIGALE are color-coded by gas-phase metallicity in solar units and ionization parameter $\log\mathrm{U}$. Only a few metallicities are shown for clarity. The FMOS-COSMOS sample is shown as gray dots with error bars. The solid black line corresponds to the \cite{Kaufmann2003} relation and the dashed-black line corresponds to \cite{Kewley2013} evaluated at $\mathrm{z}\sim1.6$. The curves of \cite{Shapley2015}, \cite{Kashino2017}, and \cite{Strom2017} are shown in green, orange, and blue, respectively. The red line represents the local-universe locus of galaxies as shown by \cite{Kewley2013}. \textbf{Right:} [SII]$\lambda\lambda$6717,31 excitation diagram ([OIII]$\lambda$5007/H$\beta$ versus [SII]$\lambda\lambda$6717,31/H$\alpha$). The current models implemented in CIGALE are color-coded by gas-phase metallicity and ionization parameter $\log\mathrm{U}$ to illustrate the coverage. Only a few metallicities are shown for clarity. The FMOS-COSMOS sample is shown as black circles with error bars. The dashed black line corresponds to the results from \cite{Kewley2001} and the blue line to \cite{Strom2017}.}
    \label{fig14}
\end{figure*}

In Fig. \ref{fig14}, the [OIII]$\lambda$5007/H$\beta$, [NII]$\lambda$6584/H$\alpha$ diagram (so-called Baldwin-Phillips-Terlevich \citep{Baldwin1981}; see also \citet{Veilleux1987}) and the [SII]$\lambda\lambda$6717,31 excitation diagrams (hereafter [NII]-BPT and [SII]-BPT) along with our photoionization models are presented color-coded by the ionization parameter and the gas-phase metallicity. In the left panel, the demarcation between star-forming galaxies and systems hosting an AGN of \cite{Kaufmann2003} from SDSS data at $\mathrm{z}=0$ is shown as a black thick line and the extreme starburst separation line of \cite{Kewley2013} at $\mathrm{z = 1.6}$ as a dashed line. The blue, green, and orange lines correspond to best-fit relations for the loci of galaxies in the [NII]-BPT at $\mathrm{z}\sim2.2$, $\mathrm{z}\sim2.3$, and $\mathrm{z}\sim1.6$ from \cite{Strom2017}, \cite{Shapley2015}, and \cite{Kashino2017}, respectively. Individual FMOS-COSMOS points show the well-known offset from the local sequence. Our FMOS-COSMOS sample with valid measurements for the four emission lines and their corresponding measured errors are represented by dots. For a smaller sample of 16 objects, we show, in the right panel, the [SII]-BPT diagram with the star-forming and AGN separation of \cite{Kewley2001} and the best fit for the loci of galaxies of \cite{Strom2017}.

The CLOUDY models cover well the star-forming region below the \cite{Kaufmann2003} line in the [NII]-BPT diagram. However, only $61\%$ of the flux ratios presented are covered pointing toward a difficulty related to the nitrogen-to-oxygen abundances more than the photoionizing field. In this diagram, $23\%$ of the objects lie in the composite region where AGN and star-forming galaxies coexist at $\mathrm{z}=0$. For the [SII]-BPT diagram models are in agreement with the locus of galaxies with valid measurements within the observational errors. The HII-region models able to predict emission line ratios above the local star-forming relation are difficult to create because of \textit{i.} nitrogen abundance underestimation; \textit{ii.} ionizing field hardness choice; \textit{iii.} gas-phase metallicity and density discrepancies or single stellar population models. The models depend strongly on the relative abundances of nitrogen and oxygen, and thus on the choice of the standard metallicity scale \citep[e.g.,][]{Nicholls2017}. The nebular scaling of nitrogen with oxygen is  problematic: while oxygen is principally produced in core-collapse supernovae in the native gas cloud from which the HII-region formed, nitrogen has both primary and secondary abundances, which are caused by delayed nucleosynthesis through hot-bottom burning and dredge-up in intermediate-mass stars as they evolve \citep{VilaCostas1993}. Increasing the nitrogen abundance has been suggested by many authors as a way to match the observations \citep[e.g.,][]{Masters2014, Masters2016, Steidel2014, Shapley2015, Yabe2015, Cowie2016, Sanders2016}. A shift by $0.2-0.4$~dex in the N/O fraction is proposed by \cite{Masters2016} while \cite{Kojima2017} and \cite{Strom2017} require $\sim0.1$~dex to cover the locus of galaxies in the BPT at $\mathrm{z} \sim 2.3$. We find that in order to cover all the sample of galaxies at $\mathrm{z} \sim 1.6,$ a shift of $0.2-0.3$~dex is necessary. [OIII]$\lambda5007$ and [NII]$\lambda6584$ have an ionization potential (IP) of $35.12$ and $14.53$~eV and are ionized by both charge transfer with protons and photoionization. Thus, [OIII]$\lambda$5007 and [NII]$\lambda$6584 abundances are less sensitive to the photoionizing field, which are determined indirectly through the HII abundance, and the initial oxygen and nitrogen abundances. [SII]$\lambda\lambda6717,37$ has an IP of $10.36$~eV, making its abundance very sensitive to the hardness of the photoionizing field in the UV. Indeed, \cite{Steidel2014, Steidel2016} applied harder radiation fields using the binary Population and Spectral Synthesis code (BPASS) and these authors were able to produce models covering larger values of $\log$([OIII]$\lambda$5007/H$\beta$), but always below the \cite{Kaufmann2003} line. Other authors point to a higher ionization parameter or electron density as the main parameter producing the offset \citep[e.g.,][]{Brinchmann2008, Kewley2013, Kewley2013b, Dopita2016, Kojima2017, Bian2020}. Nevertheless, current stellar synthesis models are unable to produce such hard radiation fields \citep{Levesque2010}. The densities from models in high-redshift galaxies ($\mathrm{z} \geq 1-2$) are higher than in low-redshift galaxies, on the order of several 10$^2$ to several 10$^3$ per cm$^{-3}$ \citep[][]{Shimakawa2015, Kashino2017}. Density (i.e., the pressure) has only secondary effects on the locus of the models in the [NII]-BPT diagram \citep{Masters2016}. 

\subsection{H\texorpdfstring{$\alpha$}, H\texorpdfstring{$\beta$}, and [OIII]\texorpdfstring{$\lambda$}5007 SED fitting }\label{subsec:fit_all}

To test the performance of SED fitting including more emission lines, we fit simultaneously the photometry and the H$\alpha$, H$\beta$, and [OIII]$\lambda$5007 fluxes to explore the impact of the ionization parameter in the dispersion of the proposed SFR-$\mathrm{L}_\mathrm{[OIII]}$ relation. Our sample of galaxies spans over a gas-phase metallicity range of $0.006 < \mathrm{Z_{gas}} < 0.016$ (see Sect. \ref{sec:Calibration}). We divide the sample in three different gas-phase metallicity bins with roughly equal number of sources and median metallicities given by ${\mathrm{Z}_\mathrm{gas}} = 0.009, {\mathrm{Z}_\mathrm{gas}} = 0.011$ and ${\mathrm{Z}_\mathrm{gas}} = 0.014$. We perform SED fitting including the UV-to-FIR photometry and the H$\alpha$, H$\beta$, and [OIII]$\lambda5007$ emission lines by fixing the gas-phase metallicity in each bin to its median value. Only two stellar metallicities of the BC03 models, $0.02$ and $0.008$, are included in the full range of gas-phase metallicity of our sample. We set the stellar metallicity to $0.02$ after checking that using $0.008$ does not affect our parameter estimation. In the case of H$\beta$ we only include measurements of the lines for objects satisfying BD~$> 2.86$ to be consistent with our models. We let the ionization parameter to vary between $-4.0<\log\mathrm{U}<-1.0$ and we use $\mathrm{n_e} = 100~\mathrm{cm}^{-3}$ consistent with the electron density derived by \citet{Kashino2017} for the FMOS-COSMOS sample.

\begin{figure}
\centering
    \includegraphics[width=0.49\textwidth,clip]{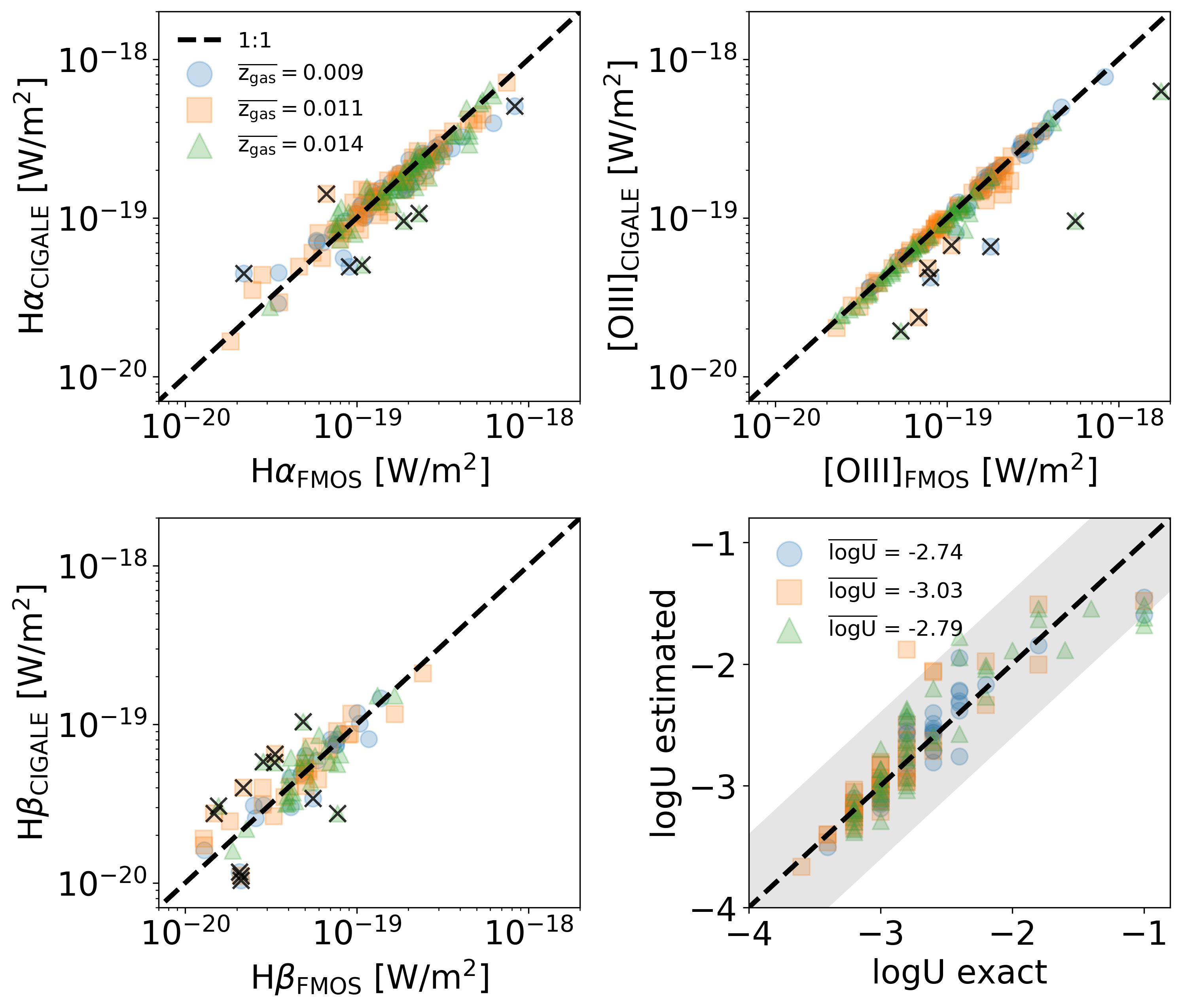}%
    \caption{Quality of the fits including H$\alpha$, H$\beta$, and [OIII]$\lambda5007$ emission lines and ionization parameter estimation. From left to right, top to bottom, we show the CIGALE fit vs. observed flux for the H$\alpha$, [OIII]$\lambda$5007, H$\beta$ emission lines. The three different gas-phase metallicity bins are presented as blue circles, orange squares, and green triangles. The black crosses correspond to excluded data with flux difference larger than $0.2$~dex. The black line corresponds to the 1:1 relation. The three emission lines are well fitted for the three different median gas-phase metallicity models. The last panel shows the estimated vs exact value for $\log\mathrm{U}$ from mock samples created with CIGALE. Symbols are the same as the legend in the first panel and the median ionization parameter value is shown. The shaded area corresponds to the standard deviation.}
    \label{fig15}
\end{figure}

We exclude from the analysis  $21$ objects with a difference between observed and fitted fluxes larger than $0.2$~dex for the three emission lines. In Fig. \ref{fig15} the observed and estimated fluxes from the three emission lines are compared. In the case of the H$\alpha$ line as compared to our previous fit in Fig. \ref{fig4} similar results are obtained and the line fit is not improved by including more emission lines. The distribution of the estimated  [OIII]$\lambda$5007fluxes is not symmetric with an excess of overestimated fluxes. H$\beta$ estimated fluxes have a standard deviation of $\sim0.13$~dex which is twice as large as that of H$\alpha$ and [OIII]$\lambda$5007 emission lines. We obtain a median attenuation from CIGALE of A$_\mathrm{H\alpha} = 1.11\pm0.18$~mag and A$_\mathrm{[OIII]} = 1.34\pm0.22$~mag ($\dagger$ as reported in Table \ref{table:3}), consistent with values derived using photometry and the H$\alpha$ emission only.
\begin{figure}
\centering
    \includegraphics[width=0.49\textwidth,clip]{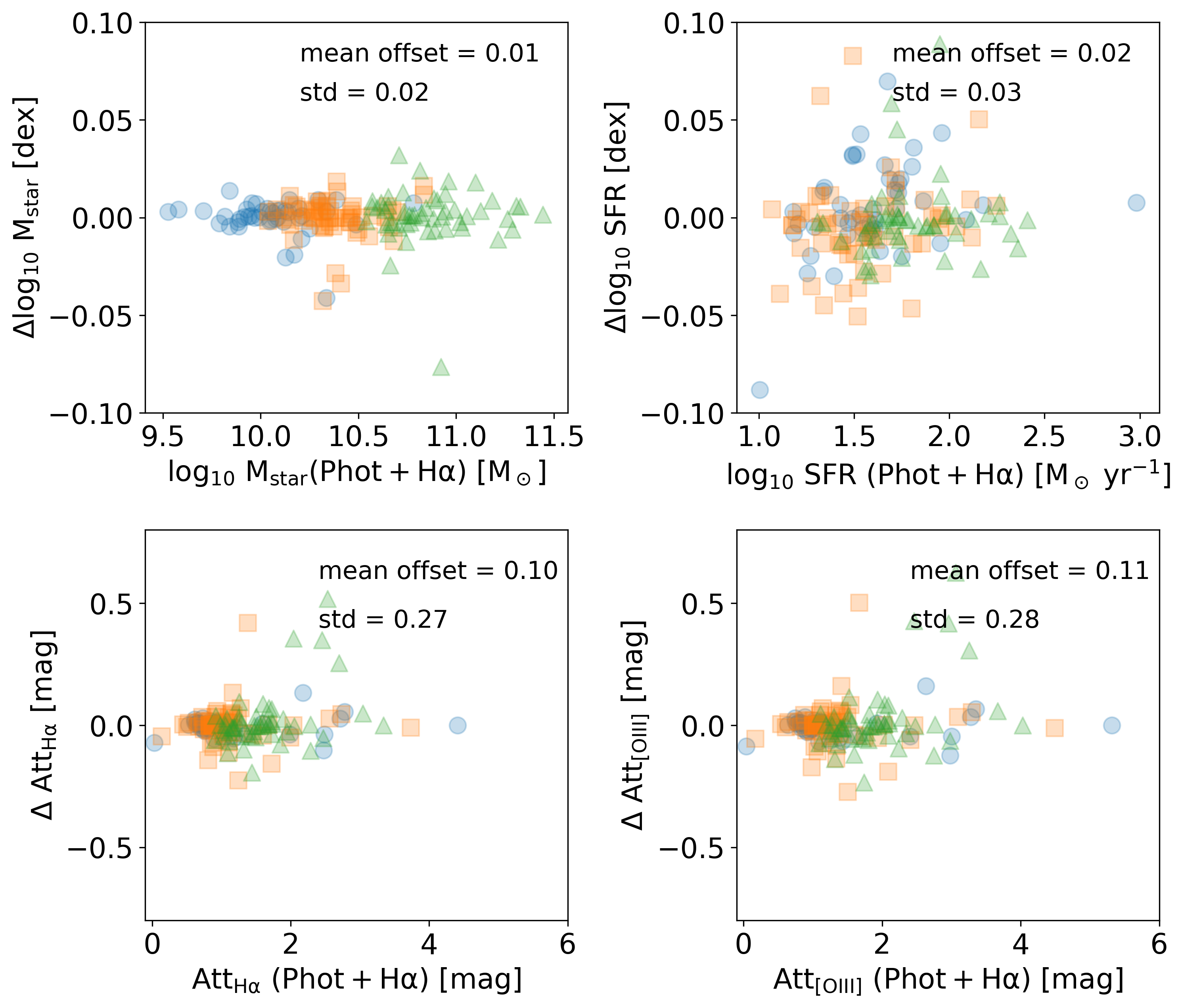}%
    \caption{Stellar mass, SFR, and H$\alpha$ and [OIII]$\lambda5007$ attenuation. We compare the derived parameters using photometry and H$\alpha$ flux in the SED fitting and adding H$\beta$ and [OIII]$\lambda5007$ fluxes. The three different gas-phase metallicity bins are presented as blue circles, orange squares, and green triangles as in Fig. \ref{fig15}. The mean offset and dispersion of the are shown for each parameter.}
    \label{fig18}
\end{figure}
Stellar mass, SFR, and attenuation in the H$\alpha$ and [OIII]$\lambda5007$ emission lines are not find to vary significantly by including more emission lines in the SED fitting process as shown in Fig. \ref{fig18}.

Using mock catalogs created with CIGALE we investigate the robustness in the estimation of the ionization parameter $\log\mathrm{U}$. The exact and estimated values of $\log\mathrm{U}$ are compared in the bottom-right panel in Fig. \ref{fig15} showing a good agreement within a $0.6$~dex ($1\sigma$) dispersion for the three different gas-phase metallicity bins guaranteeing the reliability in the estimation of the ionization parameter.

From our fit, we estimate median values of ${\log\mathrm{U_{0.009}}}\sim-2.74$, ${\log\mathrm{U_{0.011}}}\sim-3.03$, and ${\log\mathrm{U_{0.014}}}\sim-2.79,$ while including the H$\beta$, and [OIII]$\lambda5007$ lines. The 16th and 84th percentiles in logU for each gas-phase metallicity bin are ${\log\mathrm{U_{0.009}}}\sim-3.02,-2.35$, ${\log\mathrm{U_{0.011}}}\sim-3.21,-2.60$, and ${\log\mathrm{U_{0.014}}}\sim-3.14,-2.00$, respectively. \cite{Kaasinen2018} measured $\log\mathrm{U}\sim-2.72$ from an evolutionary analysis of the ionization parameter using a sample of $50$ star-forming galaxies selected from the FMOS-COSMOS catalog with DEep Imaging Multi-Object Spectrograph (DEIMOS) observations. \cite{Sanders2020} constrain the ionization parameter using BPASS grids for a MOSDEF and Keck Baryonic Structure Survey (KBSS) samples at $\mathrm{z}\sim1-3$ to $\log\mathrm{U}\sim-2.63$ and $\log\mathrm{U}\sim-2.85$, respectively. \cite{Topping2020} found that the local   $12+\log(\mathrm{O/H})$-$\log\mathrm{U}$ relationship  of \cite{PerezMontero2014} for low-redshift galaxies is still applicable at $\mathrm{z}\sim2$.  From this relation, the ionization parameter range for our sample should span over the range $-3.2<\log\mathrm{U}<-2.5,$ in agreement with our estimations.

\begin{figure}
\centering
    \includegraphics[width=0.48\textwidth,clip]{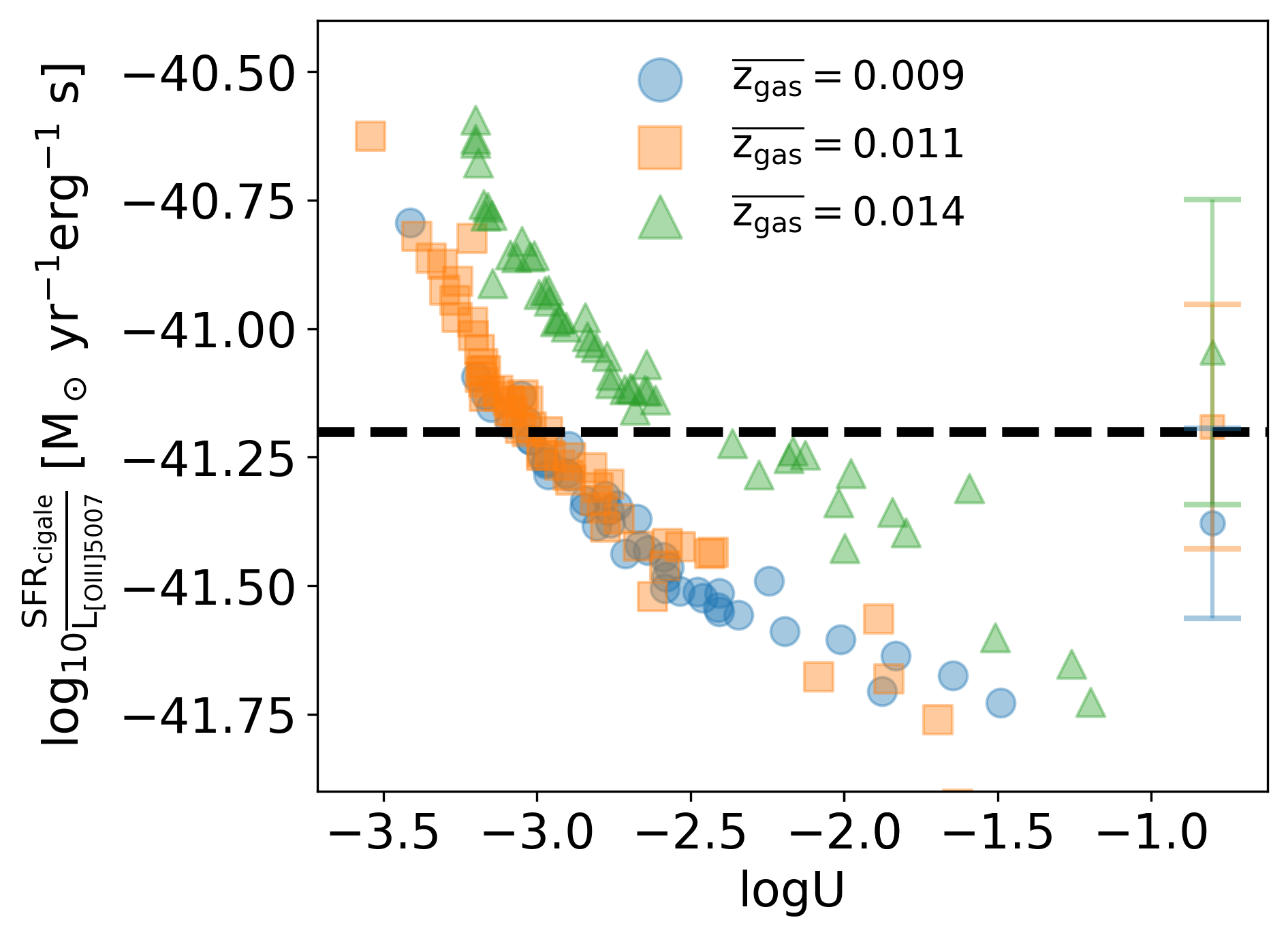}%
    \caption{SFR and L$_{\mathrm{[OIII]}5007}$ ratio vs the ionization parameter. Each metallicity bin is presented as blue dots, orange squares, and green triangles. The ionization parameter is computed with CIGALE for each fixed metallicity case. The black dashed line corresponds to the $-41.20$~[M$_\odot$ yr$^{-1} \mathrm{erg^{-1}~s}$] intercept found in Eq. \ref{eq:5_fix}. The symbols with errors represent the median values of the SFR/L$_{\mathrm{[OIII]}5007}$ ratio for each metallicity. The ionization parameter has a larger impact on the dispersion than the gas-phase metallicity.}
    \label{fig17}
\end{figure}

The $\mathrm{SFR}$-$\mathrm{L}_{\mathrm{[OIII]}\lambda5007}$ ratio can be interpreted as the H$\alpha$-[OIII]$\lambda$5007 ratio because H$\alpha$ is a tracer of SFR. The dispersion in the [OIII]$\lambda$5007/H$\alpha$ dust corrected ratio (Fig. \ref{fig9_1_1_1}) depends on both ionization parameter and gas-phase metallicity. Now we explore the influence of the ionization parameter on our previous relation between the SFR and [OIII]$\lambda$5007 (Sect. \ref{sec:Calibration}). In Fig. \ref{fig17},  we present  $\mathrm{SFR}/\mathrm{L}_{\mathrm{[OIII]}5007}$  from Eq. \ref{eq:5_fix} as a function of the ionization parameter that we derived from the fit with emission lines and the three different median metallicities in each bin. In this figure, the error bar symbols represent the median values for each bin of gas-phase metallicity with the standard deviation as the error. We measured median ratios of $(\mathrm{[OIII]}\lambda5007/\mathrm{H}\alpha)_{0.009} = 1.12$, $(\mathrm{[OIII]}\lambda5007/\mathrm{H}\alpha)_{0.011} = 0.79$ and $(\mathrm{[OIII]}\lambda5007/\mathrm{H}\alpha)_{0.014} = 0.54$ in each gas-phase metallicity bin. However similar average $\log\mathrm{U}$ values can lead to different $\mathrm{[OIII]}$/H$\alpha$ ratios due to variations with gas-phase metallicity. In Table \ref{table:4}, we show median values of the SFR-$\mathrm{L_{[OIII]\lambda5007}}$ ratio for each gas-phase metallicity in $0.5$~dex $\log\mathrm{U}$ bins for the $-3.5<\log\mathrm{U}<-2.5$ range in which the ratio seems to be quite stable. We measure a mean dispersion of $0.24$~dex in gas-phase metallicity and $1.1$~dex for the ionization parameter in the range $-3.5<\log\mathrm{U}<-2.5$. The effects of the ionization parameter on the SFR-$\mathrm{L_{[OIII]\lambda5007}}$ dispersion are dominant, but our sample covers only a range of $0.4$~dex in the gas-phase metallicity. A sample spanning over a larger range of metallicity is needed to explore in detail the relative influence of both parameters. Our results are consistent with previous studies \citep[e.g.,][]{PerezMontero2014, Kaasinen2018, Sanders2020, Topping2020}. 

\begin{table}
\caption{SFR-L$_\mathrm{[OIII]\lambda5007}$ ratio mean values in $0.5$~dex $\log\mathrm{U}$ bins for the three different metallicities.}              % title of Table
\label{table:4}      % is used to refer this table in the text
\centering                                      % used for centering table
\begin{tabular}{c | c c c }          % centered columns (4 columns)
\hline                     % inserts double horizontal lines
$\mathrm{Z_{gas}}$& $0.009$ & $0.011$ & $0.014$ \\
\hline\hline\\
& & $\log_{10} \left(\frac{\mathrm{SFR}_\mathrm{cigale}}{\mathrm{L}_{\mathrm{[OIII]}5007}}\right)$&  \\\\ % table heading
& & [M$_\odot$ yr$^{-1} \mathrm{erg^{-1}~s}$]&  \\\\ % table heading
\hline
$\log\mathrm{U}$ & & &\\
\hline\hline                                  % inserts single horizontal line
                                         % inserting body of the table
    % -1.5 -2.0         & -41.68 & -41.72 & -41.38\\
    -2.0 -2.5         & -41.54 & -41.44 & -41.25\\ 
    -2.5 -3.0         & -41.35 & -41.31 & -41.04\\ 
    -3.0 -3.5         & -41.15 & -41.10 & -40.78\\ 
    % -3.5 -4.0         & --------- & -41.63 & ---------\\
\hline                                             %inserts single line
\end{tabular}
\end{table}

In a future work, we plan to explore the effects of using SED fitting with CIGALE implementing BPASS to explore how the locus of galaxies in the [NII]-BPT is affected, leaving the abundance ratio of (N/O) as a free parameter to introduce flexibility in HII-region models and checking also different ways of modeling the emission lines. The SED fitting and HII-region model coupling remains paramount in order to perform homogeneous analysis of a sample of galaxies and break degeneracies between the different parameters involved. Fully understanding of working with spectro-photometric samples and SED fitting is needed as a preparation for new instruments such as PFS and MOONS.

% %-----------------------------------------------------------------
\section{Summary and Conclusions}\label{sec:conclusions}

In this work, we perform SED fitting using CIGALE on an FMOS-COSMOS spectro-photometric sample covering UV-to-FIR continuum emission with $21$ broad-band fluxes and emission lines at $\mathrm{z}\sim1.6$. A sample of $183$ objects was selected to have flux measurements of both H$\alpha$ and [OIII]$\lambda$5007 at $\mathrm{S/N} > 3$ in the FMOS survey. From SED fitting of both photometric and H$\alpha$ fluxes, we estimate SFR and stellar mass%covering a range of $9.5-11.5~\mathrm{M_\odot}$ and $1-3~\mathrm{M_\odot}~\mathrm{yr}^{-1}$
, and constrain dust attenuation affecting the H$\alpha$ and [OIII]$\lambda$5007 emission lines. We measure median values of the emission line attenuation of A$_\mathrm{H\alpha} = 1.16\pm0.19$~mag and A$_\mathrm{[OIII]} = 1.41\pm0.22$~mag, respectively. Both A$_\mathrm{H\alpha}$ and A$_\mathrm{[OIII]}$ increase with stellar mass with a larger attenuation correction for [OIII]$\lambda5007$ emission line as compared to H$\alpha$. We find a flatter effective attenuation curve than the Milky Way or C00 curves. A relation to obtain the attenuation for the [OIII]$\lambda$5007 line as a function of stellar mass is proposed in the same way as it exists for H$\alpha$. This relation could be useful for inferring average values of the SFR of a sample, but not for individual galaxies. The relative attenuation affecting different populations is characterized by the $\mu$ parameter in the attenuation law. We find a value of $\mu = 0.57 \pm 0.14$ consistent with different works in the literature and twice as large as compared to the original value proposed by \cite{CF00}. 

A SFR-L$_\mathrm{[OIII]\lambda5007}$ dust-corrected relation is derived. We measure a slope consistent with unity within the $2\sigma$ dispersion of the relation. We estimate a [OIII]$\lambda5007$/[OIII] $88 ~\mu$m ratio of $1.90$ for our sample of galaxies and deduce A SFR-[OIII] $88 ~\mu$m relation in agreement with previous relations that were found at both low and high redshifts, although [OIII]$\lambda5007$/[OIII] $88 ~\mu$m is strongly dependent on electron density and gas-phase metallicity. The SED fitting, including photometry, H$\alpha$, H$\beta$, and [OIII]$\lambda$5007 fluxes, is also performed with a refined grid of photoionization models and metallicities estimated from the mass-metallicity relation from \citet{Curti2020}. The variations of gas-phase metallicity and ionization parameter induce a dispersion in the SFR-L$_\mathrm{[OIII]\lambda5007}$ relation of $0.24$~dex and $1.1$~dex, respectively. The lower impact of gas-phase metallicity is likely to be due the limited range of our sample ($0.006 < \mathrm{Z_{gas}} < 0.016$) and our relation of SFR-L$_\mathrm{[OIII]\lambda5007}$ is expected to be only valid for galaxies of similar gas-phase metallicities as those studied in this work.
   
\begin{acknowledgements}
We are grateful to Daichi Kashino for sharing the FMOS private catalog, and William Bowman, Alessia Longobardi, Yannick Rhoelly, Katarzyna Małek, Miguel Figueira, and Akio Inoue for useful conversations and discussion. We thank the anonymous referee for their insightful comments to improve the quality of this work. The project has received funding from Excellence Initiative of Aix-Marseille University - AMIDEX, a French ``Investissements d'Avenir'' program. It was supported by the Programme National “Physique et Chimie du Milieu Interstellaire” (PCMI) of CNRS/INSU with INC/INP co-funded by CEA and CNES. MB acknowledges FONDECYT regular grants 1170618 and 1211000.
\end{acknowledgements}

%% The following lines are required when using BibTEX (strongly encouraged!):
\bibliographystyle{aa}  % A&A bibliography style file (aa.bst)
\bibliography{aa} % your references in file: Yourfile.bib

\begin{thebibliography}{144}
\expandafter\ifx\csname natexlab\endcsname\relax\def\natexlab#1{#1}\fi

\bibitem[{{Allen} {et~al.}(2008){Allen}, {Groves}, {Dopita}, {Sutherland}, \&
  {Kewley}}]{Allen2008}
{Allen}, M.~G., {Groves}, B.~A., {Dopita}, M.~A., {Sutherland}, R.~S., \&
  {Kewley}, L.~J. 2008, \apjs, 178, 20

\bibitem[{{{\'A}lvarez-M{\'a}rquez} {et~al.}(2019){{\'A}lvarez-M{\'a}rquez},
  {Colina}, {Marques-Chaves}, {Ceverino}, {Alonso-Herrero}, {Caputi},
  {Garc{\'\i}a-Mar{\'\i}n}, {Labiano}, {Le F{\`e}vre}, {Norgaard-Nielsen},
  {{\"O}stlin}, {P{\'e}rez-Gonz{\'a}lez}, {Pye}, {Tikkanen}, {van der Werf},
  {Walter}, \& {Wright}}]{Marquez2019}
{{\'A}lvarez-M{\'a}rquez}, J., {Colina}, L., {Marques-Chaves}, R., {et~al.}
  2019, \aap, 629, A9

\bibitem[{{Andrews} \& {Martini}(2013)}]{Andrews_Martini2013}
{Andrews}, B.~H. \& {Martini}, P. 2013, \apj, 765, 140

\bibitem[{{Andrews} {et~al.}(2018){Andrews}, {Driver}, {Davies}, {Lagos}, \&
  {Robotham}}]{Andrews2018}
{Andrews}, S.~K., {Driver}, S.~P., {Davies}, L.~J.~M., {Lagos}, C. d.~P., \&
  {Robotham}, A.~S.~G. 2018, \mnras, 474, 898

\bibitem[{{Arata} {et~al.}(2020){Arata}, {Yajima}, {Nagamine}, {Abe}, \&
  {Khochfar}}]{Arata2020}
{Arata}, S., {Yajima}, H., {Nagamine}, K., {Abe}, M., \& {Khochfar}, S. 2020,
  \mnras, 498, 5541

\bibitem[{{Arnouts} {et~al.}(1999){Arnouts}, {Cristiani}, {Moscardini},
  {Matarrese}, {Lucchin}, {Fontana}, \& {Giallongo}}]{Arnouts1999}
{Arnouts}, S., {Cristiani}, S., {Moscardini}, L., {et~al.} 1999, \mnras, 310,
  540

\bibitem[{{Asplund} {et~al.}(2009){Asplund}, {Grevesse}, {Sauval}, \&
  {Scott}}]{Asplund2009}
{Asplund}, M., {Grevesse}, N., {Sauval}, A.~J., \& {Scott}, P. 2009, \araa, 47,
  481

\bibitem[{{Baldwin} {et~al.}(1981){Baldwin}, {Phillips}, \&
  {Terlevich}}]{Baldwin1981}
{Baldwin}, J.~A., {Phillips}, M.~M., \& {Terlevich}, R. 1981, \pasp, 93, 5

\bibitem[{{Battisti} {et~al.}(2016){Battisti}, {Calzetti}, \&
  {Chary}}]{Battisti2016}
{Battisti}, A.~J., {Calzetti}, D., \& {Chary}, R.~R. 2016, \apj, 818, 13

\bibitem[{{Battisti} {et~al.}(2019){Battisti}, {da Cunha}, {Grasha}, {Salvato},
  {Daddi}, {Davies}, {Jin}, {Liu}, {Schinnerer}, {Vaccari}, \& {COSMOS
  Collaboration}}]{Battisti2019}
{Battisti}, A.~J., {da Cunha}, E., {Grasha}, K., {et~al.} 2019, \apj, 882, 61

\bibitem[{{Bian} {et~al.}(2020){Bian}, {Kewley}, {Groves}, \&
  {Dopita}}]{Bian2020}
{Bian}, F., {Kewley}, L.~J., {Groves}, B., \& {Dopita}, M.~A. 2020, \mnras,
  493, 580

\bibitem[{{Boquien} {et~al.}(2019){Boquien}, {Burgarella}, {Roehlly}, {Buat},
  {Ciesla}, {Corre}, {Inoue}, \& {Salas}}]{Boquien2019}
{Boquien}, M., {Burgarella}, D., {Roehlly}, Y., {et~al.} 2019, \aap, 622, A103

\bibitem[{{Bourne} {et~al.}(2012){Bourne}, {Maddox}, {Dunne}, {Auld}, {Baes},
  {Baldry}, {Bonfield}, {Cooray}, {Croom}, {Dariush}, {de Zotti}, {Driver},
  {Dye}, {Eales}, {Gomez}, {Gonz{\'a}lez-Nuevo}, {Hopkins}, {Ibar}, {Jarvis},
  {Lapi}, {Madore}, {Micha{\l}owski}, {Pohlen}, {Popescu}, {Rigby}, {Seibert},
  {Smith}, {Tuffs}, {van der Werf}, {Brough}, {Buttiglione}, {Cava},
  {Clements}, {Conselice}, {Fritz}, {Hopwood}, {Ivison}, {Jones}, {Kelvin},
  {Liske}, {Loveday}, {Norberg}, {Robotham}, {Rodighiero}, \&
  {Temi}}]{Bourne2012}
{Bourne}, N., {Maddox}, S.~J., {Dunne}, L., {et~al.} 2012, \mnras, 421, 3027

\bibitem[{{Bowman} {et~al.}(2019){Bowman}, {Zeimann}, {Ciardullo}, {Gronwall},
  {Schneider}, {McCarron}, {Weiss}, {Yang}, \& {Hagen}}]{Bowman2019}
{Bowman}, W.~P., {Zeimann}, G.~R., {Ciardullo}, R., {et~al.} 2019, \apj, 875,
  152

\bibitem[{{Bowman} {et~al.}(2020){Bowman}, {Zeimann}, {Nagaraj}, {Ciardullo},
  {Gronwall}, {McCarron}, {Weiss}, {Molina}, {Belles}, \&
  {Schneider}}]{Bowman2020}
{Bowman}, W.~P., {Zeimann}, G.~R., {Nagaraj}, G., {et~al.} 2020, \apj, 899, 7

\bibitem[{{Brinchmann} {et~al.}(2008){Brinchmann}, {Pettini}, \&
  {Charlot}}]{Brinchmann2008}
{Brinchmann}, J., {Pettini}, M., \& {Charlot}, S. 2008, \mnras, 385, 769

\bibitem[{{Bruzual} \& {Charlot}(2003)}]{BC03}
{Bruzual}, G. \& {Charlot}, S. 2003, \mnras, 344, 1000

\bibitem[{{Buat} {et~al.}(2018){Buat}, {Boquien}, {Ma{\l}ek}, {Corre}, {Salas},
  {Roehlly}, {Shirley}, \& {Efstathiou}}]{Buat2018}
{Buat}, V., {Boquien}, M., {Ma{\l}ek}, K., {et~al.} 2018, \aap, 619, A135

\bibitem[{{Buat} {et~al.}(2019){Buat}, {Ciesla}, {Boquien}, {Ma{\l}ek}, \&
  {Burgarella}}]{Buat2019a}
{Buat}, V., {Ciesla}, L., {Boquien}, M., {Ma{\l}ek}, K., \& {Burgarella}, D.
  2019, \aap, 632, A79

\bibitem[{{Byler} {et~al.}(2017){Byler}, {Dalcanton}, {Conroy}, \&
  {Johnson}}]{Byler2017}
{Byler}, N., {Dalcanton}, J.~J., {Conroy}, C., \& {Johnson}, B.~D. 2017, \apj,
  840, 44

\bibitem[{{Calzetti} {et~al.}(2000){Calzetti}, {Armus}, {Bohlin}, {Kinney},
  {Koornneef}, \& {Storchi-Bergmann}}]{Calzetti2000}
{Calzetti}, D., {Armus}, L., {Bohlin}, R.~C., {et~al.} 2000, \apj, 533, 682

\bibitem[{{Capak} {et~al.}(2012){Capak}, {Aussel}, {Bundy}, {Carollo}, {Chary},
  {Civano}, {Coupon}, {Diener}, {Donley}, {Dunlop}, {Elvis}, {Foucaud},
  {Green}, {Gunn}, {Hashimoto}, {Hassinger}, {Hsieh}, {Huang}, {Ilbert},
  {LeFloc'h}, {LeFevre}, {Lilly}, {Lin}, {Lin}, {Miyazaki}, {Mobasher},
  {Moriya}, {Nagao}, {Ono}, {Ouchi}, {Quimby}, {Saito}, {Salvato}, {Sand ers},
  {Schinnerer}, {Scoville}, {Shimasaku}, {Silverman}, {Smolcic}, {Strauss},
  {Surace}, {Tanaka}, {Taniguchi}, {Teplitz}, {Wang}, \& {Urata}}]{Capak2012}
{Capak}, P., {Aussel}, H., {Bundy}, K., {et~al.} 2012, {SPLASH: Spitzer Large
  Area Survey with Hyper-Suprime-Cam}, Spitzer Proposal

\bibitem[{{Caplan} \& {Deharveng}(1986)}]{Caplan1986}
{Caplan}, J. \& {Deharveng}, L. 1986, \aap, 155, 297

\bibitem[{{Cardelli} {et~al.}(1989){Cardelli}, {Clayton}, \&
  {Mathis}}]{Cardelli1989}
{Cardelli}, J.~A., {Clayton}, G.~C., \& {Mathis}, J.~S. 1989, \apj, 345, 245

\bibitem[{{Carnall} {et~al.}(2019){Carnall}, {Leja}, {Johnson}, {McLure},
  {Dunlop}, \& {Conroy}}]{Carnall2019}
{Carnall}, A.~C., {Leja}, J., {Johnson}, B.~D., {et~al.} 2019, \apj, 873, 44

\bibitem[{{Carnall} {et~al.}(2018){Carnall}, {McLure}, {Dunlop}, \&
  {Dav{\'e}}}]{Carnall2018}
{Carnall}, A.~C., {McLure}, R.~J., {Dunlop}, J.~S., \& {Dav{\'e}}, R. 2018,
  \mnras, 480, 4379

\bibitem[{{Casey}(2012)}]{Casey2012}
{Casey}, C.~M. 2012, \mnras, 425, 3094

\bibitem[{{Chabrier}(2003)}]{Chabrier2003}
{Chabrier}, G. 2003, \pasp, 115, 763

\bibitem[{{Charlot} \& {Fall}(2000)}]{CF00}
{Charlot}, S. \& {Fall}, S.~M. 2000, \apj, 539, 718

\bibitem[{{Chevallard} \& {Charlot}(2016)}]{Chevallard16}
{Chevallard}, J. \& {Charlot}, S. 2016, \mnras, 462, 1415

\bibitem[{{Chevallard} {et~al.}(2019){Chevallard}, {Curtis-Lake}, {Charlot},
  {Ferruit}, {Giardino}, {Franx}, {Maseda}, {Amorin}, {Arribas}, {Bunker},
  {Carniani}, {Husemann}, {Jakobsen}, {Maiolino}, {Pforr}, {Rawle}, {Rix},
  {Smit}, \& {Willott}}]{Chevallard2019}
{Chevallard}, J., {Curtis-Lake}, E., {Charlot}, S., {et~al.} 2019, \mnras, 483,
  2621

\bibitem[{{Ciesla} {et~al.}(2017){Ciesla}, {Elbaz}, \& {Fensch}}]{Ciesla2017}
{Ciesla}, L., {Elbaz}, D., \& {Fensch}, J. 2017, \aap, 608, A41

\bibitem[{{Civano} {et~al.}(2016){Civano}, {Marchesi}, {Comastri}, {Urry},
  {Elvis}, {Cappelluti}, {Puccetti}, {Brusa}, {Zamorani}, {Hasinger},
  {Aldcroft}, {Alexander}, {Allevato}, {Brunner}, {Capak}, {Finoguenov},
  {Fiore}, {Fruscione}, {Gilli}, {Glotfelty}, {Griffiths}, {Hao}, {Harrison},
  {Jahnke}, {Kartaltepe}, {Karim}, {LaMassa}, {Lanzuisi}, {Miyaji}, {Ranalli},
  {Salvato}, {Sargent}, {Scoville}, {Schawinski}, {Schinnerer}, {Silverman},
  {Smolcic}, {Stern}, {Toft}, {Trakhtenbrot}, {Treister}, \&
  {Vignali}}]{Civano2016}
{Civano}, F., {Marchesi}, S., {Comastri}, A., {et~al.} 2016, \apj, 819, 62

\bibitem[{{Corre} {et~al.}(2018){Corre}, {Buat}, {Basa}, {Boissier}, {Japelj},
  {Palmerio}, {Salvaterra}, {Vergani}, \& {Zafar}}]{Corre2018}
{Corre}, D., {Buat}, V., {Basa}, S., {et~al.} 2018, \aap, 617, A141

\bibitem[{{Cortese} {et~al.}(2012){Cortese}, {Ciesla}, {Boselli}, {Bianchi},
  {Gomez}, {Smith}, {Bendo}, {Eales}, {Pohlen}, {Baes}, {Corbelli}, {Davies},
  {Hughes}, {Hunt}, {Madden}, {Pierini}, {di Serego Alighieri}, {Zibetti},
  {Boquien}, {Clements}, {Cooray}, {Galametz}, {Magrini}, {Pappalardo},
  {Spinoglio}, \& {Vlahakis}}]{Cortese2012}
{Cortese}, L., {Ciesla}, L., {Boselli}, A., {et~al.} 2012, \aap, 540, A52

\bibitem[{{Cowie} {et~al.}(2016){Cowie}, {Barger}, \& {Songaila}}]{Cowie2016}
{Cowie}, L.~L., {Barger}, A.~J., \& {Songaila}, A. 2016, \apj, 817, 57

\bibitem[{{Curti} {et~al.}(2020){Curti}, {Mannucci}, {Cresci}, \&
  {Maiolino}}]{Curti2020}
{Curti}, M., {Mannucci}, F., {Cresci}, G., \& {Maiolino}, R. 2020, \mnras, 491,
  944

\bibitem[{{da Cunha} {et~al.}(2008){da Cunha}, {Charlot}, \&
  {Elbaz}}]{daCunha2008}
{da Cunha}, E., {Charlot}, S., \& {Elbaz}, D. 2008, \mnras, 388, 1595

\bibitem[{{Dale} {et~al.}(2014){Dale}, {Helou}, {Magdis}, {Armus},
  {D{\'\i}az-Santos}, \& {Shi}}]{Dale2014}
{Dale}, D.~A., {Helou}, G., {Magdis}, G.~E., {et~al.} 2014, \apj, 784, 83

\bibitem[{{De Looze} {et~al.}(2014){De Looze}, {Cormier}, {Lebouteiller},
  {Madden}, {Baes}, {Bendo}, {Boquien}, {Boselli}, {Clements}, {Cortese},
  {Cooray}, {Galametz}, {Galliano}, {Graci{\'a}-Carpio}, {Isaak}, {Karczewski},
  {Parkin}, {Pellegrini}, {R{\'e}my-Ruyer}, {Spinoglio}, {Smith}, \&
  {Sturm}}]{DeLooze2014}
{De Looze}, I., {Cormier}, D., {Lebouteiller}, V., {et~al.} 2014, \aap, 568,
  A62

\bibitem[{{Dinerstein}(1983)}]{Dinerstein1983}
{Dinerstein}, H.~L. 1983, in Planetary Nebulae, ed. L.~H. {Aller}, Vol. 103,
  79--88

\bibitem[{{Dobbels} {et~al.}(2020){Dobbels}, {Baes}, {Viaene}, {Bianchi},
  {Davies}, {Casasola}, {Clark}, {Fritz}, {Galametz}, {Galliano}, {Mosenkov},
  {Nersesian}, \& {Tr{\v{c}}ka}}]{Dobbels2020}
{Dobbels}, W., {Baes}, M., {Viaene}, S., {et~al.} 2020, \aap, 634, A57

\bibitem[{{Donley} {et~al.}(2012){Donley}, {Koekemoer}, {Brusa}, {Capak},
  {Cardamone}, {Civano}, {Ilbert}, {Impey}, {Kartaltepe}, {Miyaji}, {Salvato},
  {Sanders}, {Trump}, \& {Zamorani}}]{Donley2012}
{Donley}, J.~L., {Koekemoer}, A.~M., {Brusa}, M., {et~al.} 2012, \apj, 748, 142

\bibitem[{{Dopita} {et~al.}(2016){Dopita}, {Kewley}, {Sutherland}, \&
  {Nicholls}}]{Dopita2016}
{Dopita}, M.~A., {Kewley}, L.~J., {Sutherland}, R.~S., \& {Nicholls}, D.~C.
  2016, \apss, 361, 61

\bibitem[{{Dopita} {et~al.}(2013){Dopita}, {Sutherland}, {Nicholls}, {Kewley},
  \& {Vogt}}]{Dopita2013}
{Dopita}, M.~A., {Sutherland}, R.~S., {Nicholls}, D.~C., {Kewley}, L.~J., \&
  {Vogt}, F. P.~A. 2013, \apjs, 208, 10

\bibitem[{{Draine} {et~al.}(2014){Draine}, {Aniano}, {Krause}, {Groves},
  {Sandstrom}, {Braun}, {Leroy}, {Klaas}, {Linz}, {Rix}, {Schinnerer},
  {Schmiedeke}, \& {Walter}}]{Draine2014}
{Draine}, B.~T., {Aniano}, G., {Krause}, O., {et~al.} 2014, \apj, 780, 172

\bibitem[{{Draine} \& {Li}(2007)}]{Draine&Li2007}
{Draine}, B.~T. \& {Li}, A. 2007, \apj, 657, 810

\bibitem[{{Ellis} {et~al.}(2017){Ellis}, {Bland-Hawthorn}, {Bremer},
  {Brinchmann}, {Guzzo}, {Richard}, {Rix}, {Tolstoy}, \& {Watson}}]{Ellis2017}
{Ellis}, R.~S., {Bland-Hawthorn}, J., {Bremer}, M., {et~al.} 2017, arXiv
  e-prints, arXiv:1701.01976

\bibitem[{{Elvis} {et~al.}(2009){Elvis}, {Civano}, {Vignali}, {Puccetti},
  {Fiore}, {Cappelluti}, {Aldcroft}, {Fruscione}, {Zamorani}, {Comastri},
  {Brusa}, {Gilli}, {Miyaji}, {Damiani}, {Koekemoer}, {Finoguenov}, {Brunner},
  {Urry}, {Silverman}, {Mainieri}, {Hasinger}, {Griffiths}, {Carollo}, {Hao},
  {Guzzo}, {Blain}, {Calzetti}, {Carilli}, {Capak}, {Ettori}, {Fabbiano},
  {Impey}, {Lilly}, {Mobasher}, {Rich}, {Salvato}, {Sand ers}, {Schinnerer},
  {Scoville}, {Shopbell}, {Taylor}, {Taniguchi}, \& {Volonteri}}]{Elvis09}
{Elvis}, M., {Civano}, F., {Vignali}, C., {et~al.} 2009, \apjs, 184, 158

\bibitem[{{Ferkinhoff} {et~al.}(2010){Ferkinhoff}, {Hailey-Dunsheath},
  {Nikola}, {Parshley}, {Stacey}, {Benford}, \& {Staguhn}}]{Ferkinhoff2010}
{Ferkinhoff}, C., {Hailey-Dunsheath}, S., {Nikola}, T., {et~al.} 2010, \apjl,
  714, L147

\bibitem[{{Ferland} {et~al.}(2017){Ferland}, {Chatzikos}, {Guzm{\'a}n},
  {Lykins}, {van Hoof}, {Williams}, {Abel}, {Badnell}, {Keenan}, {Porter}, \&
  {Stancil}}]{Ferland2017}
{Ferland}, G.~J., {Chatzikos}, M., {Guzm{\'a}n}, F., {et~al.} 2017, \rmxaa, 53,
  385

\bibitem[{{Fossati} {et~al.}(2018){Fossati}, {Mendel}, {Boselli}, {Cuilland
  re}, {Vollmer}, {Boissier}, {Consolandi}, {Ferrarese}, {Gwyn}, {Amram},
  {Boquien}, {Buat}, {Burgarella}, {Cortese}, {C{\^o}t{\'e}}, {C{\^o}t{\'e}},
  {Durrell}, {Fumagalli}, {Gavazzi}, {Gomez-Lopez}, {Hensler}, {Koribalski},
  {Longobardi}, {Peng}, {Roediger}, {Sun}, \& {Toloba}}]{Fossati2018}
{Fossati}, M., {Mendel}, J.~T., {Boselli}, A., {et~al.} 2018, \aap, 614, A57

\bibitem[{{Fritz} {et~al.}(2006){Fritz}, {Franceschini}, \&
  {Hatziminaoglou}}]{Fritz2006}
{Fritz}, J., {Franceschini}, A., \& {Hatziminaoglou}, E. 2006, \mnras, 366, 767

\bibitem[{{Garn} \& {Best}(2010{\natexlab{a}})}]{GarnBest2010}
{Garn}, T. \& {Best}, P.~N. 2010{\natexlab{a}}, \mnras, 409, 421

\bibitem[{{Garn} \& {Best}(2010{\natexlab{b}})}]{Garn2010}
{Garn}, T. \& {Best}, P.~N. 2010{\natexlab{b}}, \mnras, 409, 421

\bibitem[{{Grevesse} {et~al.}(2010){Grevesse}, {Asplund}, {Sauval}, \&
  {Scott}}]{Grevesse2010}
{Grevesse}, N., {Asplund}, M., {Sauval}, A.~J., \& {Scott}, P. 2010, \apss,
  328, 179

\bibitem[{{Gutkin} {et~al.}(2016){Gutkin}, {Charlot}, \&
  {Bruzual}}]{Gutkin2016}
{Gutkin}, J., {Charlot}, S., \& {Bruzual}, G. 2016, \mnras, 462, 1757

\bibitem[{{Harikane} {et~al.}(2020){Harikane}, {Ouchi}, {Inoue}, {Matsuoka},
  {Tamura}, {Bakx}, {Fujimoto}, {Moriwaki}, {Ono}, {Nagao}, {Tadaki}, {Kojima},
  {Shibuya}, {Egami}, {Ferrara}, {Gallerani}, {Hashimoto}, {Kohno}, {Matsuda},
  {Matsuo}, {Pallottini}, {Sugahara}, \& {Vallini}}]{Harikane2020}
{Harikane}, Y., {Ouchi}, M., {Inoue}, A.~K., {et~al.} 2020, \apj, 896, 93

\bibitem[{{Hippelein} {et~al.}(2003){Hippelein}, {Maier}, {Meisenheimer},
  {Wolf}, {Fried}, {von Kuhlmann}, {K{\"u}mmel}, {Phleps}, \&
  {R{\"o}ser}}]{Hippelein2003}
{Hippelein}, H., {Maier}, C., {Meisenheimer}, K., {et~al.} 2003, \aap, 402, 65

\bibitem[{{Hurley} {et~al.}(2017){Hurley}, {Oliver}, {Betancourt}, {Clarke},
  {Cowley}, {Duivenvoorden}, {Farrah}, {Griffin}, {Lacey}, {Le Floc'h},
  {Papadopoulos}, {Sargent}, {Scudder}, {Vaccari}, {Valtchanov}, \&
  {Wang}}]{Hurley2017}
{Hurley}, P.~D., {Oliver}, S., {Betancourt}, M., {et~al.} 2017, \mnras, 464,
  885

\bibitem[{{Ibar} {et~al.}(2013){Ibar}, {Sobral}, {Best}, {Ivison}, {Smail},
  {Arumugam}, {Berta}, {B{\'e}thermin}, {Bock}, {Cava}, {Conley}, {Farrah},
  {Geach}, {Ikarashi}, {Kohno}, {Le Floc'h}, {Lutz}, {Magdis}, {Magnelli},
  {Marsden}, {Oliver}, {Page}, {Pozzi}, {Riguccini}, {Schulz}, {Seymour},
  {Smith}, {Symeonidis}, {Wang}, {Wardlow}, \& {Zemcov}}]{Ibar2013}
{Ibar}, E., {Sobral}, D., {Best}, P.~N., {et~al.} 2013, \mnras, 434, 3218

\bibitem[{{Ilbert} {et~al.}(2006){Ilbert}, {Arnouts}, {McCracken},
  {Bolzonella}, {Bertin}, {Le F{\`e}vre}, {Mellier}, {Zamorani}, {Pell{\`o}},
  {Iovino}, {Tresse}, {Le Brun}, {Bottini}, {Garilli}, {Maccagni}, {Picat},
  {Scaramella}, {Scodeggio}, {Vettolani}, {Zanichelli}, {Adami}, {Bardelli},
  {Cappi}, {Charlot}, {Ciliegi}, {Contini}, {Cucciati}, {Foucaud}, {Franzetti},
  {Gavignaud}, {Guzzo}, {Marano}, {Marinoni}, {Mazure}, {Meneux}, {Merighi},
  {Paltani}, {Pollo}, {Pozzetti}, {Radovich}, {Zucca}, {Bondi}, {Bongiorno},
  {Busarello}, {de La Torre}, {Gregorini}, {Lamareille}, {Mathez}, {Merluzzi},
  {Ripepi}, {Rizzo}, \& {Vergani}}]{Ilbert2006}
{Ilbert}, O., {Arnouts}, S., {McCracken}, H.~J., {et~al.} 2006, \aap, 457, 841

\bibitem[{{Jin} {et~al.}(2018){Jin}, {Daddi}, {Liu}, {Smol{\v{c}}i{\'c}},
  {Schinnerer}, {Calabr{\`o}}, {Gu}, {Delhaize}, {Delvecchio}, {Gao},
  {Salvato}, {Puglisi}, {Dickinson}, {Bertoldi}, {Sargent}, {Novak}, {Magdis},
  {Aretxaga}, {Wilson}, \& {Capak}}]{Jin2018}
{Jin}, S., {Daddi}, E., {Liu}, D., {et~al.} 2018, \apj, 864, 56

\bibitem[{{Johnson} {et~al.}(2021){Johnson}, {Leja}, {Conroy}, \&
  {Speagle}}]{Johnson2021}
{Johnson}, B.~D., {Leja}, J., {Conroy}, C., \& {Speagle}, J.~S. 2021, \apjs,
  254, 22

\bibitem[{{Kaasinen} {et~al.}(2018){Kaasinen}, {Kewley}, {Bian}, {Groves},
  {Kashino}, {Silverman}, \& {Kartaltepe}}]{Kaasinen2018}
{Kaasinen}, M., {Kewley}, L., {Bian}, F., {et~al.} 2018, \mnras, 477, 5568

\bibitem[{{Karim} {et~al.}(2011){Karim}, {Schinnerer},
  {Mart{\'\i}nez-Sansigre}, {Sargent}, {van der Wel}, {Rix}, {Ilbert},
  {Smol{\v{c}}i{\'c}}, {Carilli}, {Pannella}, {Koekemoer}, {Bell}, \&
  {Salvato}}]{Karim2011}
{Karim}, A., {Schinnerer}, E., {Mart{\'\i}nez-Sansigre}, A., {et~al.} 2011,
  \apj, 730, 61

\bibitem[{{Kashino} {et~al.}(2013){Kashino}, {Silverman}, {Rodighiero},
  {Renzini}, {Arimoto}, {Daddi}, {Lilly}, {Sanders}, {Kartaltepe}, {Zahid},
  {Nagao}, {Sugiyama}, {Capak}, {Carollo}, {Chu}, {Hasinger}, {Ilbert},
  {Kajisawa}, {Kewley}, {Koekemoer}, {Kova{\v{c}}}, {Le F{\`e}vre}, {Masters},
  {McCracken}, {Onodera}, {Scoville}, {Strazzullo}, {Symeonidis}, \&
  {Taniguchi}}]{kashino2013}
{Kashino}, D., {Silverman}, J.~D., {Rodighiero}, G., {et~al.} 2013, \apjl, 777,
  L8

\bibitem[{{Kashino} {et~al.}(2019){Kashino}, {Silverman}, {Sanders},
  {Kartaltepe}, {Daddi}, {Renzini}, {Rodighiero}, {Puglisi}, {Valentino},
  {Juneau}, {Arimoto}, {Nagao}, {Ilbert}, {Le F{\`e}vre}, \&
  {Koekemoer}}]{Kashino2019}
{Kashino}, D., {Silverman}, J.~D., {Sanders}, D., {et~al.} 2019, \apjs, 241, 10

\bibitem[{{Kashino} {et~al.}(2017){Kashino}, {Silverman}, {Sanders},
  {Kartaltepe}, {Daddi}, {Renzini}, {Valentino}, {Rodighiero}, {Juneau},
  {Kewley}, {Zahid}, {Arimoto}, {Nagao}, {Chu}, {Sugiyama}, {Civano}, {Ilbert},
  {Kajisawa}, {Le F{\`e}vre}, {Maier}, {Masters}, {Miyaji}, {Onodera},
  {Puglisi}, \& {Taniguchi}}]{Kashino2017}
{Kashino}, D., {Silverman}, J.~D., {Sanders}, D., {et~al.} 2017, \apj, 835, 88

\bibitem[{{Kauffmann} {et~al.}(2003){Kauffmann}, {Heckman}, {Tremonti},
  {Brinchmann}, {Charlot}, {White}, {Ridgway}, {Brinkmann}, {Fukugita}, {Hall},
  {Ivezi{\'c}}, {Richards}, \& {Schneider}}]{Kaufmann2003}
{Kauffmann}, G., {Heckman}, T.~M., {Tremonti}, C., {et~al.} 2003, \mnras, 346,
  1055

\bibitem[{{Kennicutt}(1992)}]{Kennicutt1992}
{Kennicutt}, Robert~C., J. 1992, \apj, 388, 310

\bibitem[{{Kennicutt}(1998)}]{Kennicutt1998}
{Kennicutt}, Robert~C., J. 1998, \araa, 36, 189

\bibitem[{{Kennicutt} {et~al.}(2000){Kennicutt}, {Bresolin}, {French}, \&
  {Martin}}]{Kennicutt2000}
{Kennicutt}, Robert~C., J., {Bresolin}, F., {French}, H., \& {Martin}, P. 2000,
  \apj, 537, 589

\bibitem[{Kewley {et~al.}(2013)Kewley, Dopita, Leitherer, Dav{\'{e}}, Yuan,
  Allen, Groves, \& Sutherland}]{Kewley2013b}
Kewley, L.~J., Dopita, M.~A., Leitherer, C., {et~al.} 2013, The Astrophysical
  Journal, 774, 100

\bibitem[{{Kewley} {et~al.}(2001){Kewley}, {Dopita}, {Sutherland}, {Heisler},
  \& {Trevena}}]{Kewley2001}
{Kewley}, L.~J., {Dopita}, M.~A., {Sutherland}, R.~S., {Heisler}, C.~A., \&
  {Trevena}, J. 2001, \apj, 556, 121

\bibitem[{{Kewley} {et~al.}(2013){Kewley}, {Maier}, {Yabe}, {Ohta}, {Akiyama},
  {Dopita}, \& {Yuan}}]{Kewley2013}
{Kewley}, L.~J., {Maier}, C., {Yabe}, K., {et~al.} 2013, \apjl, 774, L10

\bibitem[{{Kojima} {et~al.}(2017){Kojima}, {Ouchi}, {Nakajima}, {Shibuya},
  {Harikane}, \& {Ono}}]{Kojima2017}
{Kojima}, T., {Ouchi}, M., {Nakajima}, K., {et~al.} 2017, \pasj, 69, 44

\bibitem[{{Komatsu} {et~al.}(2011){Komatsu}, {Smith}, {Dunkley}, {Bennett},
  {Gold}, {Hinshaw}, {Jarosik}, {Larson}, {Nolta}, {Page}, {Spergel},
  {Halpern}, {Hill}, {Kogut}, {Limon}, {Meyer}, {Odegard}, {Tucker}, {Weiland},
  {Wollack}, \& {Wright}}]{Komatsu2011}
{Komatsu}, E., {Smith}, K.~M., {Dunkley}, J., {et~al.} 2011, \apjs, 192, 18

\bibitem[{{Koyama} {et~al.}(2019){Koyama}, {Shimakawa}, {Yamamura}, {Kodama},
  \& {Hayashi}}]{Koyama2019}
{Koyama}, Y., {Shimakawa}, R., {Yamamura}, I., {Kodama}, T., \& {Hayashi}, M.
  2019, \pasj, 71, 8

\bibitem[{{Laigle} {et~al.}(2016){Laigle}, {McCracken}, {Ilbert}, {Hsieh},
  {Davidzon}, {Capak}, {Hasinger}, {Silverman}, {Pichon}, {Coupon}, {Aussel},
  {Le Borgne}, {Caputi}, {Cassata}, {Chang}, {Civano}, {Dunlop}, {Fynbo},
  {Kartaltepe}, {Koekemoer}, {Le F{\`e}vre}, {Le Floc'h}, {Leauthaud}, {Lilly},
  {Lin}, {Marchesi}, {Milvang-Jensen}, {Salvato}, {Sanders}, {Scoville},
  {Smolcic}, {Stockmann}, {Taniguchi}, {Tasca}, {Toft}, {Vaccari}, \&
  {Zabl}}]{Laigle2016}
{Laigle}, C., {McCracken}, H.~J., {Ilbert}, O., {et~al.} 2016, \apjs, 224, 24

\bibitem[{{Le Floc'h} {et~al.}(2009){Le Floc'h}, {Aussel}, {Ilbert},
  {Riguccini}, {Frayer}, {Salvato}, {Arnouts}, {Surace}, {Feruglio},
  {Rodighiero}, {Capak}, {Kartaltepe}, {Heinis}, {Sheth}, {Yan}, {McCracken},
  {Thompson}, {Sanders}, {Scoville}, \& {Koekemoer}}]{LeFloch2009}
{Le Floc'h}, E., {Aussel}, H., {Ilbert}, O., {et~al.} 2009, \apj, 703, 222

\bibitem[{{Leja} {et~al.}(2019){Leja}, {Carnall}, {Johnson}, {Conroy}, \&
  {Speagle}}]{Leja2019}
{Leja}, J., {Carnall}, A.~C., {Johnson}, B.~D., {Conroy}, C., \& {Speagle},
  J.~S. 2019, \apj, 876, 3

\bibitem[{{Leja} {et~al.}(2017){Leja}, {Johnson}, {Conroy}, {van Dokkum}, \&
  {Byler}}]{Leja2017}
{Leja}, J., {Johnson}, B.~D., {Conroy}, C., {van Dokkum}, P.~G., \& {Byler}, N.
  2017, \apj, 837, 170

\bibitem[{{Levesque} {et~al.}(2010){Levesque}, {Kewley}, \&
  {Larson}}]{Levesque2010}
{Levesque}, E.~M., {Kewley}, L.~J., \& {Larson}, K.~L. 2010, \aj, 139, 712

\bibitem[{{Liu} {et~al.}(2013){Liu}, {Calzetti}, {Hong}, {Whitmore}, {Chandar},
  {O'Connell}, {Blair}, {Cohen}, {Frogel}, \& {Kim}}]{Liu2013}
{Liu}, G., {Calzetti}, D., {Hong}, S., {et~al.} 2013, \apjl, 778, L41

\bibitem[{{Lo Faro} {et~al.}(2017){Lo Faro}, {Buat}, {Roehlly},
  {Alvarez-Marquez}, {Burgarella}, {Silva}, \& {Efstathiou}}]{LoFaro2017}
{Lo Faro}, B., {Buat}, V., {Roehlly}, Y., {et~al.} 2017, \mnras, 472, 1372

\bibitem[{{Lodders}(2010)}]{Lodders2010}
{Lodders}, K. 2010, Astrophysics and Space Science Proceedings, 16, 379

\bibitem[{{Ly} {et~al.}(2007){Ly}, {Malkan}, {Kashikawa}, {Shimasaku}, {Doi},
  {Nagao}, {Iye}, {Kodama}, {Morokuma}, \& {Motohara}}]{Ly2007}
{Ly}, C., {Malkan}, M.~A., {Kashikawa}, N., {et~al.} 2007, \apj, 657, 738

\bibitem[{{Ma{\l}ek} {et~al.}(2018){Ma{\l}ek}, {Buat}, {Roehlly}, {Burgarella},
  {Hurley}, {Shirley}, {Duncan}, {Efstathiou}, {Papadopoulos}, {Vaccari},
  {Farrah}, {Marchetti}, \& {Oliver}}]{Malek2018}
{Ma{\l}ek}, K., {Buat}, V., {Roehlly}, Y., {et~al.} 2018, \aap, 620, A50

\bibitem[{{Mannucci} {et~al.}(2010){Mannucci}, {Cresci}, {Maiolino}, {Marconi},
  \& {Gnerucci}}]{Mannucci2010}
{Mannucci}, F., {Cresci}, G., {Maiolino}, R., {Marconi}, A., \& {Gnerucci}, A.
  2010, \mnras, 408, 2115

\bibitem[{{Maschietto} {et~al.}(2008){Maschietto}, {Hatch}, {Venemans},
  {R{\"o}ttgering}, {Miley}, {Overzier}, {Dopita}, {Eisenhardt}, {Kurk},
  {Meurer}, {Pentericci}, {Rosati}, {Stanford}, {van Breugel}, \&
  {Zirm}}]{Maschietto2008}
{Maschietto}, F., {Hatch}, N.~A., {Venemans}, B.~P., {et~al.} 2008, \mnras,
  389, 1223

\bibitem[{{Masters} {et~al.}(2016){Masters}, {Faisst}, \&
  {Capak}}]{Masters2016}
{Masters}, D., {Faisst}, A., \& {Capak}, P. 2016, \apj, 828, 18

\bibitem[{{Masters} {et~al.}(2014){Masters}, {McCarthy}, {Siana}, {Malkan},
  {Mobasher}, {Atek}, {Henry}, {Martin}, {Rafelski}, {Hathi}, {Scarlata},
  {Ross}, {Bunker}, {Blanc}, {Bedregal}, {Dom{\'\i}nguez}, {Colbert},
  {Teplitz}, \& {Dressler}}]{Masters2014}
{Masters}, D., {McCarthy}, P., {Siana}, B., {et~al.} 2014, \apj, 785, 153

\bibitem[{{Moriwaki} {et~al.}(2018){Moriwaki}, {Yoshida}, {Shimizu},
  {Harikane}, {Matsuda}, {Matsuo}, {Hashimoto}, {Inoue}, {Tamura}, \&
  {Nagao}}]{Moriwaki2018}
{Moriwaki}, K., {Yoshida}, N., {Shimizu}, I., {et~al.} 2018, \mnras, 481, L84

\bibitem[{{Moustakas} {et~al.}(2006){Moustakas}, {Kennicutt}, \&
  {Tremonti}}]{Moustakas2006}
{Moustakas}, J., {Kennicutt}, Robert~C., J., \& {Tremonti}, C.~A. 2006, \apj,
  642, 775

\bibitem[{{Nersesian} {et~al.}(2019){Nersesian}, {Xilouris}, {Bianchi},
  {Galliano}, {Jones}, {Baes}, {Casasola}, {Cassar{\`a}}, {Clark}, {Davies},
  {Decleir}, {Dobbels}, {De Looze}, {De Vis}, {Fritz}, {Galametz}, {Madden},
  {Mosenkov}, {Tr{\v{c}}ka}, {Verstocken}, {Viaene}, \&
  {Lianou}}]{Narsesian2019}
{Nersesian}, A., {Xilouris}, E.~M., {Bianchi}, S., {et~al.} 2019, \aap, 624,
  A80

\bibitem[{{Nicholls} {et~al.}(2017){Nicholls}, {Sutherland}, {Dopita},
  {Kewley}, \& {Groves}}]{Nicholls2017}
{Nicholls}, D.~C., {Sutherland}, R.~S., {Dopita}, M.~A., {Kewley}, L.~J., \&
  {Groves}, B.~A. 2017, \mnras, 466, 4403

\bibitem[{{Nieva} \& {Przybilla}(2012)}]{Nieva2012}
{Nieva}, M.~F. \& {Przybilla}, N. 2012, \aap, 539, A143

\bibitem[{{O'Donnell}(1994)}]{ODonnell1994}
{O'Donnell}, J.~E. 1994, \apj, 422, 158

\bibitem[{{Osterbrock}(1989)}]{Osterbrock1989}
{Osterbrock}, D.~E. 1989, {Astrophysics of gaseous nebulae and active galactic
  nuclei} (University Science Books)

\bibitem[{{Pannella} {et~al.}(2015){Pannella}, {Elbaz}, {Daddi}, {Dickinson},
  {Hwang}, {Schreiber}, {Strazzullo}, {Aussel}, {Bethermin}, {Buat},
  {Charmandaris}, {Cibinel}, {Juneau}, {Ivison}, {Le Borgne}, {Le Floc'h},
  {Leiton}, {Lin}, {Magdis}, {Morrison}, {Mullaney}, {Onodera}, {Renzini},
  {Salim}, {Sargent}, {Scott}, {Shu}, \& {Wang}}]{Pannella2015}
{Pannella}, M., {Elbaz}, D., {Daddi}, E., {et~al.} 2015, \apj, 807, 141

\bibitem[{{P{\'e}rez-Montero}(2014)}]{PerezMontero2014}
{P{\'e}rez-Montero}, E. 2014, \mnras, 441, 2663

\bibitem[{{Pettini} \& {Pagel}(2004)}]{Pettini2004}
{Pettini}, M. \& {Pagel}, B. E.~J. 2004, \mnras, 348, L59

\bibitem[{{Popescu} {et~al.}(2000){Popescu}, {Misiriotis}, {Kylafis}, {Tuffs},
  \& {Fischera}}]{Popescu2000}
{Popescu}, C.~C., {Misiriotis}, A., {Kylafis}, N.~D., {Tuffs}, R.~J., \&
  {Fischera}, J. 2000, \aap, 362, 138

\bibitem[{{Puglisi} {et~al.}(2016){Puglisi}, {Rodighiero}, {Franceschini},
  {Talia}, {Cimatti}, {Baronchelli}, {Daddi}, {Renzini}, {Schawinski},
  {Mancini}, {Silverman}, {Gruppioni}, {Lutz}, {Berta}, \&
  {Oliver}}]{Puglisi2016}
{Puglisi}, A., {Rodighiero}, G., {Franceschini}, A., {et~al.} 2016, \aap, 586,
  A83

\bibitem[{{Qin} {et~al.}(2019){Qin}, {Zheng}, {Wuyts}, {Pan}, \&
  {Ren}}]{Qin2019}
{Qin}, J., {Zheng}, X.~Z., {Wuyts}, S., {Pan}, Z., \& {Ren}, J. 2019, \apj,
  886, 28

\bibitem[{{Reddy} {et~al.}(2015){Reddy}, {Kriek}, {Shapley}, {Freeman},
  {Siana}, {Coil}, {Mobasher}, {Price}, {Sanders}, \& {Shivaei}}]{Reddy2015}
{Reddy}, N.~A., {Kriek}, M., {Shapley}, A.~E., {et~al.} 2015, \apj, 806, 259

\bibitem[{{Robotham} {et~al.}(2020){Robotham}, {Bellstedt}, {Lagos}, {Thorne},
  {Davies}, {Driver}, \& {Bravo}}]{Robotham2020}
{Robotham}, A.~S.~G., {Bellstedt}, S., {Lagos}, C. d.~P., {et~al.} 2020,
  \mnras, 495, 905

\bibitem[{{Rodighiero} {et~al.}(2011){Rodighiero}, {Daddi}, {Baronchelli},
  {Cimatti}, {Renzini}, {Aussel}, {Popesso}, {Lutz}, {Andreani}, {Berta},
  {Cava}, {Elbaz}, {Feltre}, {Fontana}, {F{\"o}rster Schreiber},
  {Franceschini}, {Genzel}, {Grazian}, {Gruppioni}, {Ilbert}, {Le Floch},
  {Magdis}, {Magliocchetti}, {Magnelli}, {Maiolino}, {McCracken}, {Nordon},
  {Poglitsch}, {Santini}, {Pozzi}, {Riguccini}, {Tacconi}, {Wuyts}, \&
  {Zamorani}}]{Rodighiero2011}
{Rodighiero}, G., {Daddi}, E., {Baronchelli}, I., {et~al.} 2011, \apjl, 739,
  L40

\bibitem[{{Sanders} {et~al.}(2021){Sanders}, {Shapley}, {Jones}, {Reddy},
  {Kriek}, {Siana}, {Coil}, {Mobasher}, {Shivaei}, {Dav{\'e}}, {Azadi},
  {Price}, {Leung}, {Freeman}, {Fetherolf}, {de Groot}, {Zick}, \&
  {Barro}}]{Sanders2021}
{Sanders}, R.~L., {Shapley}, A.~E., {Jones}, T., {et~al.} 2021, \apj, 914, 19

\bibitem[{{Sanders} {et~al.}(2016){Sanders}, {Shapley}, {Kriek}, {Reddy},
  {Freeman}, {Coil}, {Siana}, {Mobasher}, {Shivaei}, {Price}, \& {de
  Groot}}]{Sanders2016}
{Sanders}, R.~L., {Shapley}, A.~E., {Kriek}, M., {et~al.} 2016, \apj, 816, 23

\bibitem[{{Sanders} {et~al.}(2020){Sanders}, {Shapley}, {Reddy}, {Kriek},
  {Siana}, {Coil}, {Mobasher}, {Shivaei}, {Freeman}, {Azadi}, {Price}, {Leung},
  {Fetherolf}, {de Groot}, {Zick}, {Fornasini}, \& {Barro}}]{Sanders2020}
{Sanders}, R.~L., {Shapley}, A.~E., {Reddy}, N.~A., {et~al.} 2020, \mnras, 491,
  1427

\bibitem[{{Santini} {et~al.}(2014){Santini}, {Maiolino}, {Magnelli}, {Lutz},
  {Lamastra}, {Li Causi}, {Eales}, {Andreani}, {Berta}, {Buat}, {Cooray},
  {Cresci}, {Daddi}, {Farrah}, {Fontana}, {Franceschini}, {Genzel}, {Granato},
  {Grazian}, {Le Floc'h}, {Magdis}, {Magliocchetti}, {Mannucci}, {Menci},
  {Nordon}, {Oliver}, {Popesso}, {Pozzi}, {Riguccini}, {Rodighiero}, {Rosario},
  {Salvato}, {Scott}, {Silva}, {Tacconi}, {Viero}, {Wang}, {Wuyts}, \&
  {Xu}}]{Santini2014}
{Santini}, P., {Maiolino}, R., {Magnelli}, B., {et~al.} 2014, \aap, 562, A30

\bibitem[{{Sargent} {et~al.}(2012){Sargent}, {B{\'e}thermin}, {Daddi}, \&
  {Elbaz}}]{Sargent2012}
{Sargent}, M.~T., {B{\'e}thermin}, M., {Daddi}, E., \& {Elbaz}, D. 2012, \apjl,
  747, L31

\bibitem[{{Schaerer} \& {de Barros}(2009)}]{Schaerer2009}
{Schaerer}, D. \& {de Barros}, S. 2009, \aap, 502, 423

\bibitem[{{Schaerer} \& {de Barros}(2010)}]{Schaerer2010}
{Schaerer}, D. \& {de Barros}, S. 2010, \aap, 515, A73

\bibitem[{{Schreiber} {et~al.}(2015){Schreiber}, {Pannella}, {Elbaz},
  {B{\'e}thermin}, {Inami}, {Dickinson}, {Magnelli}, {Wang}, {Aussel}, {Daddi},
  {Juneau}, {Shu}, {Sargent}, {Buat}, {Faber}, {Ferguson}, {Giavalisco},
  {Koekemoer}, {Magdis}, {Morrison}, {Papovich}, {Santini}, \&
  {Scott}}]{Schreiber2015}
{Schreiber}, C., {Pannella}, M., {Elbaz}, D., {et~al.} 2015, \aap, 575, A74

\bibitem[{{Schwarz}(1978)}]{Schwarz1978}
{Schwarz}, G. 1978, Annals of Statistics, 6, 461

\bibitem[{{Scoville} {et~al.}(2007){Scoville}, {Aussel}, {Brusa}, {Capak},
  {Carollo}, {Elvis}, {Giavalisco}, {Guzzo}, {Hasinger}, {Impey}, {Kneib},
  {LeFevre}, {Lilly}, {Mobasher}, {Renzini}, {Rich}, {Sanders}, {Schinnerer},
  {Schminovich}, {Shopbell}, {Taniguchi}, \& {Tyson}}]{Scoville07}
{Scoville}, N., {Aussel}, H., {Brusa}, M., {et~al.} 2007, \apjs, 172, 1

\bibitem[{{Shapley} {et~al.}(2015){Shapley}, {Reddy}, {Kriek}, {Freeman},
  {Sanders}, {Siana}, {Coil}, {Mobasher}, {Shivaei}, {Price}, \& {de
  Groot}}]{Shapley2015}
{Shapley}, A.~E., {Reddy}, N.~A., {Kriek}, M., {et~al.} 2015, \apj, 801, 88

\bibitem[{{Shimakawa} {et~al.}(2015){Shimakawa}, {Kodama}, {Steidel}, {Tadaki},
  {Tanaka}, {Strom}, {Hayashi}, {Koyama}, {Suzuki}, \&
  {Yamamoto}}]{Shimakawa2015}
{Shimakawa}, R., {Kodama}, T., {Steidel}, C.~C., {et~al.} 2015, \mnras, 451,
  1284

\bibitem[{{Shivaei} {et~al.}(2020){Shivaei}, {Reddy}, {Rieke}, {Shapley},
  {Kriek}, {Battisti}, {Mobasher}, {Sanders}, {Fetherolf}, {Azadi}, {Coil},
  {Freeman}, {de Groot}, {Leung}, {Price}, {Siana}, \& {Zick}}]{Shivaei2020}
{Shivaei}, I., {Reddy}, N., {Rieke}, G., {et~al.} 2020, \apj, 899, 117

\bibitem[{{Silva} {et~al.}(1998){Silva}, {Granato}, {Bressan}, \&
  {Danese}}]{Silva1998}
{Silva}, L., {Granato}, G.~L., {Bressan}, A., \& {Danese}, L. 1998, \apj, 509,
  103

\bibitem[{{Silverman} {et~al.}(2015){Silverman}, {Kashino}, {Sanders},
  {Kartaltepe}, {Arimoto}, {Renzini}, {Rodighiero}, {Daddi}, {Zahid}, {Nagao},
  {Kewley}, {Lilly}, {Sugiyama}, {Baronchelli}, {Capak}, {Carollo}, {Chu},
  {Hasinger}, {Ilbert}, {Juneau}, {Kajisawa}, {Koekemoer}, {Kovac}, {Le
  F{\`e}vre}, {Masters}, {McCracken}, {Onodera}, {Schulze}, {Scoville},
  {Strazzullo}, \& {Taniguchi}}]{Silverman2015}
{Silverman}, J.~D., {Kashino}, D., {Sanders}, D., {et~al.} 2015, \apjs, 220, 12

\bibitem[{{Smith} {et~al.}(2012){Smith}, {Dunne}, {da Cunha}, {Rowlands},
  {Maddox}, {Gomez}, {Bonfield}, {Charlot}, {Driver}, {Popescu}, {Tuffs},
  {Dunlop}, {Jarvis}, {Seymour}, {Symeonidis}, {Baes}, {Bourne}, {Clements},
  {Cooray}, {De Zotti}, {Dye}, {Eales}, {Scott}, {Verma}, {van der Werf},
  {Andrae}, {Auld}, {Buttiglione}, {Cava}, {Dariush}, {Fritz}, {Hopwood},
  {Ibar}, {Ivison}, {Kelvin}, {Madore}, {Pohlen}, {Rigby}, {Robotham},
  {Seibert}, \& {Temi}}]{Smith2012}
{Smith}, D.~J.~B., {Dunne}, L., {da Cunha}, E., {et~al.} 2012, \mnras, 427, 703

\bibitem[{{Stacey} {et~al.}(2010){Stacey}, {Hailey-Dunsheath}, {Ferkinhoff},
  {Nikola}, {Parshley}, {Benford}, {Staguhn}, \& {Fiolet}}]{Stacey2010}
{Stacey}, G.~J., {Hailey-Dunsheath}, S., {Ferkinhoff}, C., {et~al.} 2010, \apj,
  724, 957

\bibitem[{{Stark} {et~al.}(2013){Stark}, {Schenker}, {Ellis}, {Robertson},
  {McLure}, \& {Dunlop}}]{Stark2013}
{Stark}, D.~P., {Schenker}, M.~A., {Ellis}, R., {et~al.} 2013, \apj, 763, 129

\bibitem[{{Steidel} {et~al.}(2014){Steidel}, {Rudie}, {Strom}, {Pettini},
  {Reddy}, {Shapley}, {Trainor}, {Erb}, {Turner}, {Konidaris}, {Kulas}, {Mace},
  {Matthews}, \& {McLean}}]{Steidel2014}
{Steidel}, C.~C., {Rudie}, G.~C., {Strom}, A.~L., {et~al.} 2014, ApJ, 795, 165

\bibitem[{{Steidel} {et~al.}(2016){Steidel}, {Strom}, {Pettini}, {Rudie},
  {Reddy}, \& {Trainor}}]{Steidel2016}
{Steidel}, C.~C., {Strom}, A.~L., {Pettini}, M., {et~al.} 2016, \apj, 826, 159

\bibitem[{{Straughn} {et~al.}(2009){Straughn}, {Pirzkal}, {Meurer}, {Cohen},
  {Windhorst}, {Malhotra}, {Rhoads}, {Gardner}, {Hathi}, {Jansen}, {Grogin},
  {Panagia}, {di Serego Alighieri}, {Gronwall}, {Walsh}, {Pasquali}, \&
  {Xu}}]{Straughn2009}
{Straughn}, A.~N., {Pirzkal}, N., {Meurer}, G.~R., {et~al.} 2009, \aj, 138,
  1022

\bibitem[{{Strom} {et~al.}(2017){Strom}, {Steidel}, {Rudie}, {Trainor},
  {Pettini}, \& {Reddy}}]{Strom2017}
{Strom}, A.~L., {Steidel}, C.~C., {Rudie}, G.~C., {et~al.} 2017, \apj, 836, 164

\bibitem[{{Tang} {et~al.}(2021){Tang}, {Stark}, {Chevallard}, {Charlot},
  {Endsley}, \& {Congiu}}]{Tang2021}
{Tang}, M., {Stark}, D.~P., {Chevallard}, J., {et~al.} 2021, \mnras, 501, 3238

\bibitem[{{Teplitz} {et~al.}(2000){Teplitz}, {Malkan}, {Steidel}, {McLean},
  {Becklin}, {Figer}, {Gilbert}, {Graham}, {Larkin}, {Levenson}, \&
  {Wilcox}}]{Teplitz2000}
{Teplitz}, H.~I., {Malkan}, M.~A., {Steidel}, C.~C., {et~al.} 2000, \apj, 542,
  18

\bibitem[{{Theios} {et~al.}(2019){Theios}, {Steidel}, {Strom}, {Rudie},
  {Trainor}, \& {Reddy}}]{Theios2019}
{Theios}, R.~L., {Steidel}, C.~C., {Strom}, A.~L., {et~al.} 2019, \apj, 871,
  128

\bibitem[{{Thorne} {et~al.}(2021){Thorne}, {Robotham}, {Davies}, {Bellstedt},
  {Driver}, {Bravo}, {Bremer}, {Holwerda}, {Hopkins}, {Lagos}, {Phillipps},
  {Siudek}, {Taylor}, \& {Wright}}]{Thorne2021}
{Thorne}, J.~E., {Robotham}, A. S.~G., {Davies}, L. J.~M., {et~al.} 2021,
  \mnras, 505, 540

\bibitem[{{Topping} {et~al.}(2020){Topping}, {Shapley}, {Reddy}, {Sanders},
  {Coil}, {Kriek}, {Mobasher}, \& {Siana}}]{Topping2020}
{Topping}, M.~W., {Shapley}, A.~E., {Reddy}, N.~A., {et~al.} 2020, \mnras, 499,
  1652

\bibitem[{{Veilleux} \& {Osterbrock}(1987)}]{Veilleux1987}
{Veilleux}, S. \& {Osterbrock}, D.~E. 1987, \apjs, 63, 295

\bibitem[{{Vila-Costas} \& {Edmunds}(1993)}]{VilaCostas1993}
{Vila-Costas}, M.~B. \& {Edmunds}, M.~G. 1993, \mnras, 265, 199

\bibitem[{{Wells} {et~al.}(2015){Wells}, {Pel}, {Glasse}, {Wright},
  {Aitink-Kroes}, {Azzollini}, {Beard}, {Brandl}, {Gallie}, {Geers}, {Glauser},
  {Hastings}, {Henning}, {Jager}, {Justtanont}, {Kruizinga}, {Lahuis}, {Lee},
  {Martinez-Delgado}, {Mart{\'\i}nez-Galarza}, {Meijers}, {Morrison},
  {M{\"u}ller}, {Nakos}, {O'Sullivan}, {Oudenhuysen}, {Parr-Burman}, {Pauwels},
  {Rohloff}, {Schmalzl}, {Sykes}, {Thelen}, {van Dishoeck}, {Vandenbussche},
  {Venema}, {Visser}, {Waters}, \& {Wright}}]{Wells2015}
{Wells}, M., {Pel}, J.~W., {Glasse}, A., {et~al.} 2015, \pasp, 127, 646

\bibitem[{{Wild} {et~al.}(2007){Wild}, {Kauffmann}, {Heckman}, {Charlot},
  {Lemson}, {Brinchmann}, {Reichard}, \& {Pasquali}}]{Wild2007}
{Wild}, V., {Kauffmann}, G., {Heckman}, T., {et~al.} 2007, \mnras, 381, 543

\bibitem[{{Wright} {et~al.}(2015){Wright}, {Wright}, {Goodson}, {Rieke},
  {Aitink-Kroes}, {Amiaux}, {Aricha-Yanguas}, {Azzollini}, {Banks},
  {Barrado-Navascues}, {Belenguer-Davila}, {Bloemmart}, {Bouchet}, {Brandl},
  {Colina}, {Detre}, {Diaz-Catala}, {Eccleston}, {Friedman},
  {Garc{\'\i}a-Mar{\'\i}n}, {G{\"u}del}, {Glasse}, {Glauser}, {Greene},
  {Groezinger}, {Grundy}, {Hastings}, {Henning}, {Hofferbert}, {Hunter},
  {Jessen}, {Justtanont}, {Karnik}, {Khorrami}, {Krause}, {Labiano}, {Lagage},
  {Langer}, {Lemke}, {Lim}, {Lorenzo-Alvarez}, {Mazy}, {McGowan}, {Meixner},
  {Morris}, {Morrison}, {M{\"u}ller}, {rgaard-Nielson}, {Olofsson},
  {O'Sullivan}, {Pel}, {Penanen}, {Petach}, {Pye}, {Ray}, {Renotte}, {Renouf},
  {Ressler}, {Samara-Ratna}, {Scheithauer}, {Schneider}, {Shaughnessy},
  {Stevenson}, {Sukhatme}, {Swinyard}, {Sykes}, {Thatcher}, {Tikkanen}, {van
  Dishoeck}, {Waelkens}, {Walker}, {Wells}, \& {Zhender}}]{Wright2015}
{Wright}, G.~S., {Wright}, D., {Goodson}, G.~B., {et~al.} 2015, \pasp, 127, 595

\bibitem[{{Yabe} {et~al.}(2015){Yabe}, {Ohta}, {Akiyama}, {Bunker}, {Dalton},
  {Ellis}, {Glazebrook}, {Goto}, {Imanishi}, {Iwamuro}, {Okada}, {Shimizu},
  {Takato}, {Tamura}, {Tonegawa}, \& {Totani}}]{Yabe2015}
{Yabe}, K., {Ohta}, K., {Akiyama}, M., {et~al.} 2015, \pasj, 67, 102

\bibitem[{{Yuan} {et~al.}(2019){Yuan}, {Burgarella}, {Corre}, {Buat},
  {Boquien}, \& {Shen}}]{Yuan2019}
{Yuan}, F.-T., {Burgarella}, D., {Corre}, D., {et~al.} 2019, \aap, 631, A123

\bibitem[{{Zahid} {et~al.}(2017){Zahid}, {Kudritzki}, {Conroy}, {Andrews}, \&
  {Ho}}]{Zahid2017}
{Zahid}, H.~J., {Kudritzki}, R.-P., {Conroy}, C., {Andrews}, B., \& {Ho}, I.~T.
  2017, \apj, 847, 18

\end{thebibliography}

\end{document}